\documentclass[%
 aps,pra,reprint,superscriptaddress, 
 showpacs,showkeys,
 longbibliography,noeprint,nopreprintnumbers,footinbib,
 amsmath,amssymb,floatfix
]{revtex4-1}
\usepackage[utf8]{inputenc}
\usepackage[pdftex]{graphicx}
\usepackage[dvipsnames]{xcolor}

\usepackage{braket} 
\usepackage[normalem]{ulem} 
\usepackage{dcolumn}
\usepackage{bm}
\usepackage{bbold} 
\usepackage{hyperref} 
\usepackage{comment}
\usepackage{dsfont}
\usepackage{bibunits}
\defaultbibliographystyle{apsrev4-1}
\defaultbibliography{hubbard-biblio}

\newcommand{\Npairs}{\ensuremath{N_\text{pairs}}}
\newcommand{\Kfin}{\ensuremath{K_f}}

\newcommand{\ii}{\textrm{i}}
\newcommand{\ee}{\textrm{e}}

\newcommand{\eqr}[1]{Eq.~(\ref{#1})}
\newcommand{\fir}[1]{Fig.~\ref{#1}}

\newcommand{\secr}[1]{Sec.~\ref{#1}}

\newcommand{\tj}{$t$-$J$\ }
\newcommand{\tja}{{\lowercase{$t$}-$J$-$\alpha$\ }}

\graphicspath{{./}}

\begin{document}

\def\mytitle{Controlling magnetic correlations in a driven Hubbard system far from half-filling}
\title{\mytitle}

\author{Hongmin Gao}
 \email{hongmin.gao@physics.ox.ac.uk}
 \affiliation{Clarendon Laboratory, University of Oxford, Parks Road, Oxford OX1 3PU, United Kingdom}

\author{Jonathan R. Coulthard}
 \affiliation{Clarendon Laboratory, University of Oxford, Parks Road, Oxford OX1 3PU, United Kingdom}

\author{Dieter Jaksch}
 \affiliation{Clarendon Laboratory, University of Oxford, Parks Road, Oxford OX1 3PU, United Kingdom}
\affiliation{center for Quantum Technologies, National University of Singapore, 3 Science Drive 2, 117543, Singapore}

\author{Jordi Mur-Petit}
 \email{jordi.murpetit@physics.ox.ac.uk}
 \affiliation{Clarendon Laboratory, University of Oxford, Parks Road, Oxford OX1 3PU, United Kingdom}

\date{\today}

\begin{abstract}
We propose using ultracold fermionic atoms trapped in a periodically shaken optical lattice as a quantum simulator of the \tj Hamiltonian, which describes the dynamics in doped antiferromagnets and is thought to be relevant to the problem of high-temperature superconductivity in the cuprates.
We show analytically that the effective Hamiltonian describing this system
for off-resonant driving
is the \tj model with additional pair hopping terms, whose parameters can all be controlled by the drive.
We then demonstrate numerically using tensor network methods for a 1D lattice that a slow modification of the driving strength allows near-adiabatic transfer of the system from the ground state of the underlying Hubbard model to the ground state of the effective \tj Hamiltonian. 
Finally, we report exact diagonalization calculations illustrating the control achievable on the dynamics of spin-singlet pairs in 2D lattices utilising this technique with current cold-atom quantum-simulation technology.
These results open new routes to explore the interplay between density and spin in strongly-correlated fermionic systems through their out-of-equilibrium dynamics.
\end{abstract}

\maketitle

\begin{bibunit}


\section{Introduction}
Thirty years after the discovery of copper-oxide high-temperature superconductors~\cite{Bednorz1986}, we still do not have a complete theoretical understanding of the nature of the low-energy physics of these materials. The essential physics at low doping is thought to be dominated by a competition between the antiferromagnetic ground state of the Heisenberg model realized at half-filling, and the hopping of single holes (or electrons) between nearest-neighbor (NN) sites for small doping, effects that are contained in the \tj model~\cite{Lee2006}. The ground state of this model features a tight competition between $d$-wave pairing correlations and a variety of inhomogeneous phases~\cite{Fradkin2015}, including stripes, checkerboard phases, and others, which appear surprisingly close to each other energetically in calculations by completely different methods~\cite{Corboz2014, Zheng2017, Huang2017science, Dodaro2017, Nocera2017, Huang2018npjqm, Jiang2018, Jiang2019}.
The difficulty to move forward on this issue makes it desirable to consider a different approach, with quantum simulation based on ultracold atom setups offering an ideal platform to unravel the generic features of the model from specific properties pertaining to any particular material.

In recent years, ultracold atoms trapped in optical lattices have become a mature platform for investigating Hamiltonians relevant for condensed-matter physics \cite{LewensBook, Gross2017}. Several laboratories have now reported quantum simulations of the fermionic Hubbard model~\cite{Greif2013, Hart2015, Cocchi2016, Parsons2016, Boll2016, Cheuk2016, Mazurenko2017, Chiu2018, Salomon2019}, including the measurement of magnetic correlations with single-site resolution; see, e.g.~, Refs.~\cite{Parsons2016, Boll2016, Cheuk2016, Mazurenko2017, Hilker2017, Chiu2018, Salomon2019}.
An interesting feature of these setups is their power to access the real-time dynamics of strongly correlated systems, in addition to a plethora of methods to measure spectral~\cite{Ernst2010,Veeravalli2008,Usui2018} and particle-correlation properties~\cite{Kollath2007,Endres2013,Streif2016}.

In this article we propose a scheme to quantum simulate the \tja model, a variant of the \tj model including terms describing the motion of spin-singlet pairs that are expected to play a significant role away from half-filling~\cite{AuerbachBook,Sorella2002,Coulthard2017}.
Our proposal is based on the principle of Floquet engineering~\cite{Bukov2015, Eckardt2017}. Specifically, we study a fermionic Hubbard model in a periodically shaken lattice \cite{Jotzu2014, Desbuquois2017, Gorg2018, Messer2018}. 
Floquet engineering has been used in recent experiments to create artificial gauge fields for neutral atoms~\cite{Aidelsburger2011, Struck2012, Tai2017}, and to realize the Haldane model of a topological insulator with both ultracold atoms~\cite{Jotzu2014} and graphene sheets~\cite{McIver2019}.
There are also theoretical proposals that use Floquet driving to control quantum magnets \cite{Mentink2015}, enhance superconducting fluctuations \cite{Coulthard2017,Dasari2018} and simulate a range of strongly correlated models, including the $t$-$J$ model \cite{Bermudez2015,Bukov2016} and the $XXZ$ model \cite{Duan2003}, among other applications \cite{Eckardt2017,Hofstetter2018}. Meanwhile, signatures of superconductivity have been observed in a range of different solid-state materials when periodically driven with an ultrafast laser pulse \cite{Kaiser2014a, Mankowsky2014, Mitrano2016}, which might also be explained by Floquet-modified Hamiltonians~\cite{Coulthard2017}. 

Here, we demonstrate that tuning the driving parameters of a Hubbard model (frequency, amplitude, directionality) provides control over all the parameters of the \tja model.
This opens the door to probing experimentally a range of hitherto unexplored regions of parameter space of this model. 
More specifically, in the usual \tj model as a limit of the repulsive Hubbard model, the superexchange interaction strength, $J$, is always much smaller than NN hopping strength, $t$, and the strength of singlet pair hopping (irrelevant at half-filling) cannot be tuned relative to $J$. The driving, however, allows us to enter unusual regimes where $J > t$ and the pair-hopping strength can be tuned, 
which lead to exotic behaviors such as an anomalous spin-charge separation regime due to the increased competition between superconducting and magnetic correlations~\cite{XXXL}.
In this work, we 
identify a driving regime that results in 
complete blocking of single-particle propagation 
accompanied with a directed coherent motion of spin-singlet pairs. 
These findings are supported by extensive numerical calculations---combining tensor network and exact diagonalization methods for one- and two-dimensional systems---exploring high-energy excitations of a low-filling fermion lattice system. 
These predictions can be readily tested utilizing current
cold-atom quantum-simulation 
setups~\cite{
Desbuquois2017, Gorg2018, Messer2018, Chiu2018, Chiu2019, Salomon2019, Vijayan20}

\section{Floquet quantum simulator for the \tja model}\label{sec:floquet-simulator}

In this section, we first show how the \tja model is engineered by periodically driving a strongly repulsive Hubbard model, then we present numerical results on slow ramping of the driving amplitude, while keeping the driving frequency constant, as a possible means to prepare a ground state of an effective \tja model in this Floquet quantum simulator. 

\subsection{Floquet engineering of the \tja Hamiltonian}\label{sec:floquet-engineering}

\begin{figure}[tb!] 
\centering
\includegraphics[width=.9\linewidth]{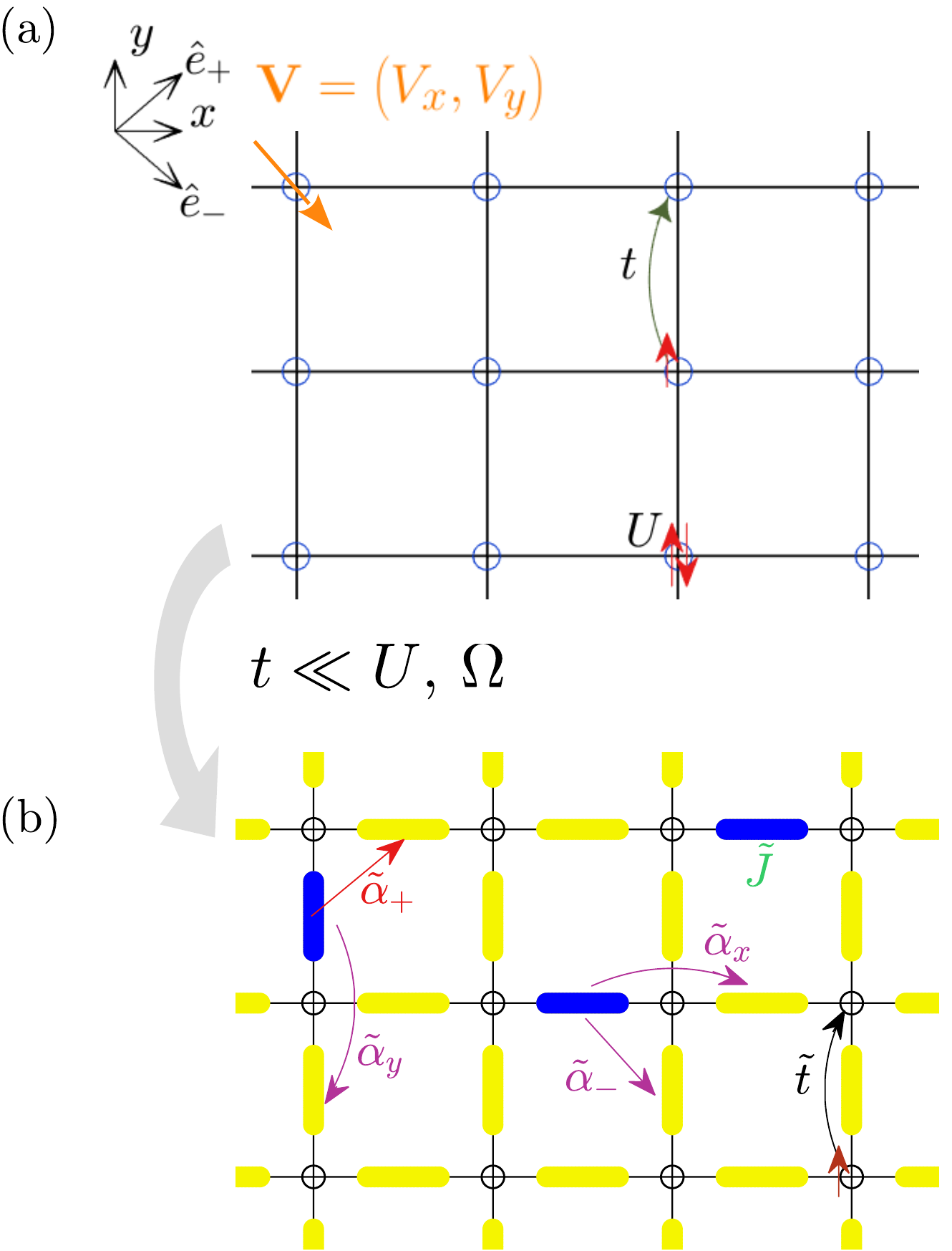}
\caption{%
 (a) Hubbard model, with hopping parameter $t$ and on-site repulsion $U$, on a driven lattice. Driving is achieved through shaking of the square lattice at frequency $\Omega$ with amplitude vector $\bm{V}$ (see text). 
 (b) When the driving frequency satisfies $t \ll \Omega, U$, the dynamics of the low-energy sector can be effectively described by the \tja model,~\eqr{eq:tja}, with an effective single-particle hopping $\tilde{t}$, superexchange energy $\tilde{J}$, and anisotropic hopping rates for spin-singlet pairs $\tilde{\alpha}_\lambda$ in the $\lambda=\{\bm{x},\bm{y},\pm\}$ directions, with  $\mathbf{e}_{\pm}=(\bm{x}\pm\bm{y})/\sqrt{2}$. Blue shapes represent lattice bonds occupied by nearest-neighbor singlet pairs, while yellow shapes represent unoccupied bonds. 
}
\label{fig:fig1-lattice}
\end{figure}

The setup we consider is sketched in Fig.~\ref{fig:fig1-lattice}(a). Fermionic atoms tunnel at a rate $t$ between the NN sites of a two-dimensional (2D) optical lattice, and have repulsive on-site interactions $U>0$. 
The system is driven by shaking the lattice; this can be implemented, e.g., by interfering two orthogonal laser beams with their reflections off mirrors mounted on piezoelectric actuators vibrating at a frequency $\Omega$~\cite{Jotzu2014,Desbuquois2017,Gorg2018,Messer2018}; lattice shaking can also be achieved by periodically modulating a superlattice in each spatial direction using acousto‐optical modulators~\cite{Eckardt2005,Struck2011,Struck2012,Parker2013,Reitter2017}. 
In the frame comoving with the lattice, the dynamics of the atoms is governed by the driven Hubbard Hamiltonian, $\hat{H} = \hat{H}_{\text{Hub}} + \hat{H}_{\text{drive}}(\tau)$, where
\begin{align}
    \hat{H}_{\text{Hub}} &= \hat{H}_{\text{hop}}(t) + \hat{H}_{\text{int}}(U) \nonumber \\
    &= -t \sum_{\langle ij \rangle \sigma} \left( \hat{c}^\dagger_{i\sigma} \hat{c}_{j\sigma} + \text{H.c.} \right) + U \sum_{i} \hat{n}_{i\uparrow}\hat{n}_{i\downarrow}
    \label{eq:Hub}
\end{align}
describes a fermionic Hubbard model, and 
\begin{equation}
 \hat{H}_{\text{drive}}(\tau)
 = \sum_{i} \left(\bm{V} \cdot \mathbf{r}_i \right) \hat{n}_i \sin \left(\Omega \tau\right) ,
 \label{eq:Hdrive}
\end{equation}
shakes the lattice with driving amplitude vector $\bm{V}=(V_x, V_y)$.
Here $\hat{c}_{i\sigma}$ is the fermionic annihilation operator for a spin-$\sigma$ fermion ($\sigma = \uparrow,\downarrow$) on site $i$ located at position vector $\mathbf{r}_i$, $\hat{n}_{i,\sigma} = \hat{c}^\dagger_{i\sigma} \hat{c}_{i\sigma}$ and $\hat{n}_i = \hat{n}_{i\uparrow}+\hat{n}_{i\downarrow}$ are the densities at site $i$. 
Physically, Eq.~\eqref{eq:Hdrive} describes an in-phase shaking of the whole lattice, in contrast with Ref.~\cite{Coulthard2017}, which considered driving out-of-phase and with different amplitudes the two sublattices (even and odd sites) of a 1D lattice.
(From now on, we set $\hbar=1$ and choose the lattice constant $a$ as our unit of length.)

\begin{figure*}[tbh] 
\centering
\includegraphics[width=0.32\linewidth]{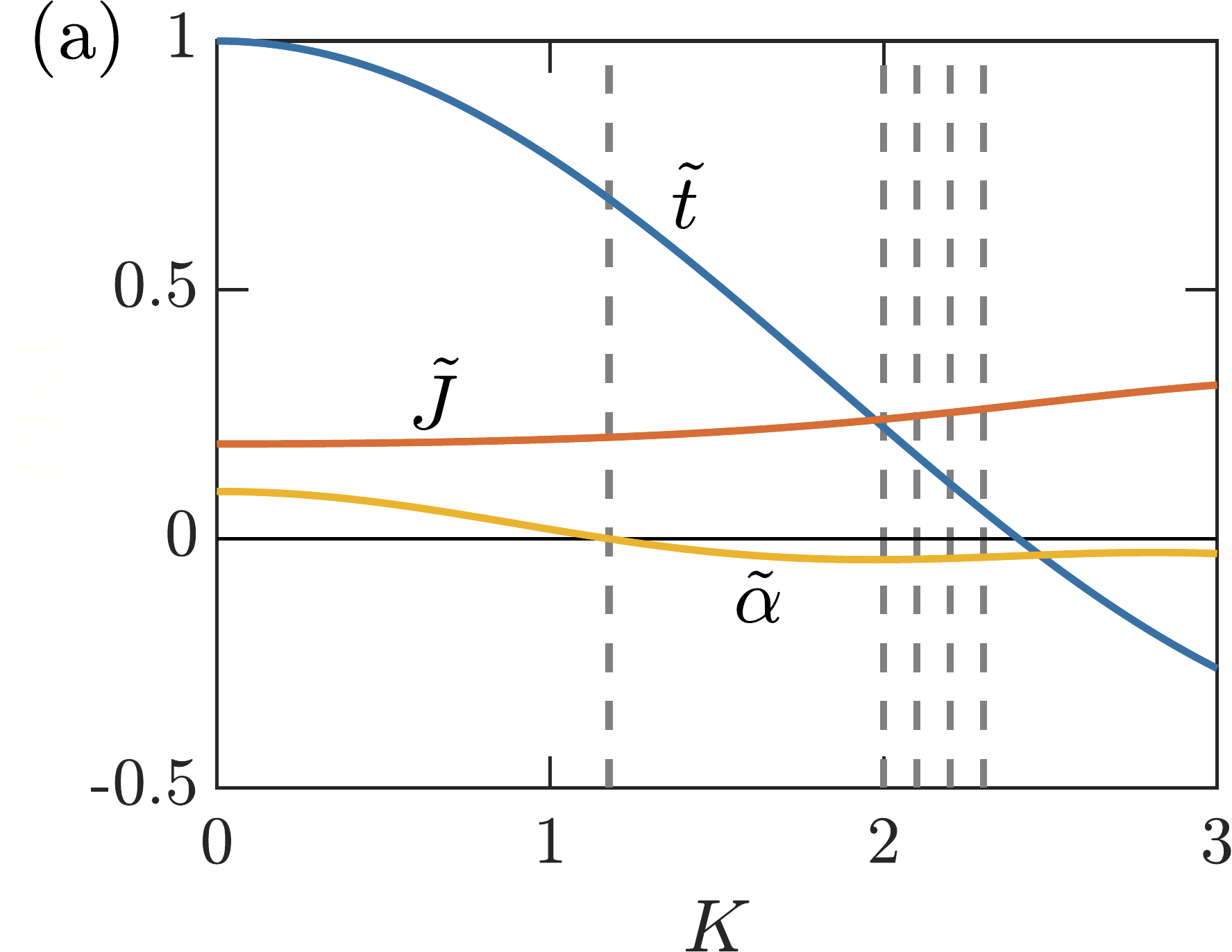} 
\includegraphics[width=0.32\linewidth]{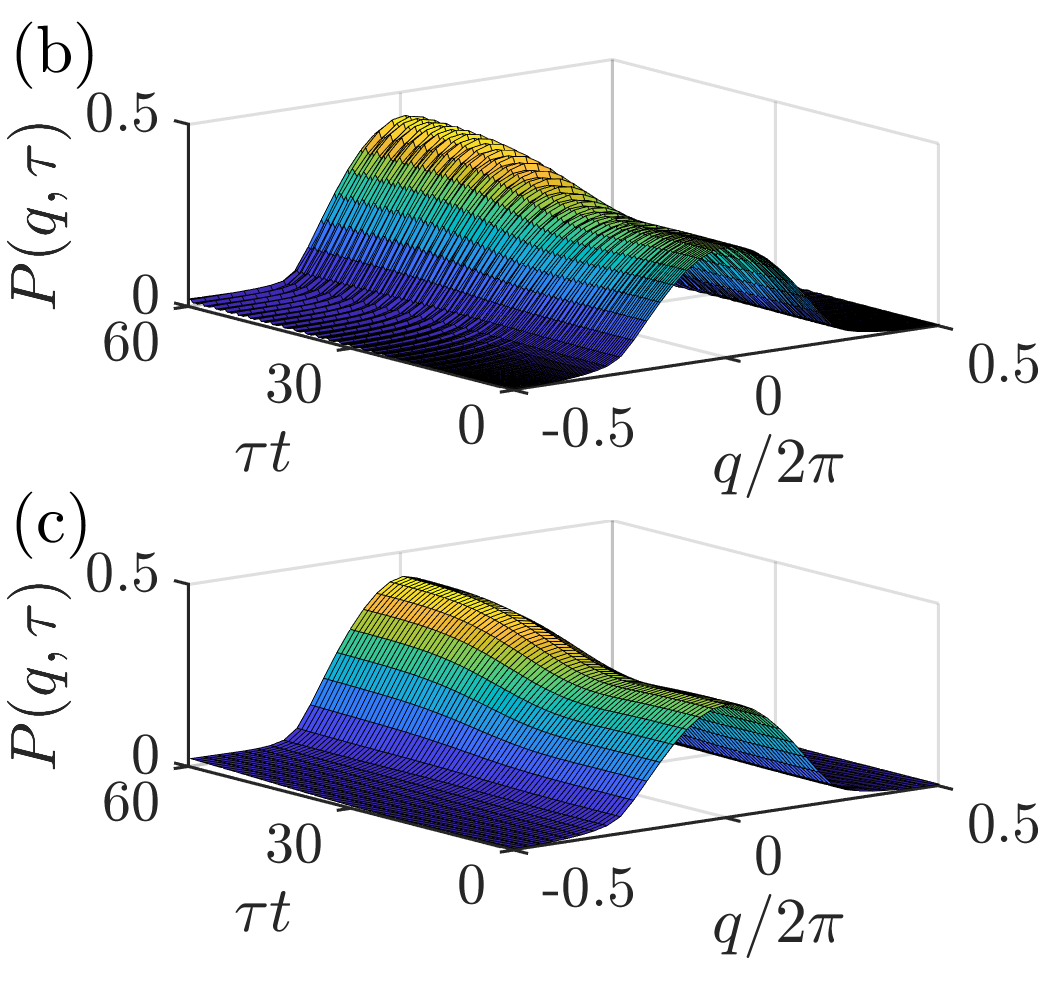}
\includegraphics[width=0.32\linewidth]{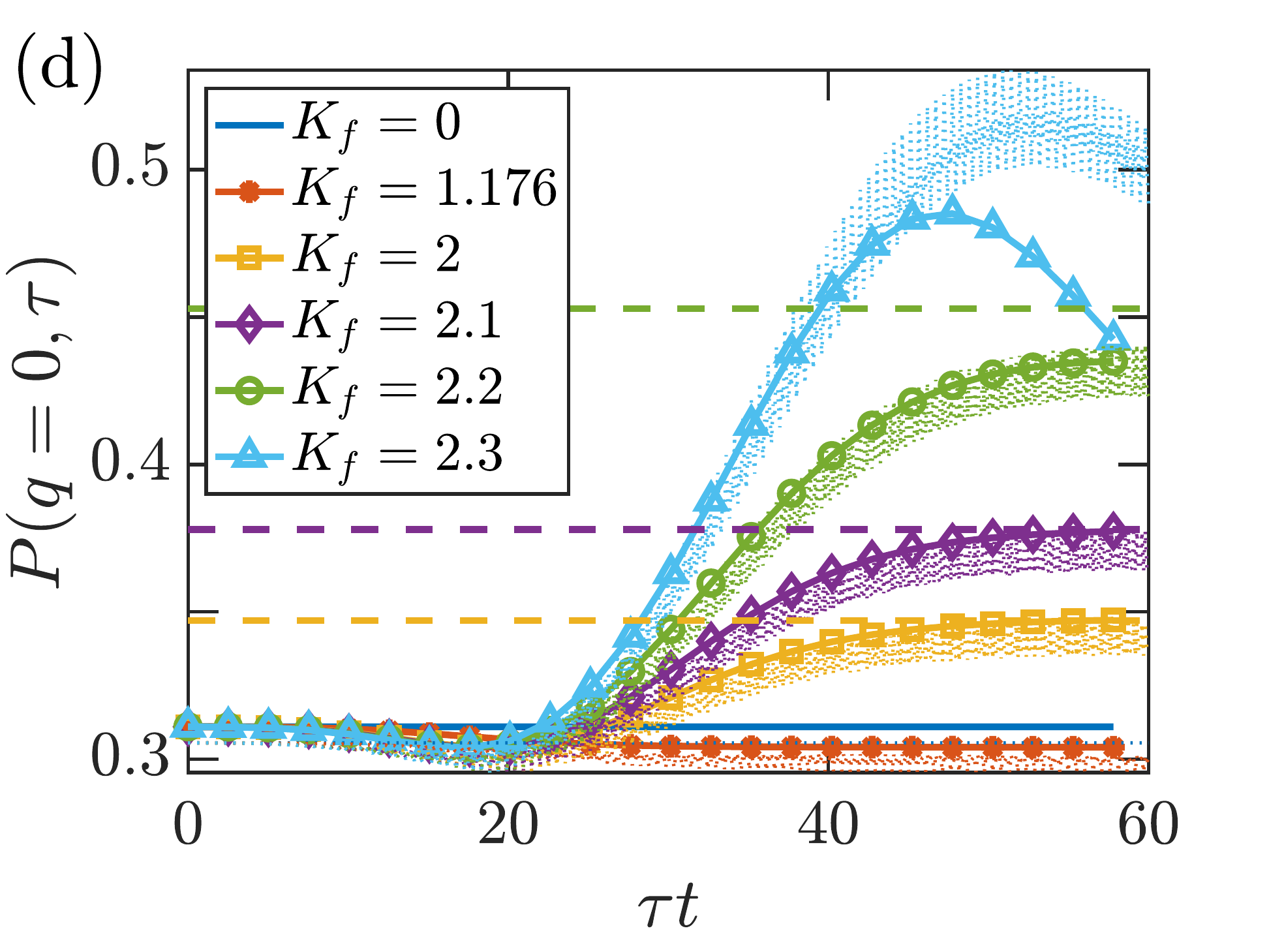} 
\caption{%
 (a) Effective parameters of the 1D \tja model for $K_x=K$.
 Vertical dashed lines indicate the values $K=K_f$ used in panel (d). 
 (b),(c) Singlet structure factor, $P(q,\tau)$, for (b) the driven Hubbard model and (c) the effective \tja model, with $U = 21t$, $\Omega = 6t$. The driving strength is smoothly ramped up to its final value, $\Kfin = 2.2$, according to
 $K(\tau) = \Kfin [ \tanh(\tau_0/\tau_{\mathrm{ramp}}) + \tanh((\tau-\tau_0)/\tau_{\mathrm{ramp}})] / [ \tanh(\tau_0/\tau_{\mathrm{ramp}}) + 1]$; in our simulations, we use $\tau_0t = 15$ and $\tau_{\mathrm{ramp}} t = 12.5$ to set the offset time and the duration of the ramp, respectively. The system size is $L=32$ sites, with eight ``up'' and eight ``down'' fermions. 
 (d) The $q=0$ component of $P(q,\tau)$ for the driven Hubbard model (dotted lines oscillating at frequency $\Omega$) and for the \tja model (solid lines with symbols) for various final driving strengths $\Kfin$. The dashed lines indicate the values of $P(q=0,\tau)$ for the ground states of the target final \tja model.
}
\label{fig:fig2-QuenchComparison}
\end{figure*}

In the limit of strong interactions, $U \gg t$, the low-energy dynamics of the static Hubbard Hamiltonian~\eqr{eq:Hub} are constrained to the manifold spanned by states without double occupancy~\cite{AuerbachBook,Lee2006}. 
One can then introduce the \tja Hamiltonian,
\begin{equation} \label{eq:tja}
 \hat{H}_{tJ\alpha}
 = \mathcal{P}_0 \Big\lbrace \hat{H}_{\text{hop}}(t) + \hat{H}_{\text{ex}}(J) 
 + \hat{H}_{\text{pair}}(\{ \alpha_{ijk} \} ) \Big\rbrace \mathcal{P}_0
\end{equation}
as the effective Hamiltonian of the system~\cite{AuerbachBook,Sorella2002,Coulthard2017}. 
Here $\mathcal{P}_0 = \prod_{i} (1-\hat{n}_{i\uparrow} \hat{n}_{i\downarrow})$ projects out states with nonzero double occupancies; 
$
\hat{H}_{\text{ex}}(J) = -\sum_{\langle ij \rangle} J \hat{b}^\dagger_{ij} \hat{b}_{ij}
$ 
is the superexchange contribution, by which NN opposite spins switch their positions, and which results in a lower energy for a spin-singlet pair straddling NN sites $i$ and $j$ [created by the operator 
$
 \hat{b}^\dagger_{ij} = \frac{1}{\sqrt{2}}
 (\hat{c}^\dagger_{i\uparrow}\hat{c}^\dagger_{j\downarrow}-\hat{c}^\dagger_{i\downarrow}\hat{c}^\dagger_{j\uparrow})
$], 
by an energy $J$ with respect to a spin-triplet pair. Finally, 
\begin{align}
  \hat{H}_{\text{pair}}(\lbrace \alpha_{ijk} \rbrace)
  &= - \sum\limits^{i\ne k}_{\langle ijk \rangle} \alpha_{ijk} \hat{b}^\dagger_{ij} \hat{b}_{jk} + \text{H.c. ,} 
  \label{eq:Hpair}
\end{align}
describes processes by which a spin-singlet hops between nearby lattice links ($\langle jk \rangle \rightarrow \langle ij \rangle$), see Fig.~\ref{fig:fig1-lattice}(b); such processes are sometimes referred to as density-dependent hoppings in the literature~\cite{footnoteJabsorbed}.

We turn now to the periodically driven Hubbard Hamiltonian, still in the strong-coupling limit, $U/t\gg 1$. When the driving frequency is high ($\Omega \gg t$) and far away from any resonance ($|U+m\Omega| \gg t \text{ }\forall m \in \mathds{Z} $), we apply perturbation theory in the Floquet basis to find that the dynamics of the system are described by an effective Hamiltonian of the form of the \tja Hamiltonian~\eqr{eq:tja}, with driving-dependent parameters~\cite{Coulthard2017} (see Appendix~\ref{apxsec: FullFloquetPertubativeAnalysis} for details on the derivation).
For the case of the square lattice in Fig.~\ref{fig:fig1-lattice}(a), the effective parameters can become anisotropic. The effective single-particle hopping rate along the $\eta=\{x,y\}$ direction is
\begin{equation}
  \tilde{t}_{\eta} = t \mathcal{J}_0(K_\eta) \:,
  \label{eq:ttilde}
\end{equation}
where $K_\eta=|V_\eta|/\Omega$, and ${\cal J}_m(K)$ is the $m$th-order Bessel function of the first kind.
The superexchange parameter between NN sites separated along $\eta=\{x,y\}$ reads 
$
  \tilde{J}_{\eta} = 4t^2 \sum_{m} \mathcal{J}_m^2(K_\eta) / (U+m\Omega) .
$ 
Finally, the pair-hopping Hamiltonian~\eqr{eq:Hpair} becomes anisotropic as well, with four different couplings, namely
\begin{align}
\tilde\alpha_{ijk} =
 \left\{ \begin{array}{ll}
   \tilde{\alpha}_\eta = 2t^2\sum_{m}\frac{\mathcal{J}_m(K_\eta)\mathcal{J}_{-m}(K_\eta)}{U+m\Omega}
   \text{,}&\mathbf{r}_i - \mathbf{r}_k \propto \bm{\eta} \\
   \tilde{\alpha}_{\pm} = 2t^2\sum_{m}\frac{\mathcal{J}_m(K_x)\mathcal{J}_{\pm m}(K_y)}{U+m\Omega}    \text{,}&\mathbf{r}_i - \mathbf{r}_k \propto \bm{e}_{\pm}
\end{array} \right..
\label{eq:alphas}
\end{align}
Here $\bm{e}_{\pm}=(\bm{x}\pm\bm{y})/\sqrt{2}$.
Equations.~\eqref{eq:ttilde} and \eqref{eq:alphas} indicate that tuning the amplitudes and frequency of the lattice driving provides control on the ratios between all parameters of the \tja model; this is illustrated in \fir{fig:fig2-QuenchComparison}(a) for a 1D model analyzed in Sec~\ref{sec:1dNumerics}. In particular, we are able to tune the pair-hopping rate $\tilde{\alpha}$ to 0, obtaining the standard \tj model, or alternatively reach the regime $J > t$, which favours the formation of nearest-neighbor singlets. 
We can also generate negative pair-hopping amplitudes, $\tilde\alpha(K \gtrsim 1.2) < 0$, which we expect will have an impact on particle transport and spin-spin correlations, by analogy to the effect of next-to-nearest-neighbor hopping in moderately-doped Hubbard systems~\cite{Jiang2019}.

\subsection{Adiabatic preparation of ground states of the \tja model\label{sec:1dNumerics}}

We demonstrate the validity and tunability of the \tja model,~\eqr{eq:tja}, as a description of the driven Hubbard model by comparing the result of evolving both in real time. 
For simplicity, we consider the case of a one-dimensional chain shaken along its length, $L$. The effective parameters of the corresponding \tja model are obtained from Eqs.~\eqref{eq:ttilde} and \eqref{eq:alphas}. The effective tunneling $\tilde{t} = t \mathcal{J}_0(K)$, the effective superexchange coupling
$\tilde{J} = 4t^2 \sum_{m} \mathcal{J}_m^2(K)/(U+m\Omega)$,
and the effective pair-hopping rate
$\tilde{\alpha} = 2t^2\sum_{m} \mathcal{J}_m(K) \mathcal{J}_{-m}(K)/(U+m\Omega)$,
with $K = |V|/\Omega$. In the limit $\Omega \ll U$, these expressions reduce to $\tilde{J} \approx J$ and $\tilde{\alpha} \approx J \mathcal{J}_{0}(2K)/2$.

For these numerical calculations, we employ tensor network methods as implemented in the open source Tensor Network Theory library \cite{AlAssam2017}. 
To demonstrate near-adiabatic transfer from the ground state of the underlying Hubbard model to the ground state of the effective \tja model,
we first compute ground states of the Hubbard model and the corresponding \tja model using density-matrix renomalization group (DMRG) calculations \cite{White1992, Schollwock2010}. 
Then, we evolve these states using the time evolving block decimation (TEBD) algorithm \cite{Vidal2004} while ramping up the driving strength as a function of time $\tau$ according to a function that smoothly interpolates from $K=0$ to various final strengths $K=\Kfin$ (see caption of \fir{fig:fig2-QuenchComparison} for details). 
For the \tja model, we determine, at each time step, the instantaneous effective parameters as 
$\tilde{t}(\tau) = \tilde{t}[K(\tau)]$, 
$\tilde{J}(\tau) = \tilde{J}[K(\tau)]$
and
$\tilde{\alpha}(\tau) = \tilde{\alpha}[K(\tau)]$,
and evolve the state with the corresponding time-dependent \tja Hamiltonian.

To characterize the states of both simulations and compare them, we compute a variety of correlations. The most significant changes due to the driving occur in the singlet-singlet correlation function,
$P_{j,k}(\tau) = \langle \hat{b}^{\dagger}_{j,j+1} \hat{b}_{k,k+1} \rangle$, 
and its discrete Fourier transform, the singlet structure factor
$
    P(q,\tau) = \frac{1}{L} \sum_{j, k} \ee^{\ii q (j-k)} P_{j,k}(\tau),
$ 
where $q$ is the dimensionless quasimomentum of the singlets. $P(q,\tau)$ can be interpreted as the momentum distribution of singlets at time $\tau$~\cite{Coulthard2017}; a narrow peak around $P(q=0,\tau)$ in the thermodynamic limit would signal off-diagonal long-range order
of the pairs, corresponding to superconducting correlations~\cite{Sorella2002}.

We present in \fir{fig:fig2-QuenchComparison}(b) the singlet structure factor as a function of $q$ and $\tau$ for the driven Hubbard model with final driving strength $\Kfin=2.2$, such that $\tilde{t}$ in the associated \tja model is suppressed relative to $\tilde{J}$ and $\tilde\alpha$ [cf.~\fir{fig:fig2-QuenchComparison}(a)].
We observe that the driving results in an increase in the magnitude of the peak around $q=0$ as time advances.
The evolution of the corresponding \tja model [\fir{fig:fig2-QuenchComparison}(c)] shows a very good agreement with these findings. This is further illustrated in \fir{fig:fig2-QuenchComparison}(d), where we show the time-dependence of the height of the peak, $P(q=0,\tau)$, for calculations of both models for a broad range of final driving strengths, $\Kfin$.
We observe that the Hubbard results oscillate rapidly, at a frequency $\Omega$; the values of the \tja simulations match them very well at the stroboscopic times $\tau = k \times 2\pi/\Omega$ ($k=0,1,\ldots$), as expected.

Additionally, we plot as dashed horizontal lines in \fir{fig:fig2-QuenchComparison}(d) the values of $P(q=0,\tau)$ computed for the \emph{ground states} of the corresponding target \tja models, i.e., those with $\tilde{t} = \tilde{t}(K_f)$, $\tilde{J} = \tilde{J}(K_f)$ and $\tilde{\alpha} = \tilde{\alpha}(K_f)$.
We find that for moderate driving strengths $\Kfin \lesssim 2.2$, a slow ramping of the driving strength results in a near-complete loading of the system into the ground state of the effective Hamiltonian. 
We note that for $\Kfin \gtrsim 2.25$, the ground state of the effective Hamiltonian at quarter filling is phase separated~\cite{Kf225footnote}; as such, we cannot reach this phase by driving. For $\Kfin = 2.2$, the proximity to phase separation underpins the slow convergence of $P(q=0,\tau)$ to the \tja ground state value. This also results in significant deviations between the driven Hubbard model and \tja\ model results for $\Kfin = 2.3$. 
We find completely analogous results for the spin structure factor, which we present in Appendix~\ref{apxsec:tjavalidity}.

\begin{figure}[tb]
    \centering
    \includegraphics[width=\linewidth]{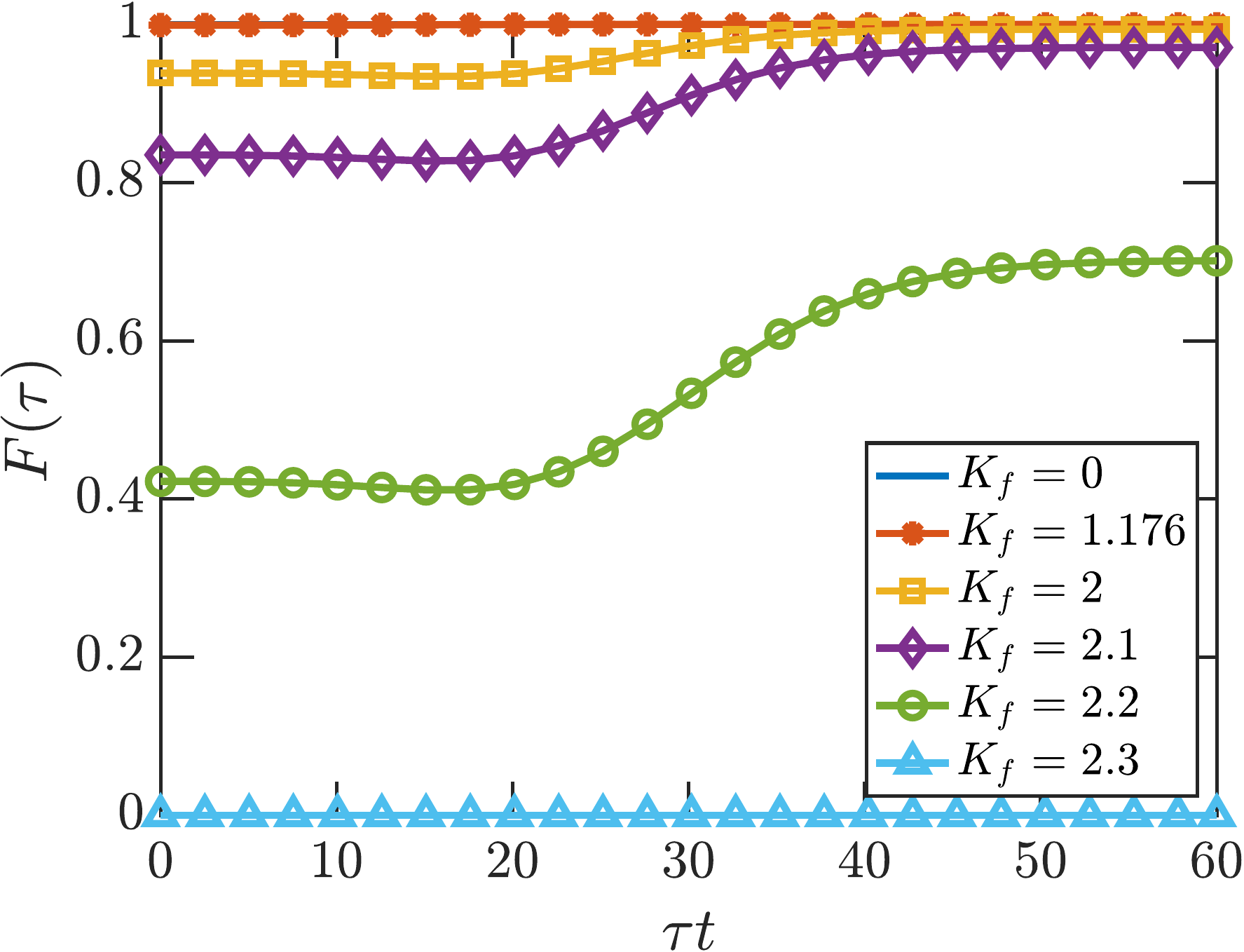}
    \caption{%
    Fidelity between the time-evolved quantum state of the \tja model relative
    and the ground state of the ``target'' \tja model, Eq.~\eqref{eq:target-fidelity}, as a function of driving time.
    Note that the lines for $K_f=0$ and $K_f=1.176$ overlap each other at the top of the plot.
    Hubbard model and drive parameters are the same as in \fir{fig:fig2-QuenchComparison}.
    }
    \label{fig:fig3-target-fidelity}
\end{figure}


As a more stringent test of the quality of ground state transfer, in \fir{fig:fig3-target-fidelity}, we examine the fidelity between the time-evolved \tja state, and the target \tja state, 
\begin{equation}
    F(\tau) = \left| \braket{ \Psi_{\tilde{t}\tilde{J}\tilde{\alpha}} | \psi(\tau) } \right|^2,
    \label{eq:target-fidelity}
\end{equation}
where $| \psi_{tj\alpha}(\tau) \rangle$ is the state at time $\tau$ evolved from the ground state of $K=0$, and $| \Psi_{\tilde{t}\tilde{J}\tilde{\alpha}} \rangle$ is the “target state", i.e., the ground state of the \tja model with effective parameters $\tilde{t}$, $\tilde{J}$ and $\tilde{\alpha}$ for the final driving strength, $K_f$, according to Eqs.~\eqref{eq:ttilde}-\eqref{eq:alphas}. For driving strengths up to $K_{f} = 2.1$, we get fidelities $F\geq 97\%$.
The fidelity for $K_f=2.2$ is reduced due to the proximity to phase separation. 
The ground state for $K_f=2.3$ is already phase separated, which explains the vanishing overlap, as explained above.
We discuss in Appendix~\ref{apx:heating} an alternative assessment of the adiabaticity of this loading procedure, based on reversing the evolution of $K(\tau)$ as in Ref.~\cite{Gorg2018}.
We remark that, given the many-body character of the system, we expect the fidelity to go down quickly with system size. In such cases, the adiabaticity of the loading process may be more readily analysed through higher-order observables like the singlet structure in Fig.~\ref{fig:fig2-QuenchComparison}, or the spin structure factor in Fig.~\ref{fig:fig9-quench-spin-structure} in  Appendix~\ref{apxsec:tjavalidity}.

We note that, in choosing the functional form of the driving ramp, $K(\tau)$, we have made no particular effort to optimise the adiabatic transfer of population to the target state, and it is likely that much more effective schemes can be found. For example, looking at \fir{fig:fig2-QuenchComparison}(a), the effective parameters increase slowly up to $K\approx 1$, and quicker later on. It may, therefore, be desirable to increase the driving strength more quickly up to $K(\tau)\approx 1$, and more slowly later, 
if one aims to create with high fidelity the ground state of the driven system. 

In summary, in this section we have demonstrated the ability (i) to engineer the \tja model by driving a Hubbard system, (ii) to control the parameters of the \tja model, and (iii) to adiabatically prepare an initial state of this model for effective parameters $\{ \tilde{t}, \tilde{J}, \tilde{\alpha} \}$ corresponding to any $K \lesssim 2.2$
by starting from the ground state of the undriven Hubbard model.
Due to computational constraints, our simulations on (iii) have focused on a 1D system; experimental observations support that quasi-adiabatic loading from the static to a driven Hubbard model can be realised in 2D lattices in similar timescales~\cite{Gorg2018}.

\section{Controlled singlet hopping on a square lattice}\label{subsec:2D}

We consider now the driven 2D system, and demonstrate that control over magnetic correlations can be achieved by tuning the anisotropy of the effective parameters ($\tilde{t}, \tilde{J}, \tilde{\alpha}$).
To this end, we build on our demonstrated capacity to engineer the \tja model and control its parameters 
(Sec.~\ref{sec:floquet-simulator}), and consider initial states with either one or two singlet pairs in the driven lattice. 
Contrary to the ground states considered in Sec.~\ref{sec:1dNumerics}, these initial states are highly non-equilibrium configurations: the single-particle hopping term quickly delocalizes the fermions (smearing the singlet) and also causes the appearance of some double occupations (``doublons''). 
We will now show that a suitable choice of lattice driving prevents the delocalization of the fermions, and leads instead to a controlled, directed hopping of the bound singlet pair.

In practice, this prediction could be tested in a three-step process: 
first fill the lattice with singlet pairs as demonstrated in Ref.~\cite{Gorg2018}; second, selectively remove the undesired pairs using existing single-atom-resolution techniques on ultracold quantum gases~\cite{Ott2016};
finally, switch on the lattice driving and monitor the system dynamics, e.g., via measuring the total number of singlets and doublons~\cite{Jotzu2014, Desbuquois2017, Gorg2018, Messer2018} or observing with single-site and single-spin resolution~\cite{Bakr2009, Sherson2010, Weitenberg2011}.

For concreteness, we focus first on driving of the square lattice with amplitudes $K_x=-K_y=K$, i.e. $\bm{V}\propto \bm{e}_-$. This choice of driving parameters leads to a particular anisotropy of the pair-hopping amplitudes that we exploit to control the hopping of singlets on the lattice; the generality of our findings is supported by analogous results for the brickwall lattice~\cite{Jotzu2014,Gorg2018,Messer2018} summarised in \secr{sec: Brickwall lattice}.
Additionally, in Appendix~\ref{apx:heating} we show that this driving does not lead to substantial heating; in particular, the system does not suffer from Floquet heating to `infinite' temperature~\cite{DAlessio2014, Lazarides2014pre, Genske2015, Herrmann2017, Weidinger2017} in the timescales under consideration.

\begin{figure*}[tb] 
 \centering
 \includegraphics[width=\linewidth]{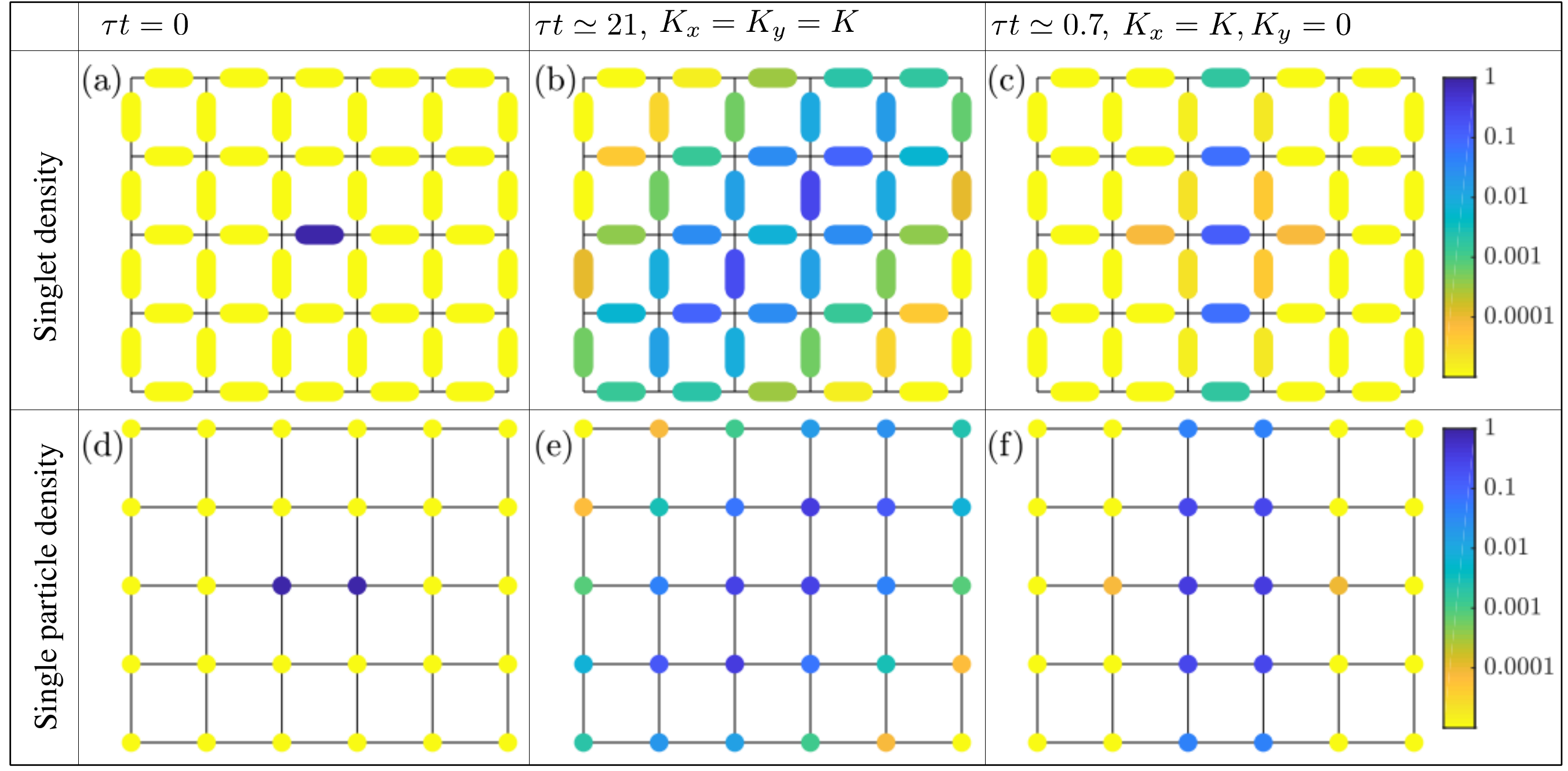}
 \caption{%
 Real space density distribution of (top row) NN singlet pairs and (bottom) particles for the driven Hubbard Hamiltonian.
 (a),(d) Initial state with one singlet pair at the lattice center.
 (b),(e) Densities after evolving for a time $\tau t \approx 21$ the Hubbard Hamiltonian ($U=50t$) driven at a frequency $\Omega=14t$, 
 with $\bm{V}\propto \bm{e}_-$ and $K \approx 2.4048$
 such that all single-particle hopping is very strongly suppressed.
 (c),(f) Densities after evolving for a short time $\tau t \approx 0.7$ with the driving along $x$-direction.
 While single-particle hopping in the $x$ direction is strongly suppressed as in (b) and (e), hopping in the $y$ direction occurs at the undriven rate $t$.
 Note that the total number of singlet pairs  drops very fast when $\bm{V}\propto \bm{x}$; 
 representative  numbers of singlet pairs for driving along $\bm{e}_-$ and $\bm{x}$ at various times are:
 $[\Npairs^{\bm{e}_-}(\tau), \Npairs^{\bm{x}}(\tau)]=[1, 1]$ at $\tau = 0$,
 [0.98, 0.29] at $\tau t \simeq 0.7$, and 
 [0.93, 0.26] at $\tau t \simeq 21$. 
 Note the logarithmic scale in the color coding, which is the same on all panels.
 }
\label{fig:fig4-SpreadMatchtja}
\end{figure*}

According to Eqs.~\eqref{eq:ttilde} and \eqref{eq:alphas}, the choice $\bm{V}\propto \bm{e}_-$ implies that single-particle hopping along the $x$ and $y$ directions is suppressed equally, with $\tilde{t} = t{\cal J}_0(K)$ in both directions, and the superexchange parameter $\tilde{J}$ is equal across all NN bonds. In contrast, the singlet hopping rates are non-zero and anisotropic:
 $\tilde{\alpha}_x = \tilde{\alpha}_y = \tilde{\alpha}_-$ while $\tilde{\alpha}_+ > |\tilde{\alpha}_-| $, so singlets are expected to spread faster along the $\bm{e}_+$ direction, i.e., perpendicular to the driving.
 [For instance, in the limit $ U \gg \Omega $ it is easy to check that $\tilde{\alpha}_- \approx \tilde{J} \mathcal{J}_0(2K)/2$ 
 and $\tilde{\alpha}_+ \approx \tilde{J}/2 \approx \tilde{\alpha}_-/{\cal J}_0(2K) > |\tilde{\alpha}_-|$.] 
Tuning the driving strength such that ${\cal J}_0(K) = 0$ ($K\approx 2.4048$) gives a particularly interesting scenario: all single-particle hopping processes are now strongly suppressed by the fast driving, whereas singlets can still move.

\subsection{One singlet pair}\label{subsec: 1 pair}
We initialize the one-singlet simulations with the pair in the center of the lattice [see \fir{fig:fig4-SpreadMatchtja}(a),(d)] and set the driving strength to $K \approx 2.4048$, at which value the single particle hopping is strongly suppressed ($\tilde{t} = 0$). Slight deviation from this value does not change the following results much providing $|\tilde{\alpha}_-| \ll |\tilde{t}|$.
We then calculate the time evolution with the driven Hubbard model by exact diagonalization.
We show snapshots corresponding to time $\tau\approx 21/t$ of the singlet pair density in~\fir{fig:fig4-SpreadMatchtja}(b), and of the single-particle density in~\fir{fig:fig4-SpreadMatchtja}(e).
We observe that the density of singlets has spread out along the $\bm{e}_+$ direction to the limits of the (finite) lattice, while spreading in other directions is much smaller. The single-particle density perfectly mirrors this behavior, which indicates there is no dynamics of the atoms beyond that contained in the singlet dynamics. In other words, the singlet remains bound throughout the evolution.
In contrast to this, if the lattice is driven along the $x$ axis ($K_x=K$, $K_y=0$), the single-particle density expands along $y$ much faster than the singlet density [Figs.~\ref{fig:fig4-SpreadMatchtja}(c) and (f)], in accordance with the prediction that $\tilde{t}=t$ for hoppings along $y$; still, propagation along $x$ is heavily suppressed.

\begin{figure}[tb] 
\centering
\includegraphics[width=\linewidth]{%
  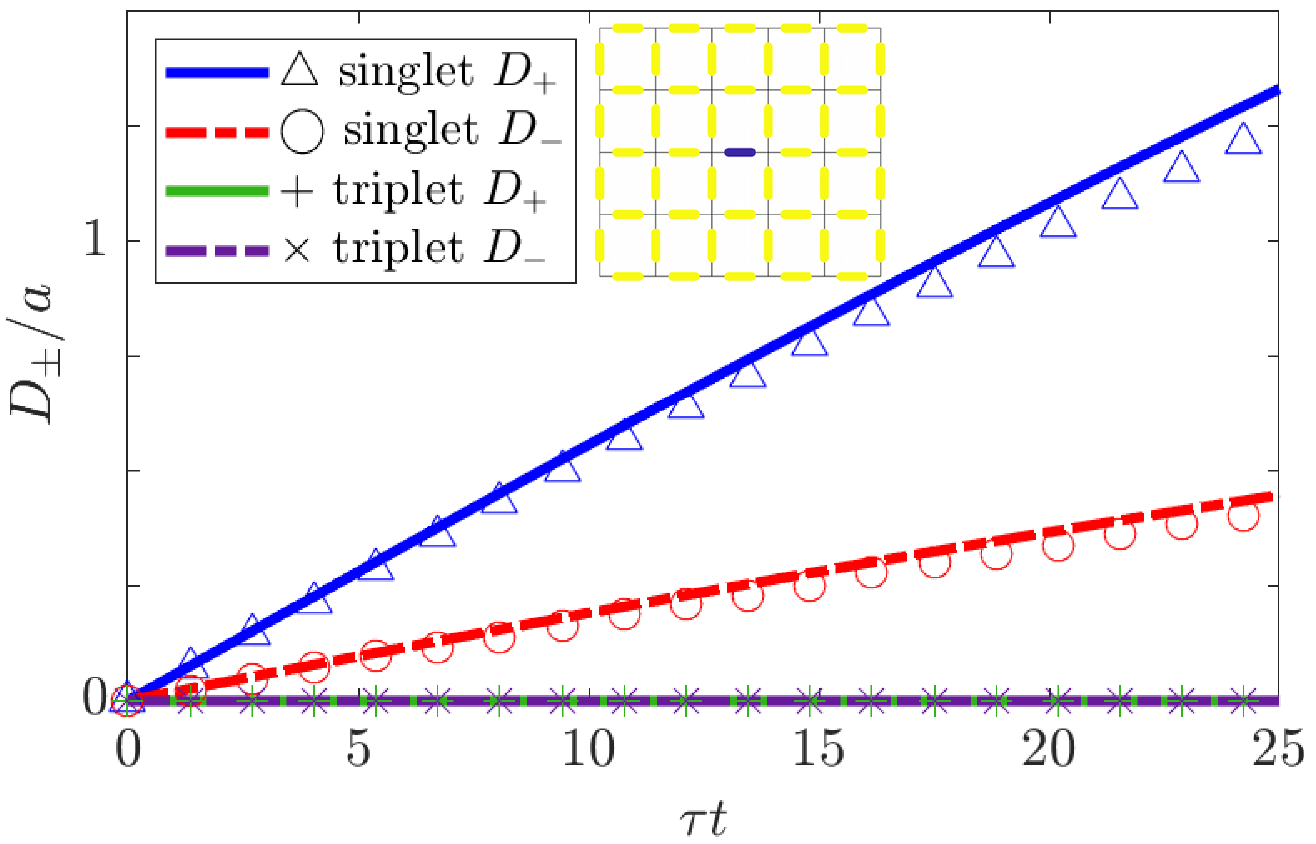}
\caption{%
 Spatial extents of the singlet and triplet density distribution along the two diagonals vs.\ driving time. The symbols are stroboscopic values from the driven Hubbard model whereas the lines are predictions using the \tja model. The inset represents the initial state, which is one singlet or triplet pair (blue ellipse) localized in the center of the square lattice, see \fir{fig:fig4-SpreadMatchtja}(a). The parameters are the same as those used in \fir{fig:fig4-SpreadMatchtja}. The \tja model captures very well the anisotropy of the expansion dynamics of the singlet pair in the driven Hubbard model, and correctly predicts that the triplet pair does not expand.}
\label{fig:fig5-2pairsarrangement}
\end{figure}

To quantitatively compare the expansion dynamics along the two diagonals for both the driven Hubbard and \tja model, we calculate the spatial extent of the singlet pair along the two diagonals, 
\begin{equation}
 D_{\pm} = \sqrt{
  \sum_{\langle ij \rangle} \langle b^\dagger_{ij} b_{ij} \rangle |(\mathbf{r}_{ij} - \mathbf{r}_0) \cdot \mathbf{e}_{\pm}|^2 /  \sum_{\langle ij \rangle} \langle b^\dagger_{ij} b_{ij} \rangle  } ,
 \label{eq:extend}
\end{equation}  
where $\mathbf{r}_{ij}$ is the position vector of the center of the bond between sites $i$ and $j$ and $\mathbf{r}_{0}$ the initial location of the singlet. (We similarly define the spatial extents for triplets.)
The results for driving along $\bm{e}_-$ (\fir{fig:fig5-2pairsarrangement}) show two distinct expansion velocities, with the expansion in the $\bm{e}_+$ direction about three times faster than in $\bm{e}_-$. 
In contrast, if we initialize the atom pair in a spin-triplet state, no expansion is observed. This agrees with the predictions from the \tja model, as the triplet can only propagate via the single-particle hopping term, which is suppressed.
These results demonstrate that tuning the driving direction and strength provides control on the speed and direction of propagation of real-space fermion pairs and their magnetic correlations.

\subsection{Two singlet pairs}
\label{subsec: 2 pairs}
\begin{figure}[tb] 
 \centering
 (a)~\raisebox{\dimexpr-\height+\baselineskip}{%
    \includegraphics[width=.84\linewidth]{%
    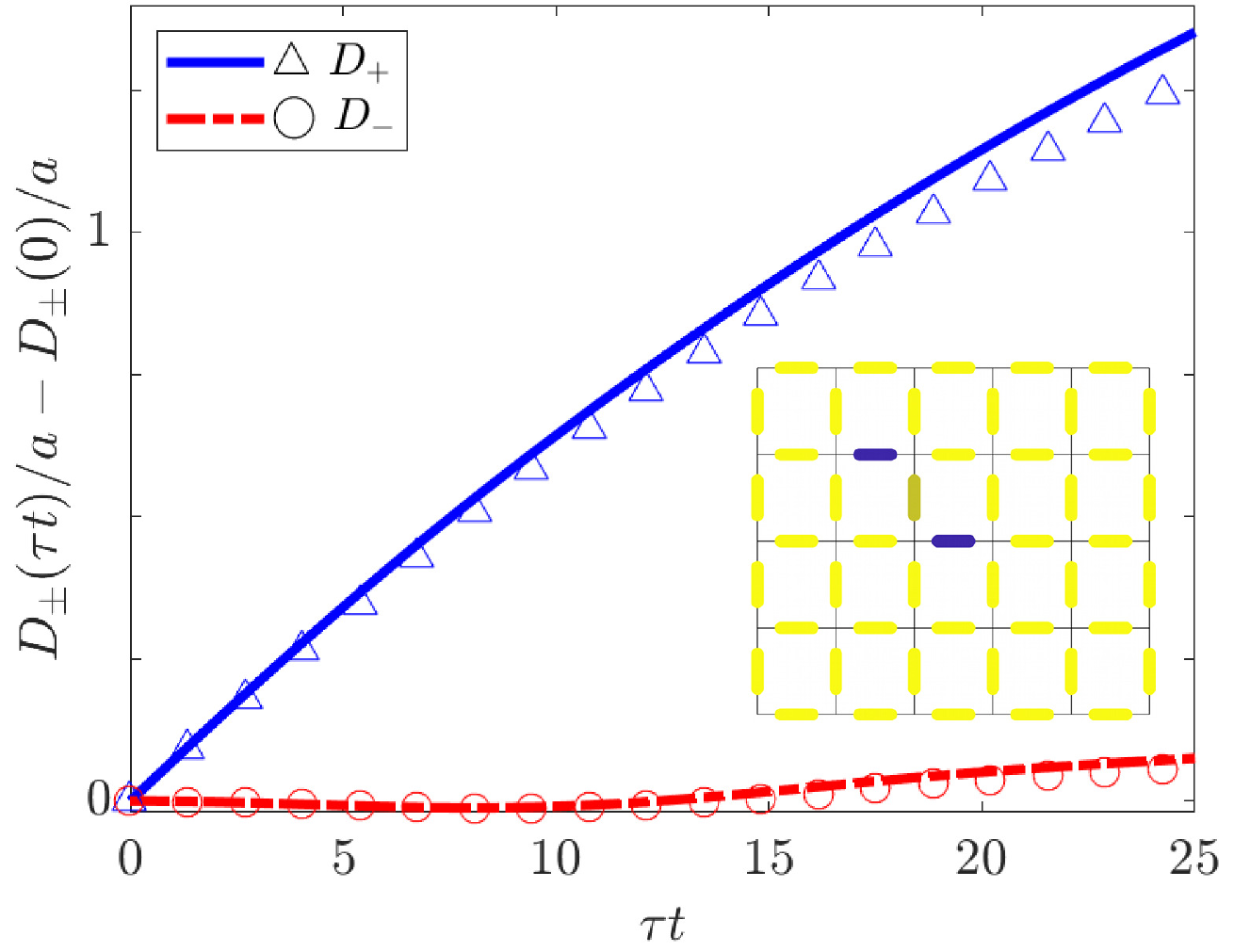}
    } 
 (b)~\raisebox{\dimexpr-\height+\baselineskip}{%
    \includegraphics[width=.84\linewidth]{%
    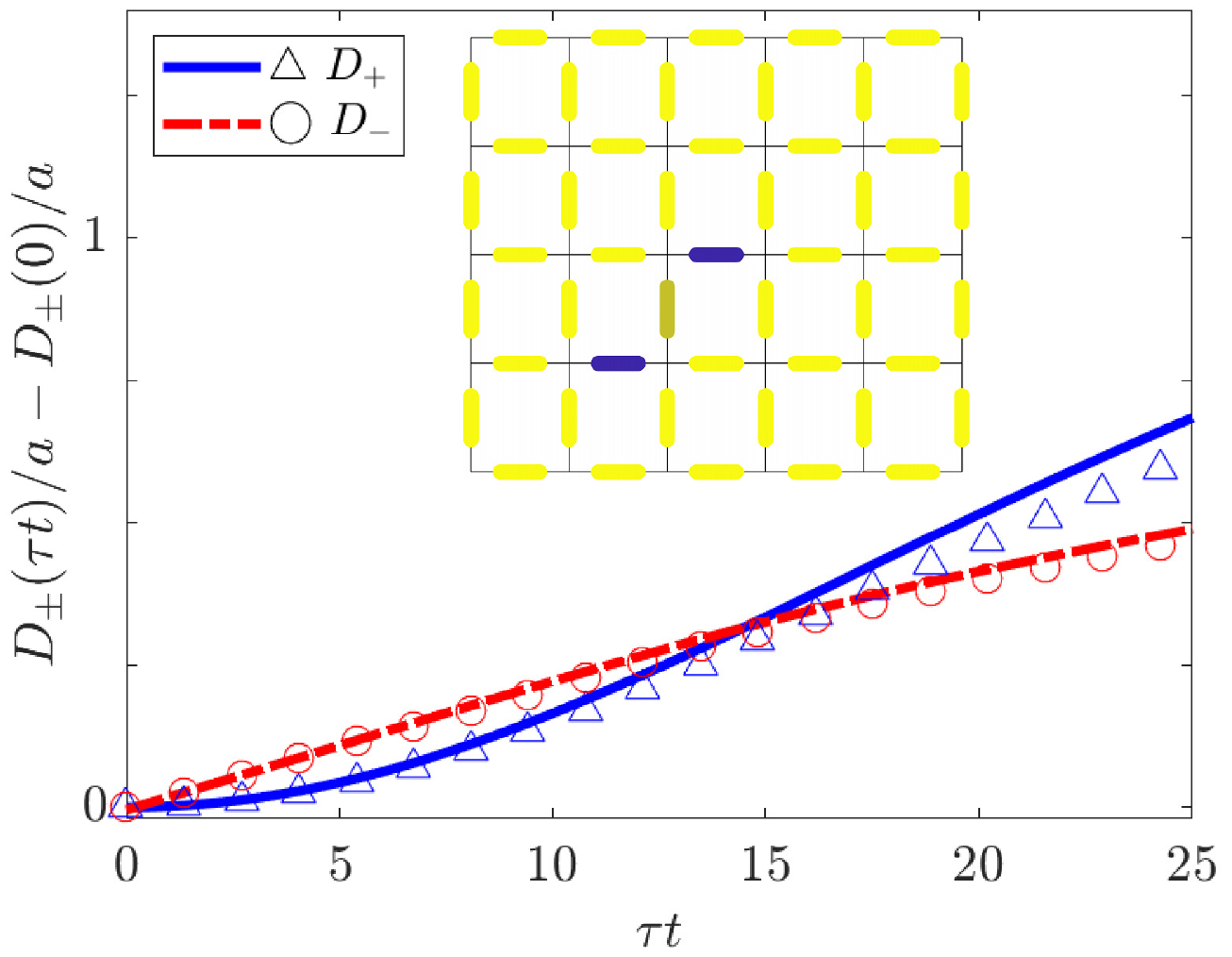}
    } 
 \caption{Spatial extents of two singlet pairs in the square lattice plotted against driving time. 
 The initial state is sketched as an inset in each panel.
 Predictions using the \tja model (solid and dashed-dotted lines) agree with stroboscopic values from the driven Hubbard model (triangles and circles). The insets illustrate the initial arrangements of the singlet pairs. Driving is along $\bm{e}_-$ in all cases.}
\label{fig:fig6-Square2pair_extend}
\end{figure}

We next consider a minimal case where we expect effects due to the interplay of pair-density and magnetic correlations to play a role. To this end, we initialize the system with two singlet pairs located on neighboring bonds, arranged next to each other along either diagonal as shown in the insets of Fig.\ref{fig:fig6-Square2pair_extend}. 

For both configurations, we place the pairs as close to each other as possible. We remark that this results in a nonzero density ($1/4$) of singlets on the bond linking the pairs; this is due to the noncommutativity of the NN singlet operators on neighboring bonds.
Then, we calculate the dynamics of the system by exact diagonalization as in Sec.~\ref{subsec: 1 pair}.

Let us comment briefly on the implications of the noncommutativity of NN singlets for lattices in dimensions higher than one. 
First, we note that two singlet pairs cannot share a lattice site due to the constraint that double occupancy is forbidden by the large $U$. Thus their closest approach will be when they are separated by a single ``linking'' lattice bond.
In this situation, one quarter of a singlet is ``created'' on the middle bond joining the two singlet pairs. If one of the two original pairs hops away, this quarter of a singlet is destroyed and we still have the same number of NN singlet pairs. However, the singlet probability amplitude in the `linking' bond can hop away just like any singlet pair; if that happens, the two original singlet pairs are lost. Therefore, when there are multiple pairs of singlets, the total number of pairs is not conserved in lattice dimensions higher than one even if the single-particle hopping is fully suppressed. This process is forbidden in 1D chains because there is no place for the central pair to hop away. 
The picture gets progressively more involved as more pairs are added to the system. Close to half-filling, due in part to the scarcity of free bonds, neither a description in terms of single particles nor in terms of singlet pairs is sufficient even if the single-particle hopping is fully suppressed; we explore that situation in~\cite{XXXL}.

We compare in Fig.~\ref{fig:fig6-Square2pair_extend} the evolution as a function of time of $D_{\pm}$, \eqr{eq:extend}, for the two initial states sketched in the insets, with the two pairs along $\bm{e}_{\pm}$, respectively.
We see that when the pairs are arranged along the driving direction [\fir{fig:fig6-Square2pair_extend}(a)], 
the spreading along the $\bm{e}_+$ direction, $D_+$, is essentially unaffected as if the pairs are independent (cf. \fir{fig:fig5-2pairsarrangement}). In contrast, the spreading along the $\bm{e}_-$ direction is significantly slower and far from ballistic compared with one-pair dynamics due to interference between the two singlets. Similarly, when the pairs are arranged perpendicular to the driving direction [\fir{fig:fig6-Square2pair_extend}(b)], 
the spreading along the $\bm{e}_-$ direction differs little from the one-pair dynamics but the spreading along the $\bm{e}_+$ direction is significantly suppressed until the pairs have moved apart ($\tau t \gtrsim 10$).

\section{Controlled singlet hopping in the brickwall lattice}
\label{sec: Brickwall lattice}

\begin{figure}[tb] 
    \centering
    \includegraphics[width=\linewidth]{%
      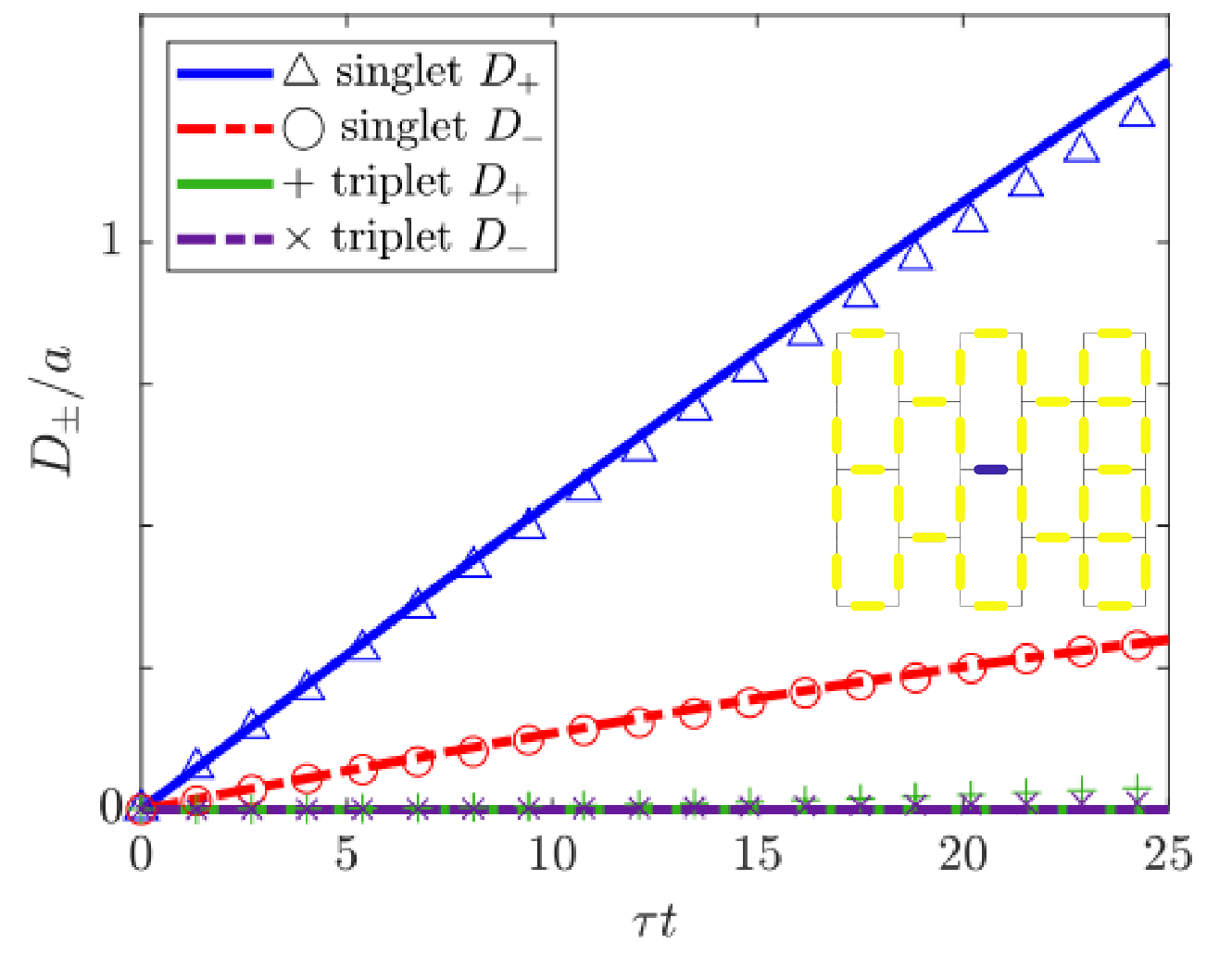}
    \caption{Spatial extents of the singlet (triplet) density distribution along the two diagonals vs.\ driving time, for a single pair localized in the center of the brickwall lattice, as sketched in the inset. 
    The symbols stand for stroboscopic values from the driven Hubbard model whereas the lines are predictions using the \tja model.
    The parameters are the same as those in Fig.~\ref{fig:fig4-SpreadMatchtja}.
    The \tja model still captures very well the anisotropy of the expansion dynamics of the pair (singlet or triplet) in the driven Hubbard model. The dynamics deviates little from what happens in a square lattice cf.\ Fig.~\ref{fig:fig5-2pairsarrangement}.
    }
    \label{fig:fig7-brickwall-1pair-extent}
\end{figure}

In this section we demonstrate that our scheme for quantum simulation of the \tja model also applies to the brickwall lattice~\cite{Jotzu2014, Desbuquois2017, Gorg2018, Messer2018}, for which Floquet heating from higher Bloch bands is significantly reduced~\cite{Messer2018} providing extended experimental run times. 
We sketch the structure of the brickwall lattice in Fig.~\ref{fig:fig7-brickwall-1pair-extent}: 
it can be seen as a flat-sided version of the honeycomb lattice, or a square lattice where every other horizontal bond has been removed. 
The fact that half of the horizontal bonds are missing means that the singlet pair-hopping processes of the \tja model along the $x$ direction are no longer possible (i.e., $\alpha_x \equiv 0$) and that there are neither single-particle nor superexchange processes on those missing bonds; all other coupling amplitudes take on the same values as on a square lattice.

\begin{figure}[tb] 
 \centering
 (a)~\raisebox{\dimexpr-\height+\baselineskip}{%
    \includegraphics[width=.9\linewidth]{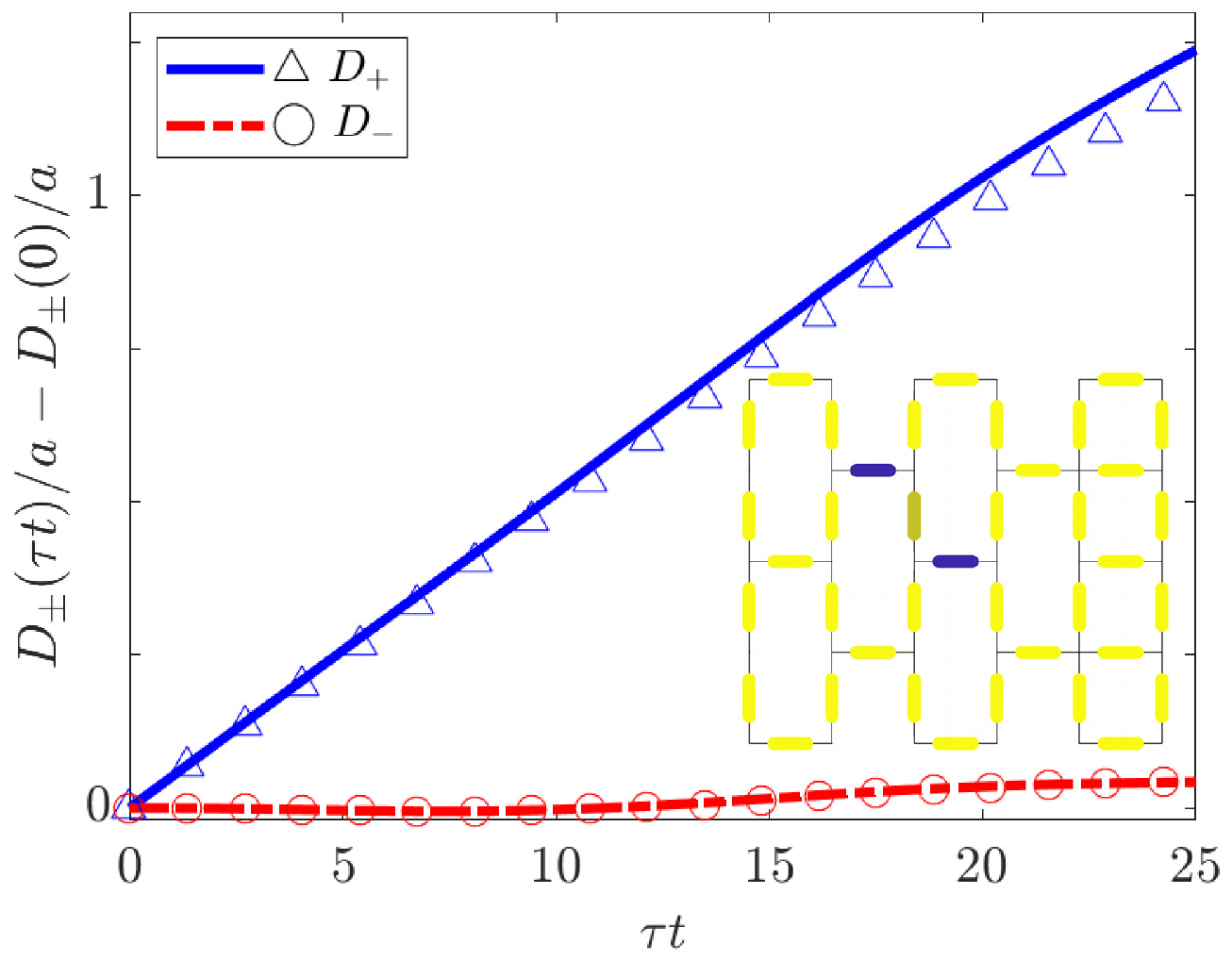}
    } 
 (b)~\raisebox{\dimexpr-\height+\baselineskip}{%
    \includegraphics[width=.9\linewidth]{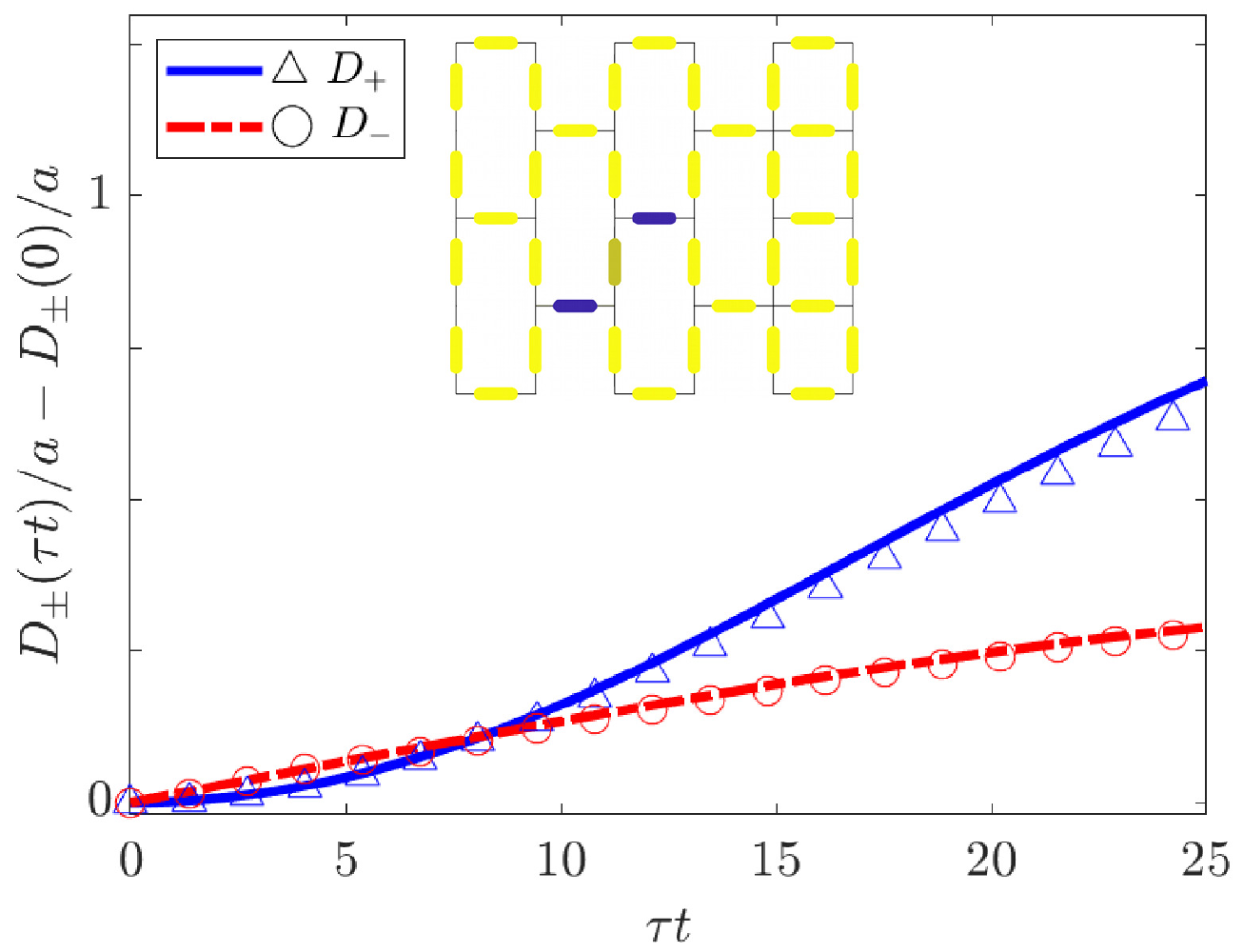}
    } 
 \caption{Spatial extents of two NN singlet pairs arranged along the two diagonals (sketched in the insets) against driving time for the brickwall lattice.
 Predictions using the \tja model (lines) agree with stroboscopic values from the driven Hubbard model (triangles and circles). The insets illustrate the initial positions of the singlet pairs.}
 \label{fig:fig8-brickwall-2pair-extent}
\end{figure}

We present in Figs.~\ref{fig:fig7-brickwall-1pair-extent} and \ref{fig:fig8-brickwall-2pair-extent} the results of our simulations in the brickwall lattice for one and two pairs respectively, using analogous initial configurations and parameters as in \secr{subsec:2D}. We find the behavior to be qualitatively the same as that in a square lattice; in particular, the \tja model still provides an effective description of the driven Hubbard model dynamics.
Figure \ref{fig:fig7-brickwall-1pair-extent} shows the results for a single pair located initially at the center of the brickwall lattice. We observe that the spreading along the $\bm{e}_+$ direction is practically the same as that for a single singlet pair on the square lattice; see \fir{fig:fig5-2pairsarrangement}. However, the spreading along the $\bm{e}_-$ direction is slightly slower, which can be attributed to the absence of the $\alpha_{x}$ terms in the brickwall lattice.

The dynamics for initial states containing two pairs is shown in~\fir{fig:fig8-brickwall-2pair-extent}.
Again the results are very similar to those for the square lattice, cf.~\fir{fig:fig6-Square2pair_extend}, the main difference being a slightly slower spreading in the direction of the driving, $\bm{e}_-$, for the case of two pairs initially aligned along $\bm{e}_+$.

\section{Summary}
In summary, we have presented a protocol to realize quantum simulations of the paradigmatic \tja model in a cold-atom setup, based on periodic driving of an optical lattice trapping a strongly-interacting fermionic gas. 
We demonstrated analytically, and corroborated with numerical simulations, that the direction and strength of the lattice driving provide access to control separately the single-particle and atom-pair(s) dynamics. In particular, we showed that one can reach a regime where only singlet pairs can propagate through the lattice, while single-particle hopping is completely suppressed.
These results point to the possibility of accessing in a controlled manner regimes where density (or charge) and spin correlations compete in new ways, which can lead to novel exotic phenomena; cf.~\cite{XXXL}.
More generally, our findings illustrate the potential of out-of-equilibrium studies to provide new insights into the interplay between the density and spin degrees of freedom in paradigmatic models of condensed-matter physics. 

Our predictions can be readily tested in cold-atom experiments, where magnetic correlations of fermionic Hubbard systems have been measured through merging pairs of nearby sites~\cite{Trotzky2010,Greif2013} and by Bragg scattering~\cite{Hart2015}, and with single-site resolution utilizing the quantum-gas-microscope technique~\cite{Parsons2016, Boll2016, Cheuk2016, Mazurenko2017, Chiu2018, Salomon2019}.
We thus expect our results will trigger new experiments 
harnessing the exquisite degree of spatial and temporal control achieved with state-of-the-art cold-atom experiments to provide a new window based on out-of-equilibrium studies to explore the nature of strongly-correlated fermionic systems.

\begin{acknowledgments}
We would like to thank T.\ Esslinger, F.\ G\"org, and M.\ Messer for useful discussions.
This work has been supported by EPSRC Grants No.\ 
EP/P01058X/1, 
No.\ EP/P009565/1, 
and No.\ EP/K038311/1  
and is partially funded by the European Research Council under the European Union’s Seventh Framework Programme (No.\ FP7/2007-2013)/ERC Grant Agreement No.\ 319286 Q-MAC.
We acknowledge the use of the University of Oxford Advanced Research Computing (ARC) facility in carrying out this work \cite{arclink}. 

HG and JRC contributed equally to this work.
\end{acknowledgments}


\appendix

\section{Derivation of the effective \lowercase{$t$}-$J$-$\alpha$ Hamiltonian using Floquet basis and perturbation theory}
\label{apxsec: FullFloquetPertubativeAnalysis}

In this section we outline how to derive the effective Hamiltonian \eqr{eq:tja} governing the stroboscopic dynamics of the driven Hubbard system using Floquet theory \cite{Shirley1965, Dunlap1986, Bukov2015}. For simplicity, we restrict the discussion here to the one-dimensional geometry, although the method generalizes straightforwardly to higher dimensions. 

The periodic driving term in \eqref{eq:Hdrive} breaks the continuous time-translation symmetry of the Hamiltonian to a discrete translation symmetry, meaning that energy is only conserved up to integer multiples of $\Omega$. Floquet's theorem states that due to the time periodicity of $\hat{H}$, there exists a complete set of solutions to the time-dependent Schr\"odinger equation,
\begin{equation}
  | \Psi_{a}(\tau) \rangle = e^{-i\epsilon_a \tau}
  |\phi_a(\tau) \rangle \:,
\end{equation} 
such that any state can be decomposed as a superposition of these solutions
\begin{equation}
 | \Psi(\tau) \rangle = \sum_{a} c_{a}
 e^{-i\epsilon_a \tau} |\phi_a(\tau) \rangle .
\end{equation}
Here $|\phi_a(\tau)\rangle = |\phi_a(\tau+T)\rangle $ are time-periodic Floquet states which are solutions to the eigenvalue equation
\begin{equation}
 (\hat{H} - i\partial_\tau ) |\phi_a (\tau) \rangle =
 \epsilon_a |\phi_a (\tau) \rangle ,
\end{equation}
with $\epsilon_a$ termed quasienergies. The quasienergy operator, $H_{Q} = (\hat{H} - i\partial_\tau )$ acts on the combined Floquet-Hilbert space $\mathcal{H}\otimes \mathcal{T}$, where $\mathcal{H}$ is the original Hilbert space and $\mathcal{T}$ is the space of square-integrable $T$-periodic functions. The scalar product in this extended space is given by
\begin{equation}
 \langle\!\langle \chi | \xi \rangle\!\rangle = \frac{1}{T}
 \int\limits^{T}_{0} \text{d}\tau \langle
 \chi(\tau)|\xi(\tau)\rangle ,
\end{equation}
where $| \xi \rangle\!\rangle$ denotes a vector in $\mathcal{H}\otimes \mathcal{T}$, and $|\xi(\tau)\rangle$ a $T$-periodic vector in $\mathcal{H}$. Notice that by extending the Hilbert space, we go from time-dependent matrix elements to time-independent ones. By choosing an appropriate time-periodic unitary transformation $\hat{R}(\tau)$, it is possible to bring $\hat{H}_{Q}$ into block-diagonal form, with the diagonal blocks identical up to an energy shift $m \Omega$ \cite{Eckardt2017}. The diagonal block, which acts only on $\mathcal{H}$, governs the stroboscopic dynamics of the system.

The choice of Floquet basis
\begin{equation}
  |a,m \rangle = |a\rangle e^{-im\Omega \tau}
\end{equation}
conveniently structures $\mathcal{H}\otimes \mathcal{T}$ into subspaces of states that contain $m$ quanta of energy $\Omega$ from the driving. When far from resonance, blocks of different $m$ are only weakly admixed by $\hat{H}_{Q}$, so that the subspace adiabatically connected to the undriven Hubbard model is the one with $m=0$. Assuming the condition $t\ll \Omega\text{, } U$ is satisfied, we perturbatively block-diagonalise $\hat{H}_{Q}$ to obtain the effective time-independent Hamiltonian.

We begin by transforming the Hamiltonian into the rotating frame with respect to the driving field. This has the effect of eliminating the explicit driving term [\eqr{eq:Hdrive} in the main text] and imprinting it as an oscillating complex phase on the hopping term, thus bounding the terms in $\hat{H}(\tau)$ which couple different “photon" sectors by $t \ll U, \Omega$. Specifically, we transform the Hamiltonian as
\begin{equation}
  \hat{H}^{R} = \ii (\partial_{\tau} \hat{R}) \hat{R}^{\dagger} + \hat{R} \hat{H} \hat{R}^{\dagger}
\end{equation}
where
$ 
  \hat{R}(\tau) = \exp\left( -\ii K \cos(\Omega \tau) \sum_{j} j \hat{n}_{j} \right).
$ 
Applying this transformation to the one-dimensional driven Hubbard model, $\hat{H} = \hat{H}_{\text{Hub}} + \hat{H}_{\text{drive}}(\tau)$, we obtain 
\begin{equation*}
  \hat{H}^{R}(\tau) 
  = U \sum_{j} \hat{n}_{j, \uparrow} \hat{n}_{j, \downarrow} 
   -t \sum_{j, \sigma} \left( \ee^{\ii K \cos(\Omega \tau)} \hat{c}^{\dagger}_{j, \sigma} \hat{c}_{j+1, \sigma} +\text{h.c.} \right).
\end{equation*}
The quasienergy operator in this basis is then
\begin{equation}
  \hat{H}_{\rm Q} = \left[\hat{H}^{R}(\tau) - \ii \partial_{\tau} \right].
\end{equation}
We then introduce the Floquet basis
\begin{equation} \label{eq:floquetbasis}
  \ket{a,n_d,m} =  \ket{a, n_d} \ee^{-\ii m (\Omega \tau + \frac{\pi}{2})}.
\end{equation}
Here $m$ is the integer ``photon number'', ${-\infty < m < \infty}$, labeling the number of excitations from the periodic driving field; $n_{d}$ is the number of doubly occupied sites in the state. The remaining label $a$ denotes an arbitrary choice of basis states consistent with the labels $n_d$ and $m$. 
When we expand $\hat{H}_{\rm Q}$ in the basis in \eqr{eq:floquetbasis}, we obtain 
\begin{equation} 
  \hat{H}_{Q} = \sum_{m,m'} \hat{H}_{m,m'} \otimes \ket{m} \bra{m'},
\end{equation}
where
\begin{align} \label{eq:tbfourier2}
 \hat{H}_{m,m'} &= 
 -t \sum_{j, \sigma} \bigg( \mathcal{J}_{m'-m} ( K ) \hat{c}^{\dagger}_{j, \sigma} \hat{c}_{j+1, \sigma} \nonumber \\
 & \quad + \mathcal{J}_{m-m'} ( K ) \hat{c}^{\dagger}_{j+1, \sigma} \hat{c}_{j, \sigma} \bigg) 
 + (\hat{H}_{U} +  m \Omega) \delta_{m,m'} \nonumber \\
 & = -t \hat{T}_{m,m'} + (\hat{H}_{U} +  m \Omega) \delta_{m,m'}
\end{align}
are blocks which act solely on $\mathcal{H}$, while $\ket{m} \bra{m'}$ acts only on $\mathcal{T}$. We note that when $t = 0$, the quasienergy operator is trivially diagonalized by states which have a well-defined  $n_{d}$, and $m$. We now examine the effect of adding a finite $t \ll U, \Omega$ as a perturbation. In general, the effective Hamiltonian within a degenerate manifold of states of a Hamiltonian $\hat{H}^{(0)}$ split by a perturbing Hamiltonian $\lambda \hat{H}^{(1)}$ is 
\begin{equation} \label{eq:geneff}
  \hat{H}_{\rm eff} = E_{n} \mathcal{P}_n + \lambda \mathcal{P}_{n} \hat{H}^{(1)} \mathcal{P}_{n} + \lambda^2 \sum_{m \neq n} \frac{\mathcal{P}_{n} \hat{H}^{(1)} \mathcal{P}_{m} \hat{H}^{(1)} \mathcal{P}_{n}}{E_{n} - E_{m}}.
\end{equation}

Here $E_{n}$ is the unperturbed energy of the $n$th degenerate manifold, which in the case of the driven Hubbard model is $E_{n_d, m} = n_d U + m \Omega$. The corresponding projector onto the $n$th degenerate manifold is $\mathcal{P}_n$. Then, making the identification
\begin{equation}
  \lambda \hat{H}^{(1)} = -t \sum_{m, m'} \hat{T}_{m, m'} \otimes \ket{m} \bra{m'},
\end{equation}
and plugging into \eqr{eq:geneff}, one obtains that the effective Hamiltonian is given by
\begin{eqnarray*}
  H_{\rm eff} 
  = -t \mathcal{P}_{0} \hat{T}_{0,0} \mathcal{P}_{0} 
  -t^2 \sum_{n_{d} > 0} \sum_{m} \frac{\mathcal{P}_{0} \hat{T}_{0,m} \mathcal{P}_{n_{d}} \hat{T}_{m,0} \mathcal{P}_{0}}{n_d U + m \Omega},
\end{eqnarray*}
which simplifies to the \tja Hamiltonian given in \eqr{eq:tja} in the main text.

We note that this breaks down when close to resonance $U \approx m\Omega$, as we no longer satisfy the condition $t \ll |U + m \Omega|$ for some choices of $m$, and states of different photon numbers become strongly admixed. 

\section{Further calculations on the validity of the driven Hubbard model as a simulator of the \tja model}
\label{apxsec:tjavalidity}

A key ingredient in quantum simulation is to accurately prepare target many-body states. 
In this work, we aim to simulate the \tja model with controllable values of the parameters $t$, $J$ and $\alpha$ by means of a driven Hubbard model.
In Sec.~\ref{sec:1dNumerics} we have considered in particular the generation of ground states of the \tja model. 
This can be achieved by meeting two criteria. First, states of the driven Hubbard model must be an accurate match for states of the time-dependent \tja model. 
Secondly, we must be able to adiabatically transfer these effective \tja states from the initial undriven regime, to the regime of interest. 
Figure \ref{fig:fig3-target-fidelity} illustrated our ability to meet the second criterion; here we provide further numerical evidence corroborating that our scheme meets the first.

\begin{figure}[t] 
\centering
\includegraphics[width=0.8\linewidth]{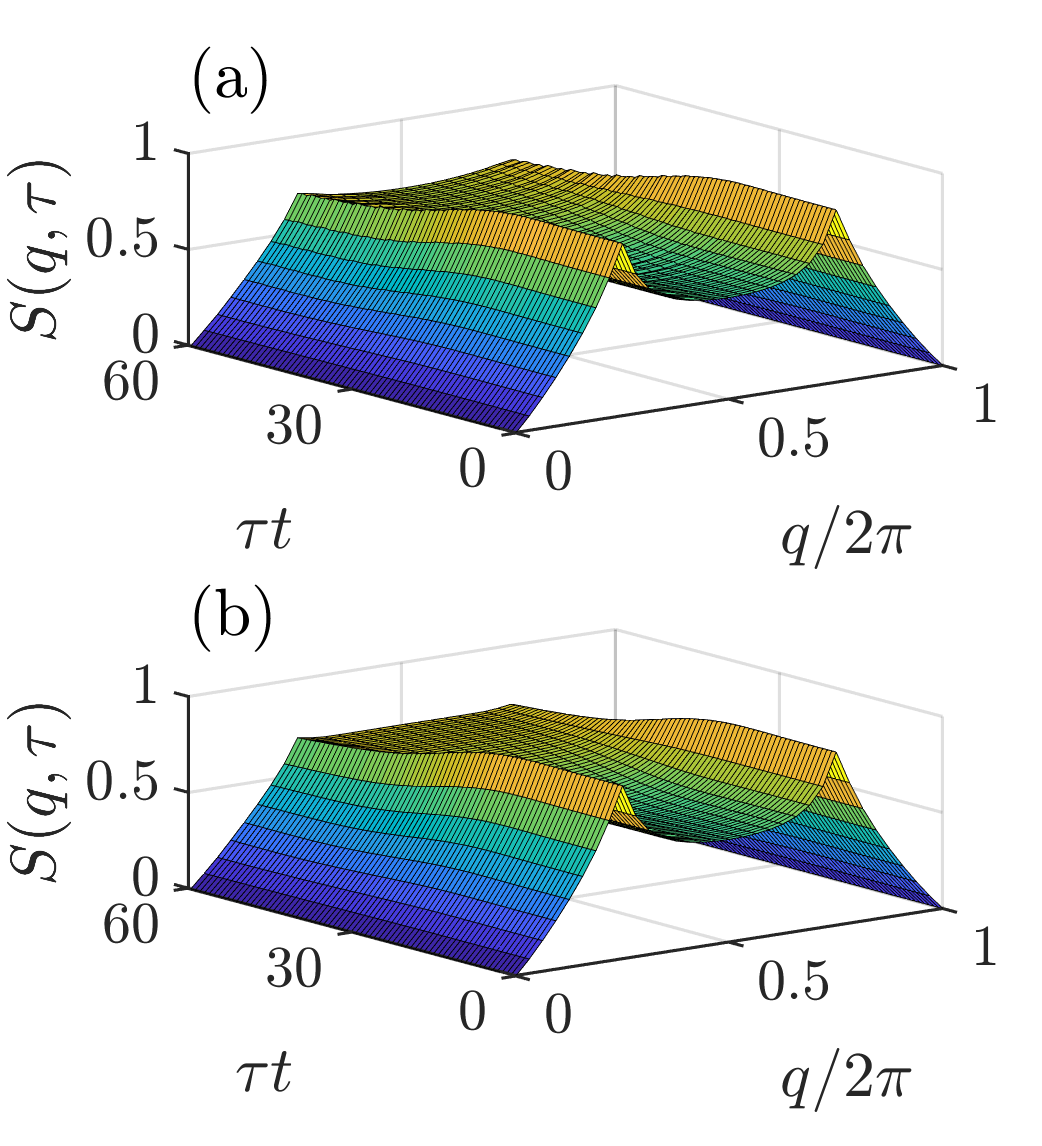}
\includegraphics[width=0.8\linewidth]{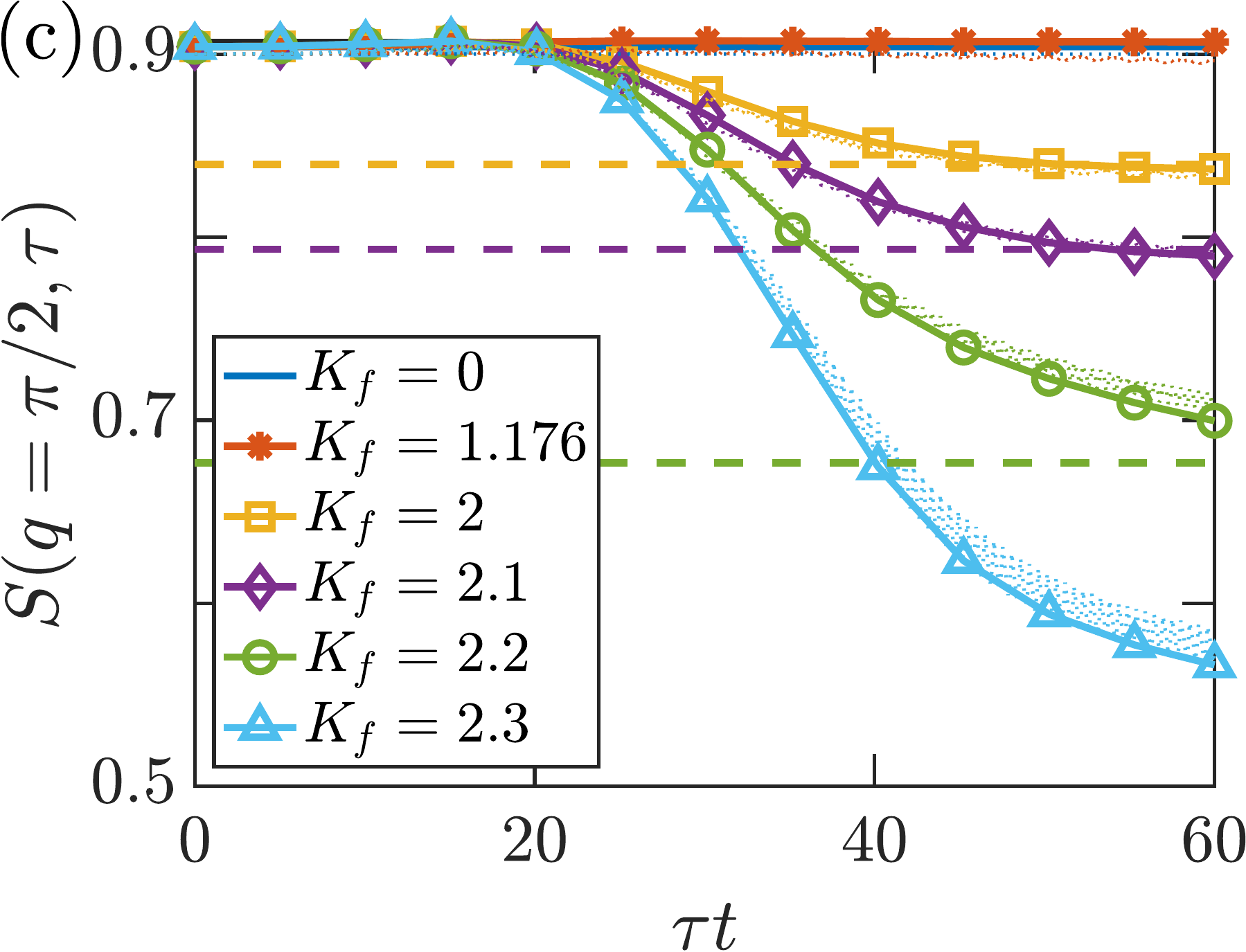} 
\caption{%
 Spin structure factor, $S(q,t)$, for the same simulation parameters as in \fir{fig:fig2-QuenchComparison}, for 
 (a) the driven Hubbard model and (b) the effective \tja model.
 (c) The $q=\pi/2$ component of $S(q,\tau)$ for the driven Hubbard model (dotted lines oscillating at frequency $\Omega$) and for the \tja model (solid lines with symbols) for various final driving strengths $\Kfin$. The dashed lines indicate the values of $S(q=\pi/2,\tau)$ for the ground states of the \tja model with the
 effective parameters $\tilde{t}$, $\tilde{J}$ and $\tilde{\alpha}$ corresponding to $K=K_f$.
}
\label{fig:fig9-quench-spin-structure}
\end{figure}

In \secr{sec:1dNumerics}, we numerically demonstrated that the singlet pairing properties of the one dimensional \tja model with time-varying parameters are accurately reproduced by the driven Hubbard model, 
see in particular the data for the singlet structure factor, $P(q,\tau)$, in Figs.~\ref{fig:fig2-QuenchComparison}(b)-\ref{fig:fig2-QuenchComparison}(d). 
Here we provide additional results on the spin structure factor,
\begin{equation}
  S(q,\tau) = \frac{1}{L} \sum_{j, k} \ee^{\ii q (j-k)} S^{z}_{j,k}(\tau),
\end{equation}
where 
$S^{z}_{j, k}(\tau) = \langle \psi(\tau)| \hat{s}^{z}_{j} \hat{s}^{z}_{k} |\psi(\tau)\rangle$, with $\ket{\psi(\tau)}$ the state of the system at time $\tau$.
We compare $S(q,\tau)$ in Figs.~\ref{fig:fig9-quench-spin-structure}(a) and (b) for (a) the driven Hubbard and (b) the effective \tja model. We see both observables are very close to each other.
Similarly, we show in \fir{fig:fig9-quench-spin-structure}(c) the value at $q=\pi/2$ as a function of time. We see that for all driving strengths considered, the time-dependent \tja model is matched by the stroboscopic value of the driven Hubbard model extremely closely.
These results corroborate that the driven Hubbard model is a valid simulator of the \tja model, and it is possible to generate the ground state of the latter for a broad range of parameters with a simple, smooth ramp of the driving strength.

\section{Heating in strongly-driven Hubbard systems} \label{apx:heating}

In this appendix we discuss heating present in the periodically driven Hubbard systems studied in the main text. We note, however, that as we are considering a single-band Hubbard model no excitations to higher Bloch bands are included.

\subsection{Driven 1D Hubbard model}\label{ssec:heating-1d}

We have discussed in Sec.~\ref{sec:1dNumerics} and Appendix~\ref{apxsec:tjavalidity} several observables indicating that it is possible to load the Hubbard system into the ground state of the effective \tja model for arbitrary $K \leq 2.1$ with fidelity $\geq 97\%$; see Fig.~\ref{fig:fig3-target-fidelity}.

As an alternative test of the quality of ground-state transfer, we perform further time-evolution simulations of the same driven Hubbard system, starting again from the ground state of the undriven Hubbard model. 
Specifically, inspired by Ref.~\cite{Gorg2018}, we compute the fidelity of the time-evolved state with respect to the initial state, 
\begin{equation}
   F_0 (\tau) = |\braket{\psi(\tau=0)|\psi(\tau)}|^2 \:.
   \label{eq:initial-fidelity}
\end{equation}
This fidelity allows one to estimate how much the system has been excited away of its ground state,  both due to Floquet heating and the finite ramping time, as a function of driving time.
In these calculations, we consider a process where the driving amplitude is smoothly ramped up to the peak value, $K_f$, before it is ramped down in the same way to zero. Specifically, we use this protocol for the driving ramp:
\begin{align}
  K(\tau)
  &= K_f \bigg[
    \tanh\left(\frac{\tau_{0}}{ \tau_{\mathrm{ramp}} }\right) 
    + \tanh\left( \frac{ \tau - \tau_{0} }{ \tau_{\mathrm{ramp}} } \right) 
  \nonumber \\
  & \quad - \tanh\left( \frac{ \tau - \tau_{\mathrm{end}} + \tau_{0} }{ \tau_{\mathrm{ramp}} } \right)
  \nonumber \\
  & \quad - \tanh\left( \frac{ \tau_{\mathrm{end}} - \tau_{0} }{ \tau_{\mathrm{ramp}} } \right) \bigg] \times
  \nonumber \\
  & \times
  \bigg[ \tanh\left( \frac{ \tau_{0} }{ \tau_{\mathrm{ramp}} }\right) + 2 \tanh\left( \frac{ \tau_{\mathrm{end}} - 2\tau_{0} }{ 2\tau_{\mathrm{ramp}} } \right) 
  \nonumber \\
  & \quad - \tanh\left( \frac{ \tau_{\mathrm{end}} - \tau_{0} }{ \tau_{\mathrm{ramp}} } \right) \bigg]^{-1},
  \label{eq:ramp-up-down}
\end{align}
For $K_f \neq 0$, we expect the fidelity $F_0$ to decrease with time until $\tau=\tau_{\mathrm{end}}/2$, when the maximum value $K=K_f$ is reached, with larger drops the larger $K_f$. 
However, for a completely adiabatic process, we expect $F_{0}(\tau_{\mathrm{end}}) = 1$.

\begin{figure}[tb]
    \centering
    \includegraphics[width=\linewidth]{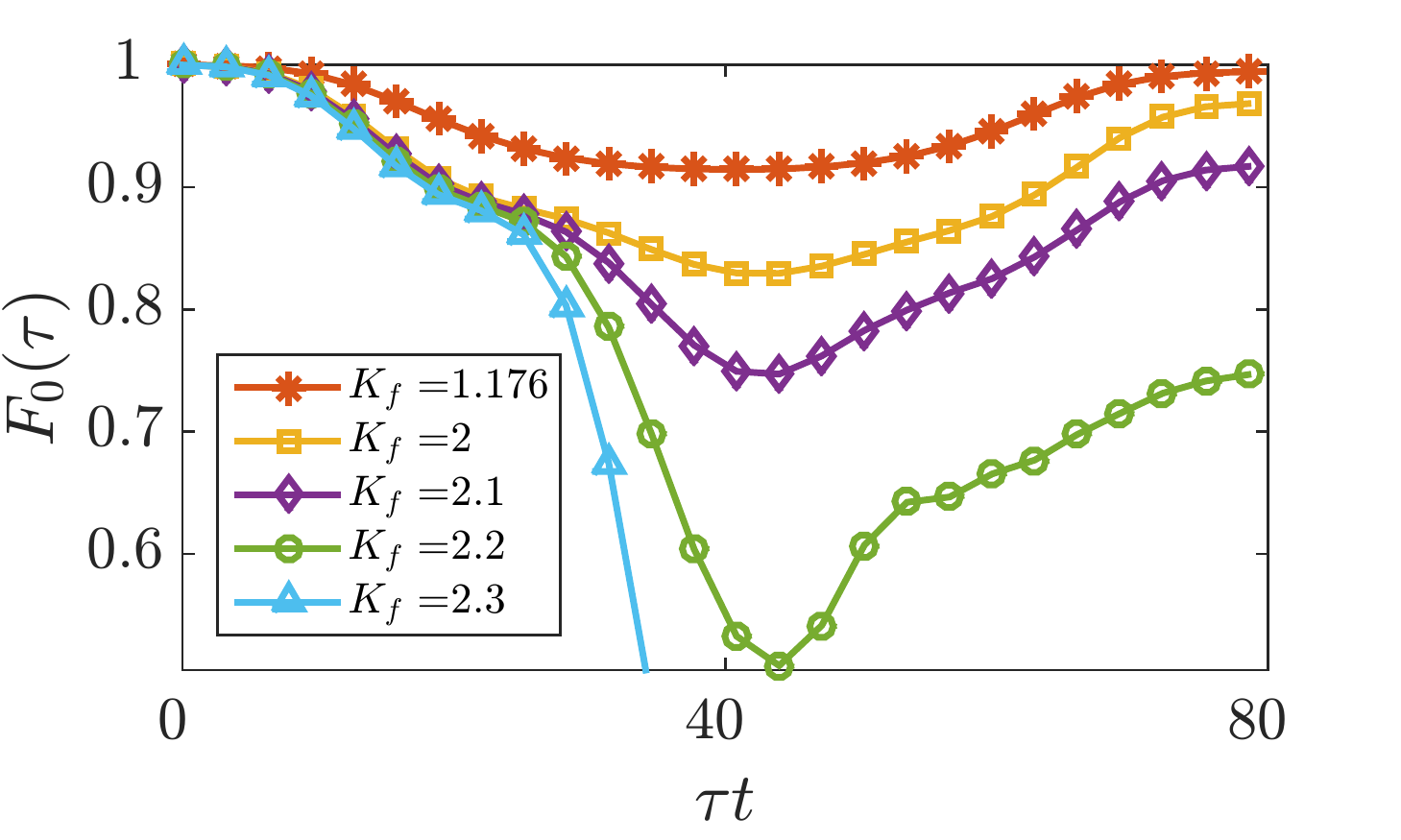}
    \caption{%
    Fidelity of the quantum state of the driven Hubbard model relative to the initial ground state, Eq.~\eqref{eq:initial-fidelity}, as the driving amplitude is ramped up to the value $K_f$, and then ramped down to zero according to Eq.~\eqref{eq:ramp-up-down},
    with $\tau_{\mathrm{end}} t = 80$. Other driving and Hubbard model parameters are the same as in \fir{fig:fig2-QuenchComparison}.
    }
    \label{fig:fig10-up-down-fidelity}
\end{figure}

The results of our calculations are shown in \fir{fig:fig10-up-down-fidelity}. We observe that, for peak driving strength $K_f \lesssim 2$, the final fidelity is $F_0 \geq 98\%$.
For $K_f = 2.1$,  $F_0$ is still $> 90\%$ at the end of the protocol; we note the slight asymmetry with respect to $\tau=\tau_{\mathrm{end}}/2$.
For $K_f = 2.2$, we obtain $F_0 \approx 75\%$ at the end of the protocol and the curve is strongly asymmetric, with the minimum fidelity occurring during the ramp-down; this suggests the system displays considerable non-adiabatic dynamics for the choice of driving parameters $\{ t_0, t_{\mathrm{ramp}}, t_{\mathrm{end}} \}$. This behavior could be improved considering longer ramp-up and -down times, $\tau_{\mathrm{ramp}}$, a longer protocol duration, $t_{\mathrm{end}}$, or more elaborate time dependences $K(\tau)$, as discussed in Sec.~\ref{sec:1dNumerics}.
Finally, the occurrence of phase separation for $K=2.3$ is reflected by the evolved state failing to return to the initial ground state (its final fidelity is $\approx 23$\%).
(We have checked that in all cases the truncation errors are smaller than the infidelities.)

\begin{figure}[tb] 
    \centering
    \includegraphics[width=.88\linewidth]{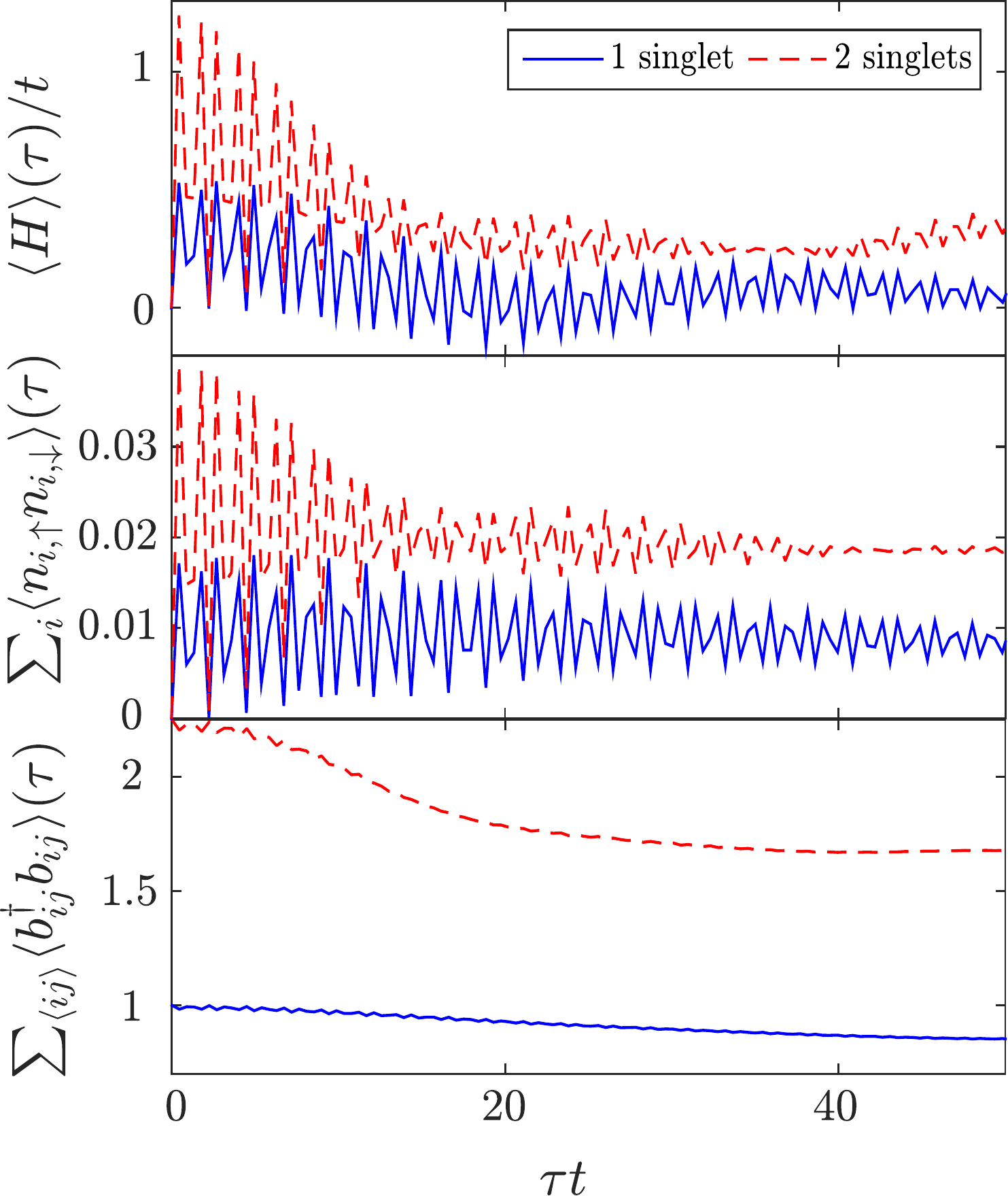}
    \caption{Observables to monitor heating in the driven Hubbard system: (top) energy, (middle) number of doublons, and (bottom) number of singlet pairs. We use the same parameters as in \fir{fig:fig4-SpreadMatchtja} and \fir{fig:fig6-Square2pair_extend}(a) for the calculations with one (blue solid lines) and two (red dashed) initial singlet pairs, respectively.
    }
    \label{fig:fig11-2Dheating}
\end{figure}

\subsection{Driven 2D Hubbard model}\label{ssec:heating-2d}

In this subsection, we discuss heating in the 2D driven Hubbard systems studied in \secr{subsec:2D}.
Specifically, we show that, on the timescales we consider, the driven Hubbard system does not suffer from ``Floquet heating'' to infinite temperature, a generic feature of driven many-body quantum systems, except for integrable or many-body-localized systems~\cite{DAlessio2014, Lazarides2014pre, Genske2015, Herrmann2017, Weidinger2017, Mori2016, Canovi2016, Abanin2017, MurPetit2018, Tindall2019}.

We show in \fir{fig:fig11-2Dheating} the evolution with time of the energy, total number of doublons, and total singlet-pair densities for the driven Hubbard model, under the same conditions as in \secr{subsec:2D}, namely, evolving an initial state with either one or two singlet pairs near the center of the lattice. Notice that we evolve the states for twice as long as we presented in \secr{subsec:2D}.

As we can see in the top panel, the energy of the system starts from $0t$ at $\tau = 0$, and after a transient and sharp increase to $\approx 1t$, it drops and stabilizes to a value $\langle H\rangle \leq 0.5t$ for a long time $10 \lesssim \tau t \lesssim 50$, showing that  the system has stopped absorbing energy from the drive.
This matches the typical way an interacting, periodically driven system heats up, with the  period $10 \lesssim \tau t \lesssim 50$,
roughly corresponding to the prethermalization plateau~\cite{Canovi2016, Weidinger2017}.

In the middle panel, we see that the total number of doublons follows a similar evolution, and stabilizes at a level well below the expected numbers of doublons for the infinite-temperature, completely order-less states ($\approx 0.033$, $0.069$ for an initial state with one and two singlets, respectively). This smooth evolution and stabilization in the number of doublons is similar to that reported for a resonantly driven Hubbard model in Ref.~\cite{Herrmann2017} 
The fact that the number of doublons remains much smaller than one supports the validity of the description with the \tja model, which neglects double occupations.

Finally, the bottom panel of \fir{fig:fig11-2Dheating} further supports that the quantum states of the driven Hubbard system is not disordered, as the number of singlet pairs remains close to the initial one for the whole evolution considered.

These results indicate that Floquet heating has not caused the systems to go to the infinite temperature states and underpins the good agreements with the effective \tja model which we have shown in the main text. 

To close, it is worth noting that these results depend sensitively on the phase of the drive. Indeed, we do observe a larger amount of energy absorbed, a larger number of doublons created, and a larger fraction of singlets destroyed, if we consider a drive
$H_{\text{drive}} \propto \cos(\Omega \tau)$, that has maximum strength at $\tau=0$, instead of 
$H_{\text{drive}} \propto \sin(\Omega \tau)$ [Eq.~\eqref{eq:Hdrive}], which turns on the driving more smoothly.

\putbib
\end{bibunit}


\begin{thebibliography}{84}%
\makeatletter
\providecommand \@ifxundefined [1]{%
 \@ifx{#1\undefined}
}%
\providecommand \@ifnum [1]{%
 \ifnum #1\expandafter \@firstoftwo
 \else \expandafter \@secondoftwo
 \fi
}%
\providecommand \@ifx [1]{%
 \ifx #1\expandafter \@firstoftwo
 \else \expandafter \@secondoftwo
 \fi
}%
\providecommand \natexlab [1]{#1}%
\providecommand \enquote  [1]{``#1''}%
\providecommand \bibnamefont  [1]{#1}%
\providecommand \bibfnamefont [1]{#1}%
\providecommand \citenamefont [1]{#1}%
\providecommand \href@noop [0]{\@secondoftwo}%
\providecommand \href [0]{\begingroup \@sanitize@url \@href}%
\providecommand \@href[1]{\@@startlink{#1}\@@href}%
\providecommand \@@href[1]{\endgroup#1\@@endlink}%
\providecommand \@sanitize@url [0]{\catcode `\\12\catcode `\$12\catcode
  `\&12\catcode `\#12\catcode `\^12\catcode `\_12\catcode `\%12\relax}%
\providecommand \@@startlink[1]{}%
\providecommand \@@endlink[0]{}%
\providecommand \url  [0]{\begingroup\@sanitize@url \@url }%
\providecommand \@url [1]{\endgroup\@href {#1}{\urlprefix }}%
\providecommand \urlprefix  [0]{URL }%
\providecommand \Eprint [0]{\href }%
\providecommand \doibase [0]{http://dx.doi.org/}%
\providecommand \selectlanguage [0]{\@gobble}%
\providecommand \bibinfo  [0]{\@secondoftwo}%
\providecommand \bibfield  [0]{\@secondoftwo}%
\providecommand \translation [1]{[#1]}%
\providecommand \BibitemOpen [0]{}%
\providecommand \bibitemStop [0]{}%
\providecommand \bibitemNoStop [0]{.\EOS\space}%
\providecommand \EOS [0]{\spacefactor3000\relax}%
\providecommand \BibitemShut  [1]{\csname bibitem#1\endcsname}%
\let\auto@bib@innerbib\@empty
\bibitem [{\citenamefont {Bednorz}\ and\ \citenamefont
  {M{\"{u}}ller}(1986)}]{Bednorz1986}%
  \BibitemOpen
  \bibfield  {author} {\bibinfo {author} {\bibfnamefont {J.~G.}\ \bibnamefont
  {Bednorz}}\ and\ \bibinfo {author} {\bibfnamefont {K.~A.}\ \bibnamefont
  {M{\"{u}}ller}},\ }\href {\doibase 10.1007/bf01303701} {\bibfield  {journal}
  {\bibinfo  {journal} {Z. Phys. B Con. Mat.}\ }\textbf {\bibinfo {volume}
  {64}},\ \bibinfo {pages} {189} (\bibinfo {year} {1986})}\BibitemShut
  {NoStop}%
\bibitem [{\citenamefont {Lee}\ \emph {et~al.}(2006)\citenamefont {Lee},
  \citenamefont {Nagaosa},\ and\ \citenamefont {Wen}}]{Lee2006}%
  \BibitemOpen
  \bibfield  {author} {\bibinfo {author} {\bibfnamefont {P.~A.}\ \bibnamefont
  {Lee}}, \bibinfo {author} {\bibfnamefont {N.}~\bibnamefont {Nagaosa}}, \ and\
  \bibinfo {author} {\bibfnamefont {X.~G.}\ \bibnamefont {Wen}},\ }\href
  {\doibase 10.1103/RevModPhys.78.17} {\bibfield  {journal} {\bibinfo
  {journal} {Rev. Mod. Phys.}\ }\textbf {\bibinfo {volume} {78}},\ \bibinfo
  {pages} {17} (\bibinfo {year} {2006})}\BibitemShut {NoStop}%
\bibitem [{\citenamefont {Fradkin}\ \emph {et~al.}(2015)\citenamefont
  {Fradkin}, \citenamefont {Kivelson},\ and\ \citenamefont
  {Tranquada}}]{Fradkin2015}%
  \BibitemOpen
  \bibfield  {author} {\bibinfo {author} {\bibfnamefont {E.}~\bibnamefont
  {Fradkin}}, \bibinfo {author} {\bibfnamefont {S.~A.}\ \bibnamefont
  {Kivelson}}, \ and\ \bibinfo {author} {\bibfnamefont {J.~M.}\ \bibnamefont
  {Tranquada}},\ }\href {\doibase 10.1103/RevModPhys.87.457} {\bibfield
  {journal} {\bibinfo  {journal} {Rev. Mod. Phys.}\ }\textbf {\bibinfo {volume}
  {87}},\ \bibinfo {pages} {457} (\bibinfo {year} {2015})}\BibitemShut
  {NoStop}%
\bibitem [{\citenamefont {Corboz}\ \emph {et~al.}(2014)\citenamefont {Corboz},
  \citenamefont {Rice},\ and\ \citenamefont {Troyer}}]{Corboz2014}%
  \BibitemOpen
  \bibfield  {author} {\bibinfo {author} {\bibfnamefont {P.}~\bibnamefont
  {Corboz}}, \bibinfo {author} {\bibfnamefont {T.~M.}\ \bibnamefont {Rice}}, \
  and\ \bibinfo {author} {\bibfnamefont {M.}~\bibnamefont {Troyer}},\ }\href
  {\doibase 10.1103/PhysRevLett.113.046402} {\bibfield  {journal} {\bibinfo
  {journal} {Phys. Rev. Lett.}\ }\textbf {\bibinfo {volume} {113}},\ \bibinfo
  {pages} {046402} (\bibinfo {year} {2014})}\BibitemShut {NoStop}%
\bibitem [{\citenamefont {Zheng}\ \emph {et~al.}(2017)\citenamefont {Zheng},
  \citenamefont {Chung}, \citenamefont {Corboz}, \citenamefont {Ehlers},
  \citenamefont {Qin}, \citenamefont {Noack}, \citenamefont {Shi},
  \citenamefont {White}, \citenamefont {Zhang},\ and\ \citenamefont
  {Chan}}]{Zheng2017}%
  \BibitemOpen
  \bibfield  {author} {\bibinfo {author} {\bibfnamefont {B.~X.}\ \bibnamefont
  {Zheng}}, \bibinfo {author} {\bibfnamefont {C.~M.}\ \bibnamefont {Chung}},
  \bibinfo {author} {\bibfnamefont {P.}~\bibnamefont {Corboz}}, \bibinfo
  {author} {\bibfnamefont {G.}~\bibnamefont {Ehlers}}, \bibinfo {author}
  {\bibfnamefont {M.~P.}\ \bibnamefont {Qin}}, \bibinfo {author} {\bibfnamefont
  {R.~M.}\ \bibnamefont {Noack}}, \bibinfo {author} {\bibfnamefont
  {H.}~\bibnamefont {Shi}}, \bibinfo {author} {\bibfnamefont {S.~R.}\
  \bibnamefont {White}}, \bibinfo {author} {\bibfnamefont {S.}~\bibnamefont
  {Zhang}}, \ and\ \bibinfo {author} {\bibfnamefont {G.~K.~L.}\ \bibnamefont
  {Chan}},\ }\href {\doibase 10.1126/science.aam7127} {\bibfield  {journal}
  {\bibinfo  {journal} {Science}\ }\textbf {\bibinfo {volume} {358}},\ \bibinfo
  {pages} {1155} (\bibinfo {year} {2017})}\BibitemShut {NoStop}%
\bibitem [{\citenamefont {Huang}\ \emph {et~al.}(2017)\citenamefont {Huang},
  \citenamefont {Mendl}, \citenamefont {Liu}, \citenamefont {Johnston},
  \citenamefont {Jiang}, \citenamefont {Moritz},\ and\ \citenamefont
  {Devereaux}}]{Huang2017science}%
  \BibitemOpen
  \bibfield  {author} {\bibinfo {author} {\bibfnamefont {E.~W.}\ \bibnamefont
  {Huang}}, \bibinfo {author} {\bibfnamefont {C.~B.}\ \bibnamefont {Mendl}},
  \bibinfo {author} {\bibfnamefont {S.}~\bibnamefont {Liu}}, \bibinfo {author}
  {\bibfnamefont {S.}~\bibnamefont {Johnston}}, \bibinfo {author}
  {\bibfnamefont {H.~C.}\ \bibnamefont {Jiang}}, \bibinfo {author}
  {\bibfnamefont {B.}~\bibnamefont {Moritz}}, \ and\ \bibinfo {author}
  {\bibfnamefont {T.~P.}\ \bibnamefont {Devereaux}},\ }\href {\doibase
  10.1126/science.aak9546} {\bibfield  {journal} {\bibinfo  {journal}
  {Science}\ }\textbf {\bibinfo {volume} {358}},\ \bibinfo {pages} {1161}
  (\bibinfo {year} {2017})}\BibitemShut {NoStop}%
\bibitem [{\citenamefont {Dodaro}\ \emph {et~al.}(2017)\citenamefont {Dodaro},
  \citenamefont {Jiang},\ and\ \citenamefont {Kivelson}}]{Dodaro2017}%
  \BibitemOpen
  \bibfield  {author} {\bibinfo {author} {\bibfnamefont {J.~F.}\ \bibnamefont
  {Dodaro}}, \bibinfo {author} {\bibfnamefont {H.~C.}\ \bibnamefont {Jiang}}, \
  and\ \bibinfo {author} {\bibfnamefont {S.~A.}\ \bibnamefont {Kivelson}},\
  }\href {\doibase 10.1103/PhysRevB.95.155116} {\bibfield  {journal} {\bibinfo
  {journal} {Phys. Rev. B}\ }\textbf {\bibinfo {volume} {95}},\ \bibinfo
  {pages} {155116} (\bibinfo {year} {2017})}\BibitemShut {NoStop}%
\bibitem [{\citenamefont {Nocera}\ \emph {et~al.}(2017)\citenamefont {Nocera},
  \citenamefont {Patel}, \citenamefont {Dagotto},\ and\ \citenamefont
  {Alvarez}}]{Nocera2017}%
  \BibitemOpen
  \bibfield  {author} {\bibinfo {author} {\bibfnamefont {A.}~\bibnamefont
  {Nocera}}, \bibinfo {author} {\bibfnamefont {N.~D.}\ \bibnamefont {Patel}},
  \bibinfo {author} {\bibfnamefont {E.}~\bibnamefont {Dagotto}}, \ and\
  \bibinfo {author} {\bibfnamefont {G.}~\bibnamefont {Alvarez}},\ }\href
  {\doibase 10.1103/PhysRevB.96.205120} {\bibfield  {journal} {\bibinfo
  {journal} {Phys. Rev. B}\ }\textbf {\bibinfo {volume} {96}},\ \bibinfo
  {pages} {205120} (\bibinfo {year} {2017})}\BibitemShut {NoStop}%
\bibitem [{\citenamefont {Huang}\ \emph {et~al.}(2018)\citenamefont {Huang},
  \citenamefont {Mendl}, \citenamefont {Jiang}, \citenamefont {Moritz},\ and\
  \citenamefont {Devereaux}}]{Huang2018npjqm}%
  \BibitemOpen
  \bibfield  {author} {\bibinfo {author} {\bibfnamefont {E.~W.}\ \bibnamefont
  {Huang}}, \bibinfo {author} {\bibfnamefont {C.~B.}\ \bibnamefont {Mendl}},
  \bibinfo {author} {\bibfnamefont {H.~C.}\ \bibnamefont {Jiang}}, \bibinfo
  {author} {\bibfnamefont {B.}~\bibnamefont {Moritz}}, \ and\ \bibinfo {author}
  {\bibfnamefont {T.~P.}\ \bibnamefont {Devereaux}},\ }\href {\doibase
  10.1038/s41535-018-0097-0} {\bibfield  {journal} {\bibinfo  {journal} {npj
  Quantum Mater.}\ }\textbf {\bibinfo {volume} {3}},\ \bibinfo {pages} {22}
  (\bibinfo {year} {2018})}\BibitemShut {NoStop}%
\bibitem [{\citenamefont {Jiang}\ \emph {et~al.}(2018)\citenamefont {Jiang},
  \citenamefont {Weng},\ and\ \citenamefont {Kivelson}}]{Jiang2018}%
  \BibitemOpen
  \bibfield  {author} {\bibinfo {author} {\bibfnamefont {H.-C.}\ \bibnamefont
  {Jiang}}, \bibinfo {author} {\bibfnamefont {Z.-Y.}\ \bibnamefont {Weng}}, \
  and\ \bibinfo {author} {\bibfnamefont {S.~A.}\ \bibnamefont {Kivelson}},\
  }\href {\doibase 10.1103/PhysRevB.98.140505} {\bibfield  {journal} {\bibinfo
  {journal} {Phys. Rev. B}\ }\textbf {\bibinfo {volume} {98}},\ \bibinfo
  {pages} {140505} (\bibinfo {year} {2018})}\BibitemShut {NoStop}%
\bibitem [{\citenamefont {Jiang}\ and\ \citenamefont
  {Devereaux}(2019)}]{Jiang2019}%
  \BibitemOpen
  \bibfield  {author} {\bibinfo {author} {\bibfnamefont {H.-C.}\ \bibnamefont
  {Jiang}}\ and\ \bibinfo {author} {\bibfnamefont {T.~P.}\ \bibnamefont
  {Devereaux}},\ }\href {\doibase 10.1126/science.aal5304} {\bibfield
  {journal} {\bibinfo  {journal} {Science}\ }\textbf {\bibinfo {volume}
  {365}},\ \bibinfo {pages} {1424} (\bibinfo {year} {2019})}\BibitemShut
  {NoStop}%
\bibitem [{\citenamefont {Lewenstein}\ \emph {et~al.}(2012)\citenamefont
  {Lewenstein}, \citenamefont {Sanpera},\ and\ \citenamefont
  {Ahufinger}}]{LewensBook}%
  \BibitemOpen
  \bibfield  {author} {\bibinfo {author} {\bibfnamefont {M.}~\bibnamefont
  {Lewenstein}}, \bibinfo {author} {\bibfnamefont {A.}~\bibnamefont {Sanpera}},
  \ and\ \bibinfo {author} {\bibfnamefont {V.}~\bibnamefont {Ahufinger}},\
  }\href {\doibase 10.1093/acprof:oso/9780199573127.001.0001} {\emph {\bibinfo
  {title} {{Ultracold atoms in optical lattices}}}}\ (\bibinfo  {publisher}
  {Oxford University Press},\ \bibinfo {address} {Oxford, UK},\ \bibinfo {year}
  {2012})\BibitemShut {NoStop}%
\bibitem [{\citenamefont {Gross}\ and\ \citenamefont
  {Bloch}(2017)}]{Gross2017}%
  \BibitemOpen
  \bibfield  {author} {\bibinfo {author} {\bibfnamefont {C.}~\bibnamefont
  {Gross}}\ and\ \bibinfo {author} {\bibfnamefont {I.}~\bibnamefont {Bloch}},\
  }\href {\doibase 10.1126/science.aal3837} {\bibfield  {journal} {\bibinfo
  {journal} {Science}\ }\textbf {\bibinfo {volume} {357}},\ \bibinfo {pages}
  {995} (\bibinfo {year} {2017})}\BibitemShut {NoStop}%
\bibitem [{\citenamefont {Greif}\ \emph {et~al.}(2013)\citenamefont {Greif},
  \citenamefont {Uehlinger}, \citenamefont {Jotzu}, \citenamefont {Tarruell},\
  and\ \citenamefont {Esslinger}}]{Greif2013}%
  \BibitemOpen
  \bibfield  {author} {\bibinfo {author} {\bibfnamefont {D.}~\bibnamefont
  {Greif}}, \bibinfo {author} {\bibfnamefont {T.}~\bibnamefont {Uehlinger}},
  \bibinfo {author} {\bibfnamefont {G.}~\bibnamefont {Jotzu}}, \bibinfo
  {author} {\bibfnamefont {L.}~\bibnamefont {Tarruell}}, \ and\ \bibinfo
  {author} {\bibfnamefont {T.}~\bibnamefont {Esslinger}},\ }\href {\doibase
  10.1126/science.1236362} {\bibfield  {journal} {\bibinfo  {journal}
  {Science}\ }\textbf {\bibinfo {volume} {340}},\ \bibinfo {pages} {1307}
  (\bibinfo {year} {2013})}\BibitemShut {NoStop}%
\bibitem [{\citenamefont {Hart}\ \emph {et~al.}(2015)\citenamefont {Hart},
  \citenamefont {Duarte}, \citenamefont {Yang}, \citenamefont {Liu},
  \citenamefont {Paiva}, \citenamefont {Khatami}, \citenamefont {Scalettar},
  \citenamefont {Trivedi}, \citenamefont {Huse},\ and\ \citenamefont
  {Hulet}}]{Hart2015}%
  \BibitemOpen
  \bibfield  {author} {\bibinfo {author} {\bibfnamefont {R.~A.}\ \bibnamefont
  {Hart}}, \bibinfo {author} {\bibfnamefont {P.~M.}\ \bibnamefont {Duarte}},
  \bibinfo {author} {\bibfnamefont {T.-L.}\ \bibnamefont {Yang}}, \bibinfo
  {author} {\bibfnamefont {X.}~\bibnamefont {Liu}}, \bibinfo {author}
  {\bibfnamefont {T.}~\bibnamefont {Paiva}}, \bibinfo {author} {\bibfnamefont
  {E.}~\bibnamefont {Khatami}}, \bibinfo {author} {\bibfnamefont {R.~T.}\
  \bibnamefont {Scalettar}}, \bibinfo {author} {\bibfnamefont {N.}~\bibnamefont
  {Trivedi}}, \bibinfo {author} {\bibfnamefont {D.~A.}\ \bibnamefont {Huse}}, \
  and\ \bibinfo {author} {\bibfnamefont {R.~G.}\ \bibnamefont {Hulet}},\ }\href
  {\doibase 10.1038/nature14223} {\bibfield  {journal} {\bibinfo  {journal}
  {Nature}\ }\textbf {\bibinfo {volume} {519}},\ \bibinfo {pages} {211}
  (\bibinfo {year} {2015})}\BibitemShut {NoStop}%
\bibitem [{\citenamefont {Cocchi}\ \emph {et~al.}(2016)\citenamefont {Cocchi},
  \citenamefont {Miller}, \citenamefont {Drewes}, \citenamefont {Koschorreck},
  \citenamefont {Pertot}, \citenamefont {Brennecke},\ and\ \citenamefont
  {K\"ohl}}]{Cocchi2016}%
  \BibitemOpen
  \bibfield  {author} {\bibinfo {author} {\bibfnamefont {E.}~\bibnamefont
  {Cocchi}}, \bibinfo {author} {\bibfnamefont {L.~A.}\ \bibnamefont {Miller}},
  \bibinfo {author} {\bibfnamefont {J.~H.}\ \bibnamefont {Drewes}}, \bibinfo
  {author} {\bibfnamefont {M.}~\bibnamefont {Koschorreck}}, \bibinfo {author}
  {\bibfnamefont {D.}~\bibnamefont {Pertot}}, \bibinfo {author} {\bibfnamefont
  {F.}~\bibnamefont {Brennecke}}, \ and\ \bibinfo {author} {\bibfnamefont
  {M.}~\bibnamefont {K\"ohl}},\ }\href {\doibase
  10.1103/PhysRevLett.116.175301} {\bibfield  {journal} {\bibinfo  {journal}
  {Phys. Rev. Lett.}\ }\textbf {\bibinfo {volume} {116}},\ \bibinfo {pages}
  {175301} (\bibinfo {year} {2016})}\BibitemShut {NoStop}%
\bibitem [{\citenamefont {Parsons}\ \emph {et~al.}(2016)\citenamefont
  {Parsons}, \citenamefont {Mazurenko}, \citenamefont {Chiu}, \citenamefont
  {Ji}, \citenamefont {Greif},\ and\ \citenamefont {Greiner}}]{Parsons2016}%
  \BibitemOpen
  \bibfield  {author} {\bibinfo {author} {\bibfnamefont {M.~F.}\ \bibnamefont
  {Parsons}}, \bibinfo {author} {\bibfnamefont {A.}~\bibnamefont {Mazurenko}},
  \bibinfo {author} {\bibfnamefont {C.~S.}\ \bibnamefont {Chiu}}, \bibinfo
  {author} {\bibfnamefont {G.}~\bibnamefont {Ji}}, \bibinfo {author}
  {\bibfnamefont {D.}~\bibnamefont {Greif}}, \ and\ \bibinfo {author}
  {\bibfnamefont {M.}~\bibnamefont {Greiner}},\ }\href {\doibase
  10.1126/science.aag1430} {\bibfield  {journal} {\bibinfo  {journal}
  {Science}\ }\textbf {\bibinfo {volume} {353}},\ \bibinfo {pages} {1253}
  (\bibinfo {year} {2016})}\BibitemShut {NoStop}%
\bibitem [{\citenamefont {Boll}\ \emph {et~al.}(2016)\citenamefont {Boll},
  \citenamefont {Hilker}, \citenamefont {Salomon}, \citenamefont {Omran},
  \citenamefont {Nespolo}, \citenamefont {Pollet}, \citenamefont {Bloch},\ and\
  \citenamefont {Gross}}]{Boll2016}%
  \BibitemOpen
  \bibfield  {author} {\bibinfo {author} {\bibfnamefont {M.}~\bibnamefont
  {Boll}}, \bibinfo {author} {\bibfnamefont {T.~A.}\ \bibnamefont {Hilker}},
  \bibinfo {author} {\bibfnamefont {G.}~\bibnamefont {Salomon}}, \bibinfo
  {author} {\bibfnamefont {A.}~\bibnamefont {Omran}}, \bibinfo {author}
  {\bibfnamefont {J.}~\bibnamefont {Nespolo}}, \bibinfo {author} {\bibfnamefont
  {L.}~\bibnamefont {Pollet}}, \bibinfo {author} {\bibfnamefont
  {I.}~\bibnamefont {Bloch}}, \ and\ \bibinfo {author} {\bibfnamefont
  {C.}~\bibnamefont {Gross}},\ }\href {\doibase 10.1126/science.aag1635}
  {\bibfield  {journal} {\bibinfo  {journal} {Science}\ }\textbf {\bibinfo
  {volume} {353}},\ \bibinfo {pages} {1257} (\bibinfo {year}
  {2016})}\BibitemShut {NoStop}%
\bibitem [{\citenamefont {Cheuk}\ \emph {et~al.}(2016)\citenamefont {Cheuk},
  \citenamefont {Nichols}, \citenamefont {Lawrence}, \citenamefont {Okan},
  \citenamefont {Zhang}, \citenamefont {Khatami}, \citenamefont {Trivedi},
  \citenamefont {Paiva}, \citenamefont {Rigol},\ and\ \citenamefont
  {Zwierlein}}]{Cheuk2016}%
  \BibitemOpen
  \bibfield  {author} {\bibinfo {author} {\bibfnamefont {L.~W.}\ \bibnamefont
  {Cheuk}}, \bibinfo {author} {\bibfnamefont {M.~A.}\ \bibnamefont {Nichols}},
  \bibinfo {author} {\bibfnamefont {K.~R.}\ \bibnamefont {Lawrence}}, \bibinfo
  {author} {\bibfnamefont {M.}~\bibnamefont {Okan}}, \bibinfo {author}
  {\bibfnamefont {H.}~\bibnamefont {Zhang}}, \bibinfo {author} {\bibfnamefont
  {E.}~\bibnamefont {Khatami}}, \bibinfo {author} {\bibfnamefont
  {N.}~\bibnamefont {Trivedi}}, \bibinfo {author} {\bibfnamefont
  {T.}~\bibnamefont {Paiva}}, \bibinfo {author} {\bibfnamefont
  {M.}~\bibnamefont {Rigol}}, \ and\ \bibinfo {author} {\bibfnamefont {M.~W.}\
  \bibnamefont {Zwierlein}},\ }\href {\doibase 10.1126/science.aag3349}
  {\bibfield  {journal} {\bibinfo  {journal} {Science}\ }\textbf {\bibinfo
  {volume} {353}},\ \bibinfo {pages} {1260} (\bibinfo {year}
  {2016})}\BibitemShut {NoStop}%
\bibitem [{\citenamefont {Mazurenko}\ \emph {et~al.}(2017)\citenamefont
  {Mazurenko}, \citenamefont {Chiu}, \citenamefont {Ji}, \citenamefont
  {Parsons}, \citenamefont {Kan{\'{a}}sz-Nagy}, \citenamefont {Schmidt},
  \citenamefont {Grusdt}, \citenamefont {Demler}, \citenamefont {Greif},\ and\
  \citenamefont {Greiner}}]{Mazurenko2017}%
  \BibitemOpen
  \bibfield  {author} {\bibinfo {author} {\bibfnamefont {A.}~\bibnamefont
  {Mazurenko}}, \bibinfo {author} {\bibfnamefont {C.~S.}\ \bibnamefont {Chiu}},
  \bibinfo {author} {\bibfnamefont {G.}~\bibnamefont {Ji}}, \bibinfo {author}
  {\bibfnamefont {M.~F.}\ \bibnamefont {Parsons}}, \bibinfo {author}
  {\bibfnamefont {M.}~\bibnamefont {Kan{\'{a}}sz-Nagy}}, \bibinfo {author}
  {\bibfnamefont {R.}~\bibnamefont {Schmidt}}, \bibinfo {author} {\bibfnamefont
  {F.}~\bibnamefont {Grusdt}}, \bibinfo {author} {\bibfnamefont
  {E.}~\bibnamefont {Demler}}, \bibinfo {author} {\bibfnamefont
  {D.}~\bibnamefont {Greif}}, \ and\ \bibinfo {author} {\bibfnamefont
  {M.}~\bibnamefont {Greiner}},\ }\href {\doibase 10.1038/nature22362}
  {\bibfield  {journal} {\bibinfo  {journal} {Nature}\ }\textbf {\bibinfo
  {volume} {545}},\ \bibinfo {pages} {462} (\bibinfo {year}
  {2017})}\BibitemShut {NoStop}%
\bibitem [{\citenamefont {Chiu}\ \emph {et~al.}(2018)\citenamefont {Chiu},
  \citenamefont {Ji}, \citenamefont {Mazurenko}, \citenamefont {Greif},\ and\
  \citenamefont {Greiner}}]{Chiu2018}%
  \BibitemOpen
  \bibfield  {author} {\bibinfo {author} {\bibfnamefont {C.~S.}\ \bibnamefont
  {Chiu}}, \bibinfo {author} {\bibfnamefont {G.}~\bibnamefont {Ji}}, \bibinfo
  {author} {\bibfnamefont {A.}~\bibnamefont {Mazurenko}}, \bibinfo {author}
  {\bibfnamefont {D.}~\bibnamefont {Greif}}, \ and\ \bibinfo {author}
  {\bibfnamefont {M.}~\bibnamefont {Greiner}},\ }\href {\doibase
  10.1103/PhysRevLett.120.243201} {\bibfield  {journal} {\bibinfo  {journal}
  {Phys. Rev. Lett.}\ }\textbf {\bibinfo {volume} {120}},\ \bibinfo {pages}
  {243201} (\bibinfo {year} {2018})}\BibitemShut {NoStop}%
\bibitem [{\citenamefont {Salomon}\ \emph {et~al.}(2019)\citenamefont
  {Salomon}, \citenamefont {Koepsell}, \citenamefont {Vijayan}, \citenamefont
  {Hilker}, \citenamefont {Nespolo}, \citenamefont {Pollet}, \citenamefont
  {Bloch},\ and\ \citenamefont {Gross}}]{Salomon2019}%
  \BibitemOpen
  \bibfield  {author} {\bibinfo {author} {\bibfnamefont {G.}~\bibnamefont
  {Salomon}}, \bibinfo {author} {\bibfnamefont {J.}~\bibnamefont {Koepsell}},
  \bibinfo {author} {\bibfnamefont {J.}~\bibnamefont {Vijayan}}, \bibinfo
  {author} {\bibfnamefont {T.~A.}\ \bibnamefont {Hilker}}, \bibinfo {author}
  {\bibfnamefont {J.}~\bibnamefont {Nespolo}}, \bibinfo {author} {\bibfnamefont
  {L.}~\bibnamefont {Pollet}}, \bibinfo {author} {\bibfnamefont
  {I.}~\bibnamefont {Bloch}}, \ and\ \bibinfo {author} {\bibfnamefont
  {C.}~\bibnamefont {Gross}},\ }\href {\doibase 10.1038/s41586-018-0778-7}
  {\bibfield  {journal} {\bibinfo  {journal} {Nature}\ }\textbf {\bibinfo
  {volume} {565}},\ \bibinfo {pages} {56} (\bibinfo {year} {2019})}\BibitemShut
  {NoStop}%
\bibitem [{\citenamefont {Hilker}\ \emph {et~al.}(2017)\citenamefont {Hilker},
  \citenamefont {Salomon}, \citenamefont {Grusdt}, \citenamefont {Omran},
  \citenamefont {Boll}, \citenamefont {Demler}, \citenamefont {Bloch},\ and\
  \citenamefont {Gross}}]{Hilker2017}%
  \BibitemOpen
  \bibfield  {author} {\bibinfo {author} {\bibfnamefont {T.~A.}\ \bibnamefont
  {Hilker}}, \bibinfo {author} {\bibfnamefont {G.}~\bibnamefont {Salomon}},
  \bibinfo {author} {\bibfnamefont {F.}~\bibnamefont {Grusdt}}, \bibinfo
  {author} {\bibfnamefont {A.}~\bibnamefont {Omran}}, \bibinfo {author}
  {\bibfnamefont {M.}~\bibnamefont {Boll}}, \bibinfo {author} {\bibfnamefont
  {E.}~\bibnamefont {Demler}}, \bibinfo {author} {\bibfnamefont
  {I.}~\bibnamefont {Bloch}}, \ and\ \bibinfo {author} {\bibfnamefont
  {C.}~\bibnamefont {Gross}},\ }\href {\doibase 10.1126/science.aam8990}
  {\bibfield  {journal} {\bibinfo  {journal} {Science}\ }\textbf {\bibinfo
  {volume} {357}},\ \bibinfo {pages} {484} (\bibinfo {year}
  {2017})}\BibitemShut {NoStop}%
\bibitem [{\citenamefont {Ernst}\ \emph {et~al.}(2010)\citenamefont {Ernst},
  \citenamefont {G{\"{o}}tze}, \citenamefont {Krauser}, \citenamefont {Pyka},
  \citenamefont {L{\"{u}}hmann}, \citenamefont {Pfannkuche},\ and\
  \citenamefont {Sengstock}}]{Ernst2010}%
  \BibitemOpen
  \bibfield  {author} {\bibinfo {author} {\bibfnamefont {P.~T.}\ \bibnamefont
  {Ernst}}, \bibinfo {author} {\bibfnamefont {S.}~\bibnamefont {G{\"{o}}tze}},
  \bibinfo {author} {\bibfnamefont {J.~S.}\ \bibnamefont {Krauser}}, \bibinfo
  {author} {\bibfnamefont {K.}~\bibnamefont {Pyka}}, \bibinfo {author}
  {\bibfnamefont {D.-S.}\ \bibnamefont {L{\"{u}}hmann}}, \bibinfo {author}
  {\bibfnamefont {D.}~\bibnamefont {Pfannkuche}}, \ and\ \bibinfo {author}
  {\bibfnamefont {K.}~\bibnamefont {Sengstock}},\ }\href {\doibase
  10.1038/nphys1476} {\bibfield  {journal} {\bibinfo  {journal} {Nature Phys.}\
  }\textbf {\bibinfo {volume} {6}},\ \bibinfo {pages} {56} (\bibinfo {year}
  {2010})}\BibitemShut {NoStop}%
\bibitem [{\citenamefont {Veeravalli}\ \emph {et~al.}(2008)\citenamefont
  {Veeravalli}, \citenamefont {Kuhnle}, \citenamefont {Dyke},\ and\
  \citenamefont {Vale}}]{Veeravalli2008}%
  \BibitemOpen
  \bibfield  {author} {\bibinfo {author} {\bibfnamefont {G.}~\bibnamefont
  {Veeravalli}}, \bibinfo {author} {\bibfnamefont {E.}~\bibnamefont {Kuhnle}},
  \bibinfo {author} {\bibfnamefont {P.}~\bibnamefont {Dyke}}, \ and\ \bibinfo
  {author} {\bibfnamefont {C.~J.}\ \bibnamefont {Vale}},\ }\href {\doibase
  10.1103/PhysRevLett.101.250403} {\bibfield  {journal} {\bibinfo  {journal}
  {Phys. Rev. Lett.}\ }\textbf {\bibinfo {volume} {101}},\ \bibinfo {pages}
  {250403} (\bibinfo {year} {2008})}\BibitemShut {NoStop}%
\bibitem [{\citenamefont {Usui}\ \emph {et~al.}(2018)\citenamefont {Usui},
  \citenamefont {Bu{\v{c}}a},\ and\ \citenamefont {Mur-Petit}}]{Usui2018}%
  \BibitemOpen
  \bibfield  {author} {\bibinfo {author} {\bibfnamefont {A.}~\bibnamefont
  {Usui}}, \bibinfo {author} {\bibfnamefont {B.}~\bibnamefont {Bu{\v{c}}a}}, \
  and\ \bibinfo {author} {\bibfnamefont {J.}~\bibnamefont {Mur-Petit}},\ }\href
  {\doibase 10.1088/1367-2630/aae418} {\bibfield  {journal} {\bibinfo
  {journal} {New J. Phys.}\ }\textbf {\bibinfo {volume} {20}},\ \bibinfo
  {pages} {103006} (\bibinfo {year} {2018})}\BibitemShut {NoStop}%
\bibitem [{\citenamefont {Kollath}\ \emph {et~al.}(2007)\citenamefont
  {Kollath}, \citenamefont {K\"ohl},\ and\ \citenamefont
  {Giamarchi}}]{Kollath2007}%
  \BibitemOpen
  \bibfield  {author} {\bibinfo {author} {\bibfnamefont {C.}~\bibnamefont
  {Kollath}}, \bibinfo {author} {\bibfnamefont {M.}~\bibnamefont {K\"ohl}}, \
  and\ \bibinfo {author} {\bibfnamefont {T.}~\bibnamefont {Giamarchi}},\ }\href
  {\doibase 10.1103/PhysRevA.76.063602} {\bibfield  {journal} {\bibinfo
  {journal} {Phys. Rev. A}\ }\textbf {\bibinfo {volume} {76}},\ \bibinfo
  {pages} {063602} (\bibinfo {year} {2007})}\BibitemShut {NoStop}%
\bibitem [{\citenamefont {Endres}\ \emph {et~al.}(2013)\citenamefont {Endres},
  \citenamefont {Cheneau}, \citenamefont {Fukuhara}, \citenamefont
  {Weitenberg}, \citenamefont {Schau{\ss}}, \citenamefont {Gross},
  \citenamefont {Mazza}, \citenamefont {Ba{\~{n}}uls}, \citenamefont {Pollet},
  \citenamefont {Bloch},\ and\ \citenamefont {Kuhr}}]{Endres2013}%
  \BibitemOpen
  \bibfield  {author} {\bibinfo {author} {\bibfnamefont {M.}~\bibnamefont
  {Endres}}, \bibinfo {author} {\bibfnamefont {M.}~\bibnamefont {Cheneau}},
  \bibinfo {author} {\bibfnamefont {T.}~\bibnamefont {Fukuhara}}, \bibinfo
  {author} {\bibfnamefont {C.}~\bibnamefont {Weitenberg}}, \bibinfo {author}
  {\bibfnamefont {P.}~\bibnamefont {Schau{\ss}}}, \bibinfo {author}
  {\bibfnamefont {C.}~\bibnamefont {Gross}}, \bibinfo {author} {\bibfnamefont
  {L.}~\bibnamefont {Mazza}}, \bibinfo {author} {\bibfnamefont {M.~C.}\
  \bibnamefont {Ba{\~{n}}uls}}, \bibinfo {author} {\bibfnamefont
  {L.}~\bibnamefont {Pollet}}, \bibinfo {author} {\bibfnamefont
  {I.}~\bibnamefont {Bloch}}, \ and\ \bibinfo {author} {\bibfnamefont
  {S.}~\bibnamefont {Kuhr}},\ }\href {\doibase 10.1007/s00340-013-5552-9}
  {\bibfield  {journal} {\bibinfo  {journal} {Appl. Phys. B}\ }\textbf
  {\bibinfo {volume} {113}},\ \bibinfo {pages} {27} (\bibinfo {year}
  {2013})}\BibitemShut {NoStop}%
\bibitem [{\citenamefont {Streif}\ \emph {et~al.}(2016)\citenamefont {Streif},
  \citenamefont {Buchleitner}, \citenamefont {Jaksch},\ and\ \citenamefont
  {Mur-Petit}}]{Streif2016}%
  \BibitemOpen
  \bibfield  {author} {\bibinfo {author} {\bibfnamefont {M.}~\bibnamefont
  {Streif}}, \bibinfo {author} {\bibfnamefont {A.}~\bibnamefont {Buchleitner}},
  \bibinfo {author} {\bibfnamefont {D.}~\bibnamefont {Jaksch}}, \ and\ \bibinfo
  {author} {\bibfnamefont {J.}~\bibnamefont {Mur-Petit}},\ }\href {\doibase
  10.1103/PhysRevA.94.053634} {\bibfield  {journal} {\bibinfo  {journal} {Phys.
  Rev. A}\ }\textbf {\bibinfo {volume} {94}},\ \bibinfo {pages} {053634}
  (\bibinfo {year} {2016})}\BibitemShut {NoStop}%
\bibitem [{\citenamefont {Auerbach}(1994)}]{AuerbachBook}%
  \BibitemOpen
  \bibfield  {author} {\bibinfo {author} {\bibfnamefont {A.}~\bibnamefont
  {Auerbach}},\ }\href@noop {} {\emph {\bibinfo {title} {{Interacting Electrons
  and Quantum Magnetism}}}}\ (\bibinfo  {publisher} {Springer-Verlag},\
  \bibinfo {address} {New York, New York, USA},\ \bibinfo {year}
  {1994})\BibitemShut {NoStop}%
\bibitem [{\citenamefont {Sorella}\ \emph {et~al.}(2002)\citenamefont
  {Sorella}, \citenamefont {Martins}, \citenamefont {Becca}, \citenamefont
  {Gazza}, \citenamefont {Capriotti}, \citenamefont {Parola},\ and\
  \citenamefont {Dagotto}}]{Sorella2002}%
  \BibitemOpen
  \bibfield  {author} {\bibinfo {author} {\bibfnamefont {S.}~\bibnamefont
  {Sorella}}, \bibinfo {author} {\bibfnamefont {G.~B.}\ \bibnamefont
  {Martins}}, \bibinfo {author} {\bibfnamefont {F.}~\bibnamefont {Becca}},
  \bibinfo {author} {\bibfnamefont {C.}~\bibnamefont {Gazza}}, \bibinfo
  {author} {\bibfnamefont {L.}~\bibnamefont {Capriotti}}, \bibinfo {author}
  {\bibfnamefont {A.}~\bibnamefont {Parola}}, \ and\ \bibinfo {author}
  {\bibfnamefont {E.}~\bibnamefont {Dagotto}},\ }\href {\doibase
  10.1103/PhysRevLett.88.117002} {\bibfield  {journal} {\bibinfo  {journal}
  {Phys. Rev. Lett.}\ }\textbf {\bibinfo {volume} {88}},\ \bibinfo {pages}
  {117002} (\bibinfo {year} {2002})}\BibitemShut {NoStop}%
\bibitem [{\citenamefont {Coulthard}\ \emph {et~al.}(2017)\citenamefont
  {Coulthard}, \citenamefont {Clark}, \citenamefont {Al-Assam}, \citenamefont
  {Cavalleri},\ and\ \citenamefont {Jaksch}}]{Coulthard2017}%
  \BibitemOpen
  \bibfield  {author} {\bibinfo {author} {\bibfnamefont {J.~R.}\ \bibnamefont
  {Coulthard}}, \bibinfo {author} {\bibfnamefont {S.~R.}\ \bibnamefont
  {Clark}}, \bibinfo {author} {\bibfnamefont {S.}~\bibnamefont {Al-Assam}},
  \bibinfo {author} {\bibfnamefont {A.}~\bibnamefont {Cavalleri}}, \ and\
  \bibinfo {author} {\bibfnamefont {D.}~\bibnamefont {Jaksch}},\ }\href
  {\doibase 10.1103/PhysRevB.96.085104} {\bibfield  {journal} {\bibinfo
  {journal} {Phys. Rev. B}\ }\textbf {\bibinfo {volume} {96}},\ \bibinfo
  {pages} {085104} (\bibinfo {year} {2017})}\BibitemShut {NoStop}%
\bibitem [{\citenamefont {Bukov}\ \emph {et~al.}(2015)\citenamefont {Bukov},
  \citenamefont {D'Alessio},\ and\ \citenamefont {Polkovnikov}}]{Bukov2015}%
  \BibitemOpen
  \bibfield  {author} {\bibinfo {author} {\bibfnamefont {M.}~\bibnamefont
  {Bukov}}, \bibinfo {author} {\bibfnamefont {L.}~\bibnamefont {D'Alessio}}, \
  and\ \bibinfo {author} {\bibfnamefont {A.}~\bibnamefont {Polkovnikov}},\
  }\href {\doibase 10.1080/00018732.2015.1055918} {\bibfield  {journal}
  {\bibinfo  {journal} {Adv. Phys.}\ }\textbf {\bibinfo {volume} {64}},\
  \bibinfo {pages} {139} (\bibinfo {year} {2015})}\BibitemShut {NoStop}%
\bibitem [{\citenamefont {Eckardt}(2017)}]{Eckardt2017}%
  \BibitemOpen
  \bibfield  {author} {\bibinfo {author} {\bibfnamefont {A.}~\bibnamefont
  {Eckardt}},\ }\href {\doibase 10.1103/RevModPhys.89.011004} {\bibfield
  {journal} {\bibinfo  {journal} {Rev. Mod. Phys.}\ }\textbf {\bibinfo {volume}
  {89}},\ \bibinfo {pages} {011004} (\bibinfo {year} {2017})}\BibitemShut
  {NoStop}%
\bibitem [{\citenamefont {Jotzu}\ \emph {et~al.}(2014)\citenamefont {Jotzu},
  \citenamefont {Messer}, \citenamefont {Desbuquois}, \citenamefont {Lebrat},
  \citenamefont {Uehlinger}, \citenamefont {Greif},\ and\ \citenamefont
  {Esslinger}}]{Jotzu2014}%
  \BibitemOpen
  \bibfield  {author} {\bibinfo {author} {\bibfnamefont {G.}~\bibnamefont
  {Jotzu}}, \bibinfo {author} {\bibfnamefont {M.}~\bibnamefont {Messer}},
  \bibinfo {author} {\bibfnamefont {R.}~\bibnamefont {Desbuquois}}, \bibinfo
  {author} {\bibfnamefont {M.}~\bibnamefont {Lebrat}}, \bibinfo {author}
  {\bibfnamefont {T.}~\bibnamefont {Uehlinger}}, \bibinfo {author}
  {\bibfnamefont {D.}~\bibnamefont {Greif}}, \ and\ \bibinfo {author}
  {\bibfnamefont {T.}~\bibnamefont {Esslinger}},\ }\href {\doibase
  10.1038/nature13915} {\bibfield  {journal} {\bibinfo  {journal} {Nature}\
  }\textbf {\bibinfo {volume} {515}},\ \bibinfo {pages} {237} (\bibinfo {year}
  {2014})}\BibitemShut {NoStop}%
\bibitem [{\citenamefont {Desbuquois}\ \emph {et~al.}(2017)\citenamefont
  {Desbuquois}, \citenamefont {Messer}, \citenamefont {G{\"{o}}rg},
  \citenamefont {Sandholzer}, \citenamefont {Jotzu},\ and\ \citenamefont
  {Esslinger}}]{Desbuquois2017}%
  \BibitemOpen
  \bibfield  {author} {\bibinfo {author} {\bibfnamefont {R.}~\bibnamefont
  {Desbuquois}}, \bibinfo {author} {\bibfnamefont {M.}~\bibnamefont {Messer}},
  \bibinfo {author} {\bibfnamefont {F.}~\bibnamefont {G{\"{o}}rg}}, \bibinfo
  {author} {\bibfnamefont {K.}~\bibnamefont {Sandholzer}}, \bibinfo {author}
  {\bibfnamefont {G.}~\bibnamefont {Jotzu}}, \ and\ \bibinfo {author}
  {\bibfnamefont {T.}~\bibnamefont {Esslinger}},\ }\href {\doibase
  10.1103/PhysRevA.96.053602} {\bibfield  {journal} {\bibinfo  {journal} {Phys.
  Rev. A}\ }\textbf {\bibinfo {volume} {96}},\ \bibinfo {pages} {053602}
  (\bibinfo {year} {2017})}\BibitemShut {NoStop}%
\bibitem [{\citenamefont {G{\"{o}}rg}\ \emph {et~al.}(2018)\citenamefont
  {G{\"{o}}rg}, \citenamefont {Messer}, \citenamefont {Sandholzer},
  \citenamefont {Jotzu}, \citenamefont {Desbuquois},\ and\ \citenamefont
  {Esslinger}}]{Gorg2018}%
  \BibitemOpen
  \bibfield  {author} {\bibinfo {author} {\bibfnamefont {F.}~\bibnamefont
  {G{\"{o}}rg}}, \bibinfo {author} {\bibfnamefont {M.}~\bibnamefont {Messer}},
  \bibinfo {author} {\bibfnamefont {K.}~\bibnamefont {Sandholzer}}, \bibinfo
  {author} {\bibfnamefont {G.}~\bibnamefont {Jotzu}}, \bibinfo {author}
  {\bibfnamefont {R.}~\bibnamefont {Desbuquois}}, \ and\ \bibinfo {author}
  {\bibfnamefont {T.}~\bibnamefont {Esslinger}},\ }\href {\doibase
  10.1038/nature25135} {\bibfield  {journal} {\bibinfo  {journal} {Nature}\
  }\textbf {\bibinfo {volume} {553}},\ \bibinfo {pages} {481} (\bibinfo {year}
  {2018})}\BibitemShut {NoStop}%
\bibitem [{\citenamefont {Messer}\ \emph {et~al.}(2018)\citenamefont {Messer},
  \citenamefont {Sandholzer}, \citenamefont {G\"org}, \citenamefont {Minguzzi},
  \citenamefont {Desbuquois},\ and\ \citenamefont {Esslinger}}]{Messer2018}%
  \BibitemOpen
  \bibfield  {author} {\bibinfo {author} {\bibfnamefont {M.}~\bibnamefont
  {Messer}}, \bibinfo {author} {\bibfnamefont {K.}~\bibnamefont {Sandholzer}},
  \bibinfo {author} {\bibfnamefont {F.}~\bibnamefont {G\"org}}, \bibinfo
  {author} {\bibfnamefont {J.}~\bibnamefont {Minguzzi}}, \bibinfo {author}
  {\bibfnamefont {R.}~\bibnamefont {Desbuquois}}, \ and\ \bibinfo {author}
  {\bibfnamefont {T.}~\bibnamefont {Esslinger}},\ }\href {\doibase
  10.1103/PhysRevLett.121.233603} {\bibfield  {journal} {\bibinfo  {journal}
  {Phys. Rev. Lett.}\ }\textbf {\bibinfo {volume} {121}},\ \bibinfo {pages}
  {233603} (\bibinfo {year} {2018})}\BibitemShut {NoStop}%
\bibitem [{\citenamefont {Aidelsburger}\ \emph {et~al.}(2011)\citenamefont
  {Aidelsburger}, \citenamefont {Atala}, \citenamefont {Nascimb\`ene},
  \citenamefont {Trotzky}, \citenamefont {Chen},\ and\ \citenamefont
  {Bloch}}]{Aidelsburger2011}%
  \BibitemOpen
  \bibfield  {author} {\bibinfo {author} {\bibfnamefont {M.}~\bibnamefont
  {Aidelsburger}}, \bibinfo {author} {\bibfnamefont {M.}~\bibnamefont {Atala}},
  \bibinfo {author} {\bibfnamefont {S.}~\bibnamefont {Nascimb\`ene}}, \bibinfo
  {author} {\bibfnamefont {S.}~\bibnamefont {Trotzky}}, \bibinfo {author}
  {\bibfnamefont {Y.-A.}\ \bibnamefont {Chen}}, \ and\ \bibinfo {author}
  {\bibfnamefont {I.}~\bibnamefont {Bloch}},\ }\href {\doibase
  10.1103/PhysRevLett.107.255301} {\bibfield  {journal} {\bibinfo  {journal}
  {Phys. Rev. Lett.}\ }\textbf {\bibinfo {volume} {107}},\ \bibinfo {pages}
  {255301} (\bibinfo {year} {2011})}\BibitemShut {NoStop}%
\bibitem [{\citenamefont {Struck}\ \emph {et~al.}(2012)\citenamefont {Struck},
  \citenamefont {Olschl{\"{a}}ger}, \citenamefont {Weinberg}, \citenamefont
  {Hauke}, \citenamefont {Simonet}, \citenamefont {Eckardt}, \citenamefont
  {Lewenstein}, \citenamefont {Sengstock},\ and\ \citenamefont
  {Windpassinger}}]{Struck2012}%
  \BibitemOpen
  \bibfield  {author} {\bibinfo {author} {\bibfnamefont {J.}~\bibnamefont
  {Struck}}, \bibinfo {author} {\bibfnamefont {C.}~\bibnamefont
  {Olschl{\"{a}}ger}}, \bibinfo {author} {\bibfnamefont {M.}~\bibnamefont
  {Weinberg}}, \bibinfo {author} {\bibfnamefont {P.}~\bibnamefont {Hauke}},
  \bibinfo {author} {\bibfnamefont {J.}~\bibnamefont {Simonet}}, \bibinfo
  {author} {\bibfnamefont {A.}~\bibnamefont {Eckardt}}, \bibinfo {author}
  {\bibfnamefont {M.}~\bibnamefont {Lewenstein}}, \bibinfo {author}
  {\bibfnamefont {K.}~\bibnamefont {Sengstock}}, \ and\ \bibinfo {author}
  {\bibfnamefont {P.}~\bibnamefont {Windpassinger}},\ }\href {\doibase
  10.1103/PhysRevLett.108.225304} {\bibfield  {journal} {\bibinfo  {journal}
  {Phys. Rev. Lett.}\ }\textbf {\bibinfo {volume} {108}},\ \bibinfo {pages}
  {225304} (\bibinfo {year} {2012})}\BibitemShut {NoStop}%
\bibitem [{\citenamefont {Tai}\ \emph {et~al.}(2017)\citenamefont {Tai},
  \citenamefont {Lukin}, \citenamefont {Rispoli}, \citenamefont {Schittko},
  \citenamefont {Menke}, \citenamefont {Borgnia}, \citenamefont {Preiss},
  \citenamefont {Grusdt}, \citenamefont {Kaufman},\ and\ \citenamefont
  {Greiner}}]{Tai2017}%
  \BibitemOpen
  \bibfield  {author} {\bibinfo {author} {\bibfnamefont {M.~E.}\ \bibnamefont
  {Tai}}, \bibinfo {author} {\bibfnamefont {A.}~\bibnamefont {Lukin}}, \bibinfo
  {author} {\bibfnamefont {M.}~\bibnamefont {Rispoli}}, \bibinfo {author}
  {\bibfnamefont {R.}~\bibnamefont {Schittko}}, \bibinfo {author}
  {\bibfnamefont {T.}~\bibnamefont {Menke}}, \bibinfo {author} {\bibfnamefont
  {D.}~\bibnamefont {Borgnia}}, \bibinfo {author} {\bibfnamefont {P.~M.}\
  \bibnamefont {Preiss}}, \bibinfo {author} {\bibfnamefont {F.}~\bibnamefont
  {Grusdt}}, \bibinfo {author} {\bibfnamefont {A.~M.}\ \bibnamefont {Kaufman}},
  \ and\ \bibinfo {author} {\bibfnamefont {M.}~\bibnamefont {Greiner}},\ }\href
  {\doibase 10.1038/nature22811} {\bibfield  {journal} {\bibinfo  {journal}
  {Nature}\ }\textbf {\bibinfo {volume} {546}},\ \bibinfo {pages} {519}
  (\bibinfo {year} {2017})}\BibitemShut {NoStop}%
\bibitem [{\citenamefont {McIver}\ \emph {et~al.}(2019)\citenamefont {McIver},
  \citenamefont {Schulte}, \citenamefont {Stein}, \citenamefont {Matsuyama},
  \citenamefont {Jotzu}, \citenamefont {Meier},\ and\ \citenamefont
  {Cavalleri}}]{McIver2019}%
  \BibitemOpen
  \bibfield  {author} {\bibinfo {author} {\bibfnamefont {J.~W.}\ \bibnamefont
  {McIver}}, \bibinfo {author} {\bibfnamefont {B.}~\bibnamefont {Schulte}},
  \bibinfo {author} {\bibfnamefont {F.-U.}\ \bibnamefont {Stein}}, \bibinfo
  {author} {\bibfnamefont {T.}~\bibnamefont {Matsuyama}}, \bibinfo {author}
  {\bibfnamefont {G.}~\bibnamefont {Jotzu}}, \bibinfo {author} {\bibfnamefont
  {G.}~\bibnamefont {Meier}}, \ and\ \bibinfo {author} {\bibfnamefont
  {A.}~\bibnamefont {Cavalleri}},\ }\href {\doibase 10.1038/s41567-019-0698-y}
  {\bibfield  {journal} {\bibinfo  {journal} {Nature Phys.}\ }\textbf {\bibinfo
  {volume} {16}},\ \bibinfo {pages} {38} (\bibinfo {year} {2019})}\BibitemShut
  {NoStop}%
\bibitem [{\citenamefont {Mentink}\ \emph {et~al.}(2015)\citenamefont
  {Mentink}, \citenamefont {Balzer},\ and\ \citenamefont
  {Eckstein}}]{Mentink2015}%
  \BibitemOpen
  \bibfield  {author} {\bibinfo {author} {\bibfnamefont {J.~H.}\ \bibnamefont
  {Mentink}}, \bibinfo {author} {\bibfnamefont {K.}~\bibnamefont {Balzer}}, \
  and\ \bibinfo {author} {\bibfnamefont {M.}~\bibnamefont {Eckstein}},\ }\href
  {\doibase 10.1038/ncomms7708} {\bibfield  {journal} {\bibinfo  {journal}
  {Nature Commun.}\ }\textbf {\bibinfo {volume} {6}},\ \bibinfo {pages} {6708}
  (\bibinfo {year} {2015})}\BibitemShut {NoStop}%
\bibitem [{\citenamefont {Dasari}\ and\ \citenamefont
  {Eckstein}(2018)}]{Dasari2018}%
  \BibitemOpen
  \bibfield  {author} {\bibinfo {author} {\bibfnamefont {N.}~\bibnamefont
  {Dasari}}\ and\ \bibinfo {author} {\bibfnamefont {M.}~\bibnamefont
  {Eckstein}},\ }\href {\doibase 10.1103/PhysRevB.98.235149} {\bibfield
  {journal} {\bibinfo  {journal} {Phys. Rev. B}\ }\textbf {\bibinfo {volume}
  {98}},\ \bibinfo {pages} {235149} (\bibinfo {year} {2018})}\BibitemShut
  {NoStop}%
\bibitem [{\citenamefont {Bermudez}\ and\ \citenamefont
  {Porras}(2015)}]{Bermudez2015}%
  \BibitemOpen
  \bibfield  {author} {\bibinfo {author} {\bibfnamefont {A.}~\bibnamefont
  {Bermudez}}\ and\ \bibinfo {author} {\bibfnamefont {D.}~\bibnamefont
  {Porras}},\ }\href {\doibase 10.1088/1367-2630/17/10/103021} {\bibfield
  {journal} {\bibinfo  {journal} {New J. Phys.}\ }\textbf {\bibinfo {volume}
  {17}},\ \bibinfo {pages} {103021} (\bibinfo {year} {2015})}\BibitemShut
  {NoStop}%
\bibitem [{\citenamefont {Bukov}\ \emph {et~al.}(2016)\citenamefont {Bukov},
  \citenamefont {Kolodrubetz},\ and\ \citenamefont {Polkovnikov}}]{Bukov2016}%
  \BibitemOpen
  \bibfield  {author} {\bibinfo {author} {\bibfnamefont {M.}~\bibnamefont
  {Bukov}}, \bibinfo {author} {\bibfnamefont {M.}~\bibnamefont {Kolodrubetz}},
  \ and\ \bibinfo {author} {\bibfnamefont {A.}~\bibnamefont {Polkovnikov}},\
  }\href {\doibase 10.1103/PhysRevLett.116.125301} {\bibfield  {journal}
  {\bibinfo  {journal} {Phys. Rev. Lett.}\ }\textbf {\bibinfo {volume} {116}},\
  \bibinfo {pages} {125301} (\bibinfo {year} {2016})}\BibitemShut {NoStop}%
\bibitem [{\citenamefont {Duan}\ \emph {et~al.}(2003)\citenamefont {Duan},
  \citenamefont {Demler},\ and\ \citenamefont {Lukin}}]{Duan2003}%
  \BibitemOpen
  \bibfield  {author} {\bibinfo {author} {\bibfnamefont {L.-M.}\ \bibnamefont
  {Duan}}, \bibinfo {author} {\bibfnamefont {E.}~\bibnamefont {Demler}}, \ and\
  \bibinfo {author} {\bibfnamefont {M.~D.}\ \bibnamefont {Lukin}},\ }\href
  {\doibase 10.1103/PhysRevLett.91.090402} {\bibfield  {journal} {\bibinfo
  {journal} {Phys. Rev. Lett.}\ }\textbf {\bibinfo {volume} {91}},\ \bibinfo
  {pages} {090402} (\bibinfo {year} {2003})}\BibitemShut {NoStop}%
\bibitem [{\citenamefont {Hofstetter}\ and\ \citenamefont
  {Qin}(2018)}]{Hofstetter2018}%
  \BibitemOpen
  \bibfield  {author} {\bibinfo {author} {\bibfnamefont {W.}~\bibnamefont
  {Hofstetter}}\ and\ \bibinfo {author} {\bibfnamefont {T.}~\bibnamefont
  {Qin}},\ }\href {\doibase 10.1088/1361-6455/aaa31b} {\bibfield  {journal}
  {\bibinfo  {journal} {J. Phys. B}\ }\textbf {\bibinfo {volume} {51}},\
  \bibinfo {pages} {082001} (\bibinfo {year} {2018})}\BibitemShut {NoStop}%
\bibitem [{\citenamefont {Kaiser}\ \emph {et~al.}(2014)\citenamefont {Kaiser},
  \citenamefont {Hunt}, \citenamefont {Nicoletti}, \citenamefont {Hu},
  \citenamefont {Gierz}, \citenamefont {Liu}, \citenamefont {Le~Tacon},
  \citenamefont {Loew}, \citenamefont {Haug}, \citenamefont {Keimer},\ and\
  \citenamefont {Cavalleri}}]{Kaiser2014a}%
  \BibitemOpen
  \bibfield  {author} {\bibinfo {author} {\bibfnamefont {S.}~\bibnamefont
  {Kaiser}}, \bibinfo {author} {\bibfnamefont {C.~R.}\ \bibnamefont {Hunt}},
  \bibinfo {author} {\bibfnamefont {D.}~\bibnamefont {Nicoletti}}, \bibinfo
  {author} {\bibfnamefont {W.}~\bibnamefont {Hu}}, \bibinfo {author}
  {\bibfnamefont {I.}~\bibnamefont {Gierz}}, \bibinfo {author} {\bibfnamefont
  {H.~Y.}\ \bibnamefont {Liu}}, \bibinfo {author} {\bibfnamefont
  {M.}~\bibnamefont {Le~Tacon}}, \bibinfo {author} {\bibfnamefont
  {T.}~\bibnamefont {Loew}}, \bibinfo {author} {\bibfnamefont {D.}~\bibnamefont
  {Haug}}, \bibinfo {author} {\bibfnamefont {B.}~\bibnamefont {Keimer}}, \ and\
  \bibinfo {author} {\bibfnamefont {A.}~\bibnamefont {Cavalleri}},\ }\href
  {\doibase 10.1103/PhysRevB.89.184516} {\bibfield  {journal} {\bibinfo
  {journal} {Phys. Rev. B}\ }\textbf {\bibinfo {volume} {89}},\ \bibinfo
  {pages} {184516} (\bibinfo {year} {2014})}\BibitemShut {NoStop}%
\bibitem [{\citenamefont {Mankowsky}\ \emph {et~al.}(2014)\citenamefont
  {Mankowsky}, \citenamefont {Subedi}, \citenamefont {F\"{o}rst}, \citenamefont
  {Mariager}, \citenamefont {Chollet}, \citenamefont {Lemke}, \citenamefont
  {Robinson}, \citenamefont {Glownia}, \citenamefont {Minitti}, \citenamefont
  {Frano}, \citenamefont {Fechner}, \citenamefont {Spaldin}, \citenamefont
  {Loew}, \citenamefont {Keimer}, \citenamefont {Georges},\ and\ \citenamefont
  {Cavalleri}}]{Mankowsky2014}%
  \BibitemOpen
  \bibfield  {author} {\bibinfo {author} {\bibfnamefont {R.}~\bibnamefont
  {Mankowsky}}, \bibinfo {author} {\bibfnamefont {A.}~\bibnamefont {Subedi}},
  \bibinfo {author} {\bibfnamefont {M.}~\bibnamefont {F\"{o}rst}}, \bibinfo
  {author} {\bibfnamefont {S.~O.}\ \bibnamefont {Mariager}}, \bibinfo {author}
  {\bibfnamefont {M.}~\bibnamefont {Chollet}}, \bibinfo {author} {\bibfnamefont
  {H.~T.}\ \bibnamefont {Lemke}}, \bibinfo {author} {\bibfnamefont {J.~S.}\
  \bibnamefont {Robinson}}, \bibinfo {author} {\bibfnamefont {J.~M.}\
  \bibnamefont {Glownia}}, \bibinfo {author} {\bibfnamefont {M.~P.}\
  \bibnamefont {Minitti}}, \bibinfo {author} {\bibfnamefont {A.}~\bibnamefont
  {Frano}}, \bibinfo {author} {\bibfnamefont {M.}~\bibnamefont {Fechner}},
  \bibinfo {author} {\bibfnamefont {N.~A.}\ \bibnamefont {Spaldin}}, \bibinfo
  {author} {\bibfnamefont {T.}~\bibnamefont {Loew}}, \bibinfo {author}
  {\bibfnamefont {B.}~\bibnamefont {Keimer}}, \bibinfo {author} {\bibfnamefont
  {A.}~\bibnamefont {Georges}}, \ and\ \bibinfo {author} {\bibfnamefont
  {A.}~\bibnamefont {Cavalleri}},\ }\href {\doibase 10.1038/nature13875}
  {\bibfield  {journal} {\bibinfo  {journal} {Nature}\ }\textbf {\bibinfo
  {volume} {516}},\ \bibinfo {pages} {71} (\bibinfo {year} {2014})}\BibitemShut
  {NoStop}%
\bibitem [{\citenamefont {Mitrano}\ \emph {et~al.}(2016)\citenamefont
  {Mitrano}, \citenamefont {Cantaluppi}, \citenamefont {Nicoletti},
  \citenamefont {Kaiser}, \citenamefont {Perucchi}, \citenamefont {Lupi},
  \citenamefont {Di~Pietro}, \citenamefont {Pontiroli}, \citenamefont
  {Ricc{\`o}}, \citenamefont {Clark}, \citenamefont {Jaksch},\ and\
  \citenamefont {Cavalleri}}]{Mitrano2016}%
  \BibitemOpen
  \bibfield  {author} {\bibinfo {author} {\bibfnamefont {M.}~\bibnamefont
  {Mitrano}}, \bibinfo {author} {\bibfnamefont {A.}~\bibnamefont {Cantaluppi}},
  \bibinfo {author} {\bibfnamefont {D.}~\bibnamefont {Nicoletti}}, \bibinfo
  {author} {\bibfnamefont {S.}~\bibnamefont {Kaiser}}, \bibinfo {author}
  {\bibfnamefont {A.}~\bibnamefont {Perucchi}}, \bibinfo {author}
  {\bibfnamefont {S.}~\bibnamefont {Lupi}}, \bibinfo {author} {\bibfnamefont
  {P.}~\bibnamefont {Di~Pietro}}, \bibinfo {author} {\bibfnamefont
  {D.}~\bibnamefont {Pontiroli}}, \bibinfo {author} {\bibfnamefont
  {M.}~\bibnamefont {Ricc{\`o}}}, \bibinfo {author} {\bibfnamefont {S.~R.}\
  \bibnamefont {Clark}}, \bibinfo {author} {\bibfnamefont {D.}~\bibnamefont
  {Jaksch}}, \ and\ \bibinfo {author} {\bibfnamefont {A.}~\bibnamefont
  {Cavalleri}},\ }\href {http://dx.doi.org/10.1038/nature16522} {\bibfield
  {journal} {\bibinfo  {journal} {Nature}\ }\textbf {\bibinfo {volume} {530}},\
  \bibinfo {pages} {461} (\bibinfo {year} {2016})}\BibitemShut {NoStop}%
\bibitem [{\citenamefont {Gao}\ \emph {et~al.}(2020)\citenamefont {Gao},
  \citenamefont {Coulthard}, \citenamefont {Jaksch},\ and\ \citenamefont
  {Mur-Petit}}]{XXXL}%
  \BibitemOpen
  \bibfield  {author} {\bibinfo {author} {\bibfnamefont {H.}~\bibnamefont
  {Gao}}, \bibinfo {author} {\bibfnamefont {J.~R.}\ \bibnamefont {Coulthard}},
  \bibinfo {author} {\bibfnamefont {D.}~\bibnamefont {Jaksch}}, \ and\ \bibinfo
  {author} {\bibfnamefont {J.}~\bibnamefont {Mur-Petit}},\ }\href@noop {} {\
  (\bibinfo {year} {2020})},\ \bibinfo {note} {to be submitted}\BibitemShut
  {NoStop}%
\bibitem [{\citenamefont {Chiu}\ \emph {et~al.}(2019)\citenamefont {Chiu},
  \citenamefont {Ji}, \citenamefont {Bohrdt}, \citenamefont {Xu}, \citenamefont
  {Knap}, \citenamefont {Demler}, \citenamefont {Grusdt}, \citenamefont
  {Greiner},\ and\ \citenamefont {Greif}}]{Chiu2019}%
  \BibitemOpen
  \bibfield  {author} {\bibinfo {author} {\bibfnamefont {C.~S.}\ \bibnamefont
  {Chiu}}, \bibinfo {author} {\bibfnamefont {G.}~\bibnamefont {Ji}}, \bibinfo
  {author} {\bibfnamefont {A.}~\bibnamefont {Bohrdt}}, \bibinfo {author}
  {\bibfnamefont {M.}~\bibnamefont {Xu}}, \bibinfo {author} {\bibfnamefont
  {M.}~\bibnamefont {Knap}}, \bibinfo {author} {\bibfnamefont {E.}~\bibnamefont
  {Demler}}, \bibinfo {author} {\bibfnamefont {F.}~\bibnamefont {Grusdt}},
  \bibinfo {author} {\bibfnamefont {M.}~\bibnamefont {Greiner}}, \ and\
  \bibinfo {author} {\bibfnamefont {D.}~\bibnamefont {Greif}},\ }\href
  {\doibase 10.1126/science.aav3587} {\bibfield  {journal} {\bibinfo  {journal}
  {Science}\ }\textbf {\bibinfo {volume} {365}},\ \bibinfo {pages} {251}
  (\bibinfo {year} {2019})}\BibitemShut {NoStop}%
\bibitem [{\citenamefont {Vijayan}\ \emph {et~al.}(2020)\citenamefont
  {Vijayan}, \citenamefont {Sompet}, \citenamefont {Salomon}, \citenamefont
  {Koepsell}, \citenamefont {Hirthe}, \citenamefont {Bohrdt}, \citenamefont
  {Grusdt}, \citenamefont {Bloch},\ and\ \citenamefont {Gross}}]{Vijayan20}%
  \BibitemOpen
  \bibfield  {author} {\bibinfo {author} {\bibfnamefont {J.}~\bibnamefont
  {Vijayan}}, \bibinfo {author} {\bibfnamefont {P.}~\bibnamefont {Sompet}},
  \bibinfo {author} {\bibfnamefont {G.}~\bibnamefont {Salomon}}, \bibinfo
  {author} {\bibfnamefont {J.}~\bibnamefont {Koepsell}}, \bibinfo {author}
  {\bibfnamefont {S.}~\bibnamefont {Hirthe}}, \bibinfo {author} {\bibfnamefont
  {A.}~\bibnamefont {Bohrdt}}, \bibinfo {author} {\bibfnamefont
  {F.}~\bibnamefont {Grusdt}}, \bibinfo {author} {\bibfnamefont
  {I.}~\bibnamefont {Bloch}}, \ and\ \bibinfo {author} {\bibfnamefont
  {C.}~\bibnamefont {Gross}},\ }\href {\doibase 10.1126/science.aay2354}
  {\bibfield  {journal} {\bibinfo  {journal} {Science}\ }\textbf {\bibinfo
  {volume} {367}},\ \bibinfo {pages} {186} (\bibinfo {year}
  {2020})}\BibitemShut {NoStop}%
\bibitem [{\citenamefont {Eckardt}\ \emph {et~al.}(2005)\citenamefont
  {Eckardt}, \citenamefont {Weiss},\ and\ \citenamefont
  {Holthaus}}]{Eckardt2005}%
  \BibitemOpen
  \bibfield  {author} {\bibinfo {author} {\bibfnamefont {A.}~\bibnamefont
  {Eckardt}}, \bibinfo {author} {\bibfnamefont {C.}~\bibnamefont {Weiss}}, \
  and\ \bibinfo {author} {\bibfnamefont {M.}~\bibnamefont {Holthaus}},\ }\href
  {\doibase 10.1103/PhysRevLett.95.260404} {\bibfield  {journal} {\bibinfo
  {journal} {Phys. Rev. Lett.}\ }\textbf {\bibinfo {volume} {95}},\ \bibinfo
  {pages} {260404} (\bibinfo {year} {2005})}\BibitemShut {NoStop}%
\bibitem [{\citenamefont {Struck}\ \emph {et~al.}(2011)\citenamefont {Struck},
  \citenamefont {{\"{O}}lschl{\"{a}}ger}, \citenamefont {{Le Targat}},
  \citenamefont {Soltan-Panahi}, \citenamefont {Eckardt}, \citenamefont
  {Lewenstein}, \citenamefont {Windpassinger},\ and\ \citenamefont
  {Sengstock}}]{Struck2011}%
  \BibitemOpen
  \bibfield  {author} {\bibinfo {author} {\bibfnamefont {J.}~\bibnamefont
  {Struck}}, \bibinfo {author} {\bibfnamefont {C.}~\bibnamefont
  {{\"{O}}lschl{\"{a}}ger}}, \bibinfo {author} {\bibfnamefont {R.}~\bibnamefont
  {{Le Targat}}}, \bibinfo {author} {\bibfnamefont {P.}~\bibnamefont
  {Soltan-Panahi}}, \bibinfo {author} {\bibfnamefont {A.}~\bibnamefont
  {Eckardt}}, \bibinfo {author} {\bibfnamefont {M.}~\bibnamefont {Lewenstein}},
  \bibinfo {author} {\bibfnamefont {P.}~\bibnamefont {Windpassinger}}, \ and\
  \bibinfo {author} {\bibfnamefont {K.}~\bibnamefont {Sengstock}},\ }\href
  {\doibase 10.1126/science.1207239} {\bibfield  {journal} {\bibinfo  {journal}
  {Science}\ }\textbf {\bibinfo {volume} {333}},\ \bibinfo {pages} {996}
  (\bibinfo {year} {2011})}\BibitemShut {NoStop}%
\bibitem [{\citenamefont {Parker}\ \emph {et~al.}(2013)\citenamefont {Parker},
  \citenamefont {Ha},\ and\ \citenamefont {Chin}}]{Parker2013}%
  \BibitemOpen
  \bibfield  {author} {\bibinfo {author} {\bibfnamefont {C.~V.}\ \bibnamefont
  {Parker}}, \bibinfo {author} {\bibfnamefont {L.~C.}\ \bibnamefont {Ha}}, \
  and\ \bibinfo {author} {\bibfnamefont {C.}~\bibnamefont {Chin}},\ }\href
  {\doibase 10.1038/nphys2789} {\bibfield  {journal} {\bibinfo  {journal}
  {Nature Phys.}\ }\textbf {\bibinfo {volume} {9}},\ \bibinfo {pages} {769}
  (\bibinfo {year} {2013})}\BibitemShut {NoStop}%
\bibitem [{\citenamefont {Reitter}\ \emph {et~al.}(2017)\citenamefont
  {Reitter}, \citenamefont {N{\"{a}}ger}, \citenamefont {Wintersperger},
  \citenamefont {Str{\"{a}}ter}, \citenamefont {Bloch}, \citenamefont
  {Eckardt},\ and\ \citenamefont {Schneider}}]{Reitter2017}%
  \BibitemOpen
  \bibfield  {author} {\bibinfo {author} {\bibfnamefont {M.}~\bibnamefont
  {Reitter}}, \bibinfo {author} {\bibfnamefont {J.}~\bibnamefont
  {N{\"{a}}ger}}, \bibinfo {author} {\bibfnamefont {K.}~\bibnamefont
  {Wintersperger}}, \bibinfo {author} {\bibfnamefont {C.}~\bibnamefont
  {Str{\"{a}}ter}}, \bibinfo {author} {\bibfnamefont {I.}~\bibnamefont
  {Bloch}}, \bibinfo {author} {\bibfnamefont {A.}~\bibnamefont {Eckardt}}, \
  and\ \bibinfo {author} {\bibfnamefont {U.}~\bibnamefont {Schneider}},\ }\href
  {\doibase 10.1103/PhysRevLett.119.200402} {\bibfield  {journal} {\bibinfo
  {journal} {Phys. Rev. Lett.}\ }\textbf {\bibinfo {volume} {119}},\ \bibinfo
  {pages} {200402} (\bibinfo {year} {2017})}\BibitemShut {NoStop}%
\bibitem [{foo()}]{footnoteJabsorbed}%
  \BibitemOpen
  \href@noop {} {}\bibinfo {note} {We absorbed a factor $J$ into the rates
  $\alpha_{ijk}$ as compared with other works, e.g., Refs.~\cite{Ammon1995,
  Coulthard2018}}\BibitemShut {NoStop}%
\bibitem [{\citenamefont {Al-Assam}\ \emph {et~al.}(2017)\citenamefont
  {Al-Assam}, \citenamefont {Clark},\ and\ \citenamefont
  {Jaksch}}]{AlAssam2017}%
  \BibitemOpen
  \bibfield  {author} {\bibinfo {author} {\bibfnamefont {S.}~\bibnamefont
  {Al-Assam}}, \bibinfo {author} {\bibfnamefont {S.~R.}\ \bibnamefont {Clark}},
  \ and\ \bibinfo {author} {\bibfnamefont {D.}~\bibnamefont {Jaksch}},\ }\href
  {\doibase 10.1088/1742-5468/aa7df3} {\bibfield  {journal} {\bibinfo
  {journal} {J. Stat. Mech.: Theor. Exp.}\ }\textbf {\bibinfo {volume}
  {2017}},\ \bibinfo {pages} {093102} (\bibinfo {year} {2017})}\BibitemShut
  {NoStop}%
\bibitem [{\citenamefont {White}(1992)}]{White1992}%
  \BibitemOpen
  \bibfield  {author} {\bibinfo {author} {\bibfnamefont {S.~R.}\ \bibnamefont
  {White}},\ }\href {\doibase 10.1103/PhysRevLett.69.2863} {\bibfield
  {journal} {\bibinfo  {journal} {Phys. Rev. Lett.}\ }\textbf {\bibinfo
  {volume} {69}},\ \bibinfo {pages} {2863} (\bibinfo {year}
  {1992})}\BibitemShut {NoStop}%
\bibitem [{\citenamefont {Schollwöck}(2011)}]{Schollwock2010}%
  \BibitemOpen
  \bibfield  {author} {\bibinfo {author} {\bibfnamefont {U.}~\bibnamefont
  {Schollwöck}},\ }\href {\doibase https://doi.org/10.1016/j.aop.2010.09.012}
  {\bibfield  {journal} {\bibinfo  {journal} {Ann. Phys. (N.Y.)}\ }\textbf
  {\bibinfo {volume} {326}},\ \bibinfo {pages} {96} (\bibinfo {year}
  {2011})}\BibitemShut {NoStop}%
\bibitem [{\citenamefont {Vidal}(2004)}]{Vidal2004}%
  \BibitemOpen
  \bibfield  {author} {\bibinfo {author} {\bibfnamefont {G.}~\bibnamefont
  {Vidal}},\ }\href {\doibase 10.1103/PhysRevLett.93.040502} {\bibfield
  {journal} {\bibinfo  {journal} {Phys. Rev. Lett.}\ }\textbf {\bibinfo
  {volume} {93}},\ \bibinfo {pages} {040502} (\bibinfo {year}
  {2004})}\BibitemShut {NoStop}%
\bibitem [{Kf2()}]{Kf225footnote}%
  \BibitemOpen
  \href@noop {} {}\bibinfo {note} {We checked this with DMRG calculations as
  was done in Ref.~\cite{Coulthard2018}. We also note that the parameters for
  which phase separation occurs depend on the fermion density.}\BibitemShut
  {Stop}%
\bibitem [{\citenamefont {Ott}(2016)}]{Ott2016}%
  \BibitemOpen
  \bibfield  {author} {\bibinfo {author} {\bibfnamefont {H.}~\bibnamefont
  {Ott}},\ }\href {\doibase 10.1088/0034-4885/79/5/054401} {\bibfield
  {journal} {\bibinfo  {journal} {Rep. Prog. Phys.}\ }\textbf {\bibinfo
  {volume} {79}},\ \bibinfo {pages} {054401} (\bibinfo {year}
  {2016})}\BibitemShut {NoStop}%
\bibitem [{\citenamefont {Bakr}\ \emph {et~al.}(2009)\citenamefont {Bakr},
  \citenamefont {Gillen}, \citenamefont {Peng}, \citenamefont {F{\"{o}}lling},\
  and\ \citenamefont {Greiner}}]{Bakr2009}%
  \BibitemOpen
  \bibfield  {author} {\bibinfo {author} {\bibfnamefont {W.~S.}\ \bibnamefont
  {Bakr}}, \bibinfo {author} {\bibfnamefont {J.~I.}\ \bibnamefont {Gillen}},
  \bibinfo {author} {\bibfnamefont {A.}~\bibnamefont {Peng}}, \bibinfo {author}
  {\bibfnamefont {S.}~\bibnamefont {F{\"{o}}lling}}, \ and\ \bibinfo {author}
  {\bibfnamefont {M.}~\bibnamefont {Greiner}},\ }\href {\doibase
  10.1038/nature08482} {\bibfield  {journal} {\bibinfo  {journal} {Nature}\
  }\textbf {\bibinfo {volume} {462}},\ \bibinfo {pages} {74} (\bibinfo {year}
  {2009})}\BibitemShut {NoStop}%
\bibitem [{\citenamefont {Sherson}\ \emph {et~al.}(2010)\citenamefont
  {Sherson}, \citenamefont {Weitenberg}, \citenamefont {Endres}, \citenamefont
  {Cheneau}, \citenamefont {Bloch},\ and\ \citenamefont {Kuhr}}]{Sherson2010}%
  \BibitemOpen
  \bibfield  {author} {\bibinfo {author} {\bibfnamefont {J.~F.}\ \bibnamefont
  {Sherson}}, \bibinfo {author} {\bibfnamefont {C.}~\bibnamefont {Weitenberg}},
  \bibinfo {author} {\bibfnamefont {M.}~\bibnamefont {Endres}}, \bibinfo
  {author} {\bibfnamefont {M.}~\bibnamefont {Cheneau}}, \bibinfo {author}
  {\bibfnamefont {I.}~\bibnamefont {Bloch}}, \ and\ \bibinfo {author}
  {\bibfnamefont {S.}~\bibnamefont {Kuhr}},\ }\href {\doibase
  10.1038/nature09378} {\bibfield  {journal} {\bibinfo  {journal} {Nature}\
  }\textbf {\bibinfo {volume} {467}},\ \bibinfo {pages} {68} (\bibinfo {year}
  {2010})}\BibitemShut {NoStop}%
\bibitem [{\citenamefont {Weitenberg}\ \emph {et~al.}(2011)\citenamefont
  {Weitenberg}, \citenamefont {Endres}, \citenamefont {Sherson}, \citenamefont
  {Cheneau}, \citenamefont {Schau}, \citenamefont {Fukuhara}, \citenamefont
  {Bloch},\ and\ \citenamefont {Kuhr}}]{Weitenberg2011}%
  \BibitemOpen
  \bibfield  {author} {\bibinfo {author} {\bibfnamefont {C.}~\bibnamefont
  {Weitenberg}}, \bibinfo {author} {\bibfnamefont {M.}~\bibnamefont {Endres}},
  \bibinfo {author} {\bibfnamefont {J.~F.}\ \bibnamefont {Sherson}}, \bibinfo
  {author} {\bibfnamefont {M.}~\bibnamefont {Cheneau}}, \bibinfo {author}
  {\bibfnamefont {P.}~\bibnamefont {Schau}}, \bibinfo {author} {\bibfnamefont
  {T.}~\bibnamefont {Fukuhara}}, \bibinfo {author} {\bibfnamefont
  {I.}~\bibnamefont {Bloch}}, \ and\ \bibinfo {author} {\bibfnamefont
  {S.}~\bibnamefont {Kuhr}},\ }\href {\doibase 10.1038/nature09827} {\bibfield
  {journal} {\bibinfo  {journal} {Nature}\ }\textbf {\bibinfo {volume} {471}},\
  \bibinfo {pages} {319} (\bibinfo {year} {2011})}\BibitemShut {NoStop}%
\bibitem [{\citenamefont {D'Alessio}\ and\ \citenamefont
  {Rigol}(2014)}]{DAlessio2014}%
  \BibitemOpen
  \bibfield  {author} {\bibinfo {author} {\bibfnamefont {L.}~\bibnamefont
  {D'Alessio}}\ and\ \bibinfo {author} {\bibfnamefont {M.}~\bibnamefont
  {Rigol}},\ }\href {\doibase 10.1103/PhysRevX.4.041048} {\bibfield  {journal}
  {\bibinfo  {journal} {Phys. Rev. X}\ }\textbf {\bibinfo {volume} {4}},\
  \bibinfo {pages} {041048} (\bibinfo {year} {2014})}\BibitemShut {NoStop}%
\bibitem [{\citenamefont {Lazarides}\ \emph {et~al.}(2014)\citenamefont
  {Lazarides}, \citenamefont {Das},\ and\ \citenamefont
  {Moessner}}]{Lazarides2014pre}%
  \BibitemOpen
  \bibfield  {author} {\bibinfo {author} {\bibfnamefont {A.}~\bibnamefont
  {Lazarides}}, \bibinfo {author} {\bibfnamefont {A.}~\bibnamefont {Das}}, \
  and\ \bibinfo {author} {\bibfnamefont {R.}~\bibnamefont {Moessner}},\ }\href
  {\doibase 10.1103/PhysRevE.90.012110} {\bibfield  {journal} {\bibinfo
  {journal} {Phys. Rev. E}\ }\textbf {\bibinfo {volume} {90}},\ \bibinfo
  {pages} {012110} (\bibinfo {year} {2014})}\BibitemShut {NoStop}%
\bibitem [{\citenamefont {Genske}\ and\ \citenamefont
  {Rosch}(2015)}]{Genske2015}%
  \BibitemOpen
  \bibfield  {author} {\bibinfo {author} {\bibfnamefont {M.}~\bibnamefont
  {Genske}}\ and\ \bibinfo {author} {\bibfnamefont {A.}~\bibnamefont {Rosch}},\
  }\href {\doibase 10.1103/PhysRevA.92.062108} {\bibfield  {journal} {\bibinfo
  {journal} {Phys. Rev. A}\ }\textbf {\bibinfo {volume} {92}},\ \bibinfo
  {pages} {062108} (\bibinfo {year} {2015})}\BibitemShut {NoStop}%
\bibitem [{\citenamefont {Herrmann}\ \emph {et~al.}(2017)\citenamefont
  {Herrmann}, \citenamefont {Murakami}, \citenamefont {Eckstein},\ and\
  \citenamefont {Werner}}]{Herrmann2017}%
  \BibitemOpen
  \bibfield  {author} {\bibinfo {author} {\bibfnamefont {A.}~\bibnamefont
  {Herrmann}}, \bibinfo {author} {\bibfnamefont {Y.}~\bibnamefont {Murakami}},
  \bibinfo {author} {\bibfnamefont {M.}~\bibnamefont {Eckstein}}, \ and\
  \bibinfo {author} {\bibfnamefont {P.}~\bibnamefont {Werner}},\ }\href
  {\doibase 10.1209/0295-5075/120/57001} {\bibfield  {journal} {\bibinfo
  {journal} {EPL}\ }\textbf {\bibinfo {volume} {120}},\ \bibinfo {pages}
  {57001} (\bibinfo {year} {2017})}\BibitemShut {NoStop}%
\bibitem [{\citenamefont {Weidinger}\ and\ \citenamefont
  {Knap}(2017)}]{Weidinger2017}%
  \BibitemOpen
  \bibfield  {author} {\bibinfo {author} {\bibfnamefont {S.~A.}\ \bibnamefont
  {Weidinger}}\ and\ \bibinfo {author} {\bibfnamefont {M.}~\bibnamefont
  {Knap}},\ }\href {\doibase 10.1038/srep45382} {\bibfield  {journal} {\bibinfo
   {journal} {Sci. Rep.}\ }\textbf {\bibinfo {volume} {7}},\ \bibinfo {pages}
  {45382} (\bibinfo {year} {2017})}\BibitemShut {NoStop}%
\bibitem [{\citenamefont {Trotzky}\ \emph {et~al.}(2010)\citenamefont
  {Trotzky}, \citenamefont {Chen}, \citenamefont {Schnorrberger}, \citenamefont
  {Cheinet},\ and\ \citenamefont {Bloch}}]{Trotzky2010}%
  \BibitemOpen
  \bibfield  {author} {\bibinfo {author} {\bibfnamefont {S.}~\bibnamefont
  {Trotzky}}, \bibinfo {author} {\bibfnamefont {Y.-A.}\ \bibnamefont {Chen}},
  \bibinfo {author} {\bibfnamefont {U.}~\bibnamefont {Schnorrberger}}, \bibinfo
  {author} {\bibfnamefont {P.}~\bibnamefont {Cheinet}}, \ and\ \bibinfo
  {author} {\bibfnamefont {I.}~\bibnamefont {Bloch}},\ }\href {\doibase
  10.1103/PhysRevLett.105.265303} {\bibfield  {journal} {\bibinfo  {journal}
  {Phys. Rev. Lett.}\ }\textbf {\bibinfo {volume} {105}},\ \bibinfo {pages}
  {265303} (\bibinfo {year} {2010})}\BibitemShut {NoStop}%
\bibitem [{arc()}]{arclink}%
  \BibitemOpen
  \href@noop {} {}\bibinfo {note}
  {Http://dx.doi.org/10.5281/zenodo.22558}\BibitemShut {NoStop}%
\bibitem [{\citenamefont {Shirley}(1965)}]{Shirley1965}%
  \BibitemOpen
  \bibfield  {author} {\bibinfo {author} {\bibfnamefont {J.~H.}\ \bibnamefont
  {Shirley}},\ }\href {\doibase 10.1103/PhysRev.138.B979} {\bibfield  {journal}
  {\bibinfo  {journal} {Phys. Rev.}\ }\textbf {\bibinfo {volume} {138}},\
  \bibinfo {pages} {B979} (\bibinfo {year} {1965})}\BibitemShut {NoStop}%
\bibitem [{\citenamefont {Dunlap}\ and\ \citenamefont
  {Kenkre}(1986)}]{Dunlap1986}%
  \BibitemOpen
  \bibfield  {author} {\bibinfo {author} {\bibfnamefont {D.~H.}\ \bibnamefont
  {Dunlap}}\ and\ \bibinfo {author} {\bibfnamefont {V.~M.}\ \bibnamefont
  {Kenkre}},\ }\href {\doibase 10.1103/PhysRevB.34.3625} {\bibfield  {journal}
  {\bibinfo  {journal} {Phys. Rev. B}\ }\textbf {\bibinfo {volume} {34}},\
  \bibinfo {pages} {3625} (\bibinfo {year} {1986})}\BibitemShut {NoStop}%
\bibitem [{\citenamefont {Mori}\ \emph {et~al.}(2016)\citenamefont {Mori},
  \citenamefont {Kuwahara},\ and\ \citenamefont {Saito}}]{Mori2016}%
  \BibitemOpen
  \bibfield  {author} {\bibinfo {author} {\bibfnamefont {T.}~\bibnamefont
  {Mori}}, \bibinfo {author} {\bibfnamefont {T.}~\bibnamefont {Kuwahara}}, \
  and\ \bibinfo {author} {\bibfnamefont {K.}~\bibnamefont {Saito}},\ }\href
  {\doibase 10.1103/PhysRevLett.116.120401} {\bibfield  {journal} {\bibinfo
  {journal} {Phys. Rev. Lett.}\ }\textbf {\bibinfo {volume} {116}},\ \bibinfo
  {pages} {120401} (\bibinfo {year} {2016})}\BibitemShut {NoStop}%
\bibitem [{\citenamefont {Canovi}\ \emph {et~al.}(2016)\citenamefont {Canovi},
  \citenamefont {Kollar},\ and\ \citenamefont {Eckstein}}]{Canovi2016}%
  \BibitemOpen
  \bibfield  {author} {\bibinfo {author} {\bibfnamefont {E.}~\bibnamefont
  {Canovi}}, \bibinfo {author} {\bibfnamefont {M.}~\bibnamefont {Kollar}}, \
  and\ \bibinfo {author} {\bibfnamefont {M.}~\bibnamefont {Eckstein}},\ }\href
  {\doibase 10.1103/PhysRevE.93.012130} {\bibfield  {journal} {\bibinfo
  {journal} {Phys. Rev. E}\ }\textbf {\bibinfo {volume} {93}},\ \bibinfo
  {pages} {012130} (\bibinfo {year} {2016})}\BibitemShut {NoStop}%
\bibitem [{\citenamefont {Abanin}\ \emph {et~al.}(2017)\citenamefont {Abanin},
  \citenamefont {De~Roeck}, \citenamefont {Ho},\ and\ \citenamefont
  {Huveneers}}]{Abanin2017}%
  \BibitemOpen
  \bibfield  {author} {\bibinfo {author} {\bibfnamefont {D.}~\bibnamefont
  {Abanin}}, \bibinfo {author} {\bibfnamefont {W.}~\bibnamefont {De~Roeck}},
  \bibinfo {author} {\bibfnamefont {W.~W.}\ \bibnamefont {Ho}}, \ and\ \bibinfo
  {author} {\bibfnamefont {F.}~\bibnamefont {Huveneers}},\ }\href {\doibase
  10.1007/s00220-017-2930-x} {\bibfield  {journal} {\bibinfo  {journal}
  {Commun. Math. Phys.}\ }\textbf {\bibinfo {volume} {354}},\ \bibinfo {pages}
  {809} (\bibinfo {year} {2017})}\BibitemShut {NoStop}%
\bibitem [{\citenamefont {Mur-Petit}\ \emph {et~al.}(2018)\citenamefont
  {Mur-Petit}, \citenamefont {Rela{\~n}o}, \citenamefont {Molina},\ and\
  \citenamefont {Jaksch}}]{MurPetit2018}%
  \BibitemOpen
  \bibfield  {author} {\bibinfo {author} {\bibfnamefont {J.}~\bibnamefont
  {Mur-Petit}}, \bibinfo {author} {\bibfnamefont {A.}~\bibnamefont
  {Rela{\~n}o}}, \bibinfo {author} {\bibfnamefont {R.~A.}\ \bibnamefont
  {Molina}}, \ and\ \bibinfo {author} {\bibfnamefont {D.}~\bibnamefont
  {Jaksch}},\ }\href {\doibase 10.1038/s41467-018-04407-1} {\bibfield
  {journal} {\bibinfo  {journal} {Nature Commun.}\ }\textbf {\bibinfo {volume}
  {9}},\ \bibinfo {pages} {2006} (\bibinfo {year} {2018})}\BibitemShut
  {NoStop}%
\bibitem [{\citenamefont {Tindall}\ \emph {et~al.}(2019)\citenamefont
  {Tindall}, \citenamefont {Bu{\v{c}}a}, \citenamefont {Coulthard},\ and\
  \citenamefont {Jaksch}}]{Tindall2019}%
  \BibitemOpen
  \bibfield  {author} {\bibinfo {author} {\bibfnamefont {J.}~\bibnamefont
  {Tindall}}, \bibinfo {author} {\bibfnamefont {B.}~\bibnamefont {Bu{\v{c}}a}},
  \bibinfo {author} {\bibfnamefont {J.~R.}\ \bibnamefont {Coulthard}}, \ and\
  \bibinfo {author} {\bibfnamefont {D.}~\bibnamefont {Jaksch}},\ }\href
  {\doibase 10.1103/PhysRevLett.123.030603} {\bibfield  {journal} {\bibinfo
  {journal} {Phys. Rev. Lett.}\ }\textbf {\bibinfo {volume} {123}},\ \bibinfo
  {pages} {030603} (\bibinfo {year} {2019})}\BibitemShut {NoStop}%
\bibitem [{\citenamefont {Ammon}\ \emph {et~al.}(1995)\citenamefont {Ammon},
  \citenamefont {Troyer},\ and\ \citenamefont {Tsunetsugu}}]{Ammon1995}%
  \BibitemOpen
  \bibfield  {author} {\bibinfo {author} {\bibfnamefont {B.}~\bibnamefont
  {Ammon}}, \bibinfo {author} {\bibfnamefont {M.}~\bibnamefont {Troyer}}, \
  and\ \bibinfo {author} {\bibfnamefont {H.}~\bibnamefont {Tsunetsugu}},\
  }\href {\doibase 10.1103/PhysRevB.52.629} {\bibfield  {journal} {\bibinfo
  {journal} {Phys. Rev. B}\ }\textbf {\bibinfo {volume} {52}},\ \bibinfo
  {pages} {629} (\bibinfo {year} {1995})}\BibitemShut {NoStop}%
\bibitem [{\citenamefont {Coulthard}\ \emph {et~al.}(2018)\citenamefont
  {Coulthard}, \citenamefont {Clark},\ and\ \citenamefont
  {Jaksch}}]{Coulthard2018}%
  \BibitemOpen
  \bibfield  {author} {\bibinfo {author} {\bibfnamefont {J.~R.}\ \bibnamefont
  {Coulthard}}, \bibinfo {author} {\bibfnamefont {S.~R.}\ \bibnamefont
  {Clark}}, \ and\ \bibinfo {author} {\bibfnamefont {D.}~\bibnamefont
  {Jaksch}},\ }\href {\doibase 10.1103/PhysRevB.98.035116} {\bibfield
  {journal} {\bibinfo  {journal} {Phys. Rev. B}\ }\textbf {\bibinfo {volume}
  {98}},\ \bibinfo {pages} {035116} (\bibinfo {year} {2018})}\BibitemShut
  {NoStop}%
\end{thebibliography}%


\begin{thebibliography}{0}%
\makeatletter
\providecommand \@ifxundefined [1]{%
 \@ifx{#1\undefined}
}%
\providecommand \@ifnum [1]{%
 \ifnum #1\expandafter \@firstoftwo
 \else \expandafter \@secondoftwo
 \fi
}%
\providecommand \@ifx [1]{%
 \ifx #1\expandafter \@firstoftwo
 \else \expandafter \@secondoftwo
 \fi
}%
\providecommand \natexlab [1]{#1}%
\providecommand \enquote  [1]{``#1''}%
\providecommand \bibnamefont  [1]{#1}%
\providecommand \bibfnamefont [1]{#1}%
\providecommand \citenamefont [1]{#1}%
\providecommand \href@noop [0]{\@secondoftwo}%
\providecommand \href [0]{\begingroup \@sanitize@url \@href}%
\providecommand \@href[1]{\@@startlink{#1}\@@href}%
\providecommand \@@href[1]{\endgroup#1\@@endlink}%
\providecommand \@sanitize@url [0]{\catcode `\\12\catcode `\$12\catcode
  `\&12\catcode `\#12\catcode `\^12\catcode `\_12\catcode `\%12\relax}%
\providecommand \@@startlink[1]{}%
\providecommand \@@endlink[0]{}%
\providecommand \url  [0]{\begingroup\@sanitize@url \@url }%
\providecommand \@url [1]{\endgroup\@href {#1}{\urlprefix }}%
\providecommand \urlprefix  [0]{URL }%
\providecommand \Eprint [0]{\href }%
\providecommand \doibase [0]{http://dx.doi.org/}%
\providecommand \selectlanguage [0]{\@gobble}%
\providecommand \bibinfo  [0]{\@secondoftwo}%
\providecommand \bibfield  [0]{\@secondoftwo}%
\providecommand \translation [1]{[#1]}%
\providecommand \BibitemOpen [0]{}%
\providecommand \bibitemStop [0]{}%
\providecommand \bibitemNoStop [0]{.\EOS\space}%
\providecommand \EOS [0]{\spacefactor3000\relax}%
\providecommand \BibitemShut  [1]{\csname bibitem#1\endcsname}%
\let\auto@bib@innerbib\@empty
\end{thebibliography}%
\end{document}